\documentclass[useAMS,usenatbib]{mn2e}
\usepackage{amsmath,graphicx,txfonts}
\defcitealias{har02}{Paper I}
\defcitealias{har06}{Paper II}
\defcitealias{mcl07}{Paper IV}

\title[Globular clusters in NGC 5128]
{Structural parameters for globular clusters in NGC 5128. III. \\
ACS surface-brightness profiles and model fits} 

\author[D. E. McLaughlin et al.]
{%
Dean E. McLaughlin,$^{1}$\thanks{E-mail:
       dem@astro.keele.ac.uk (DEM);
       pbarmby@uwo.ca (PB);
       harris@physics.mcmaster.ca (WEH);
       dforbes@astro.swin.edu.au (DAF);
       glharris@astro.waterloo.ca (GLHH)}
Pauline Barmby,$^{2}$\footnotemark[1]
William E. Harris,$^{3}$\footnotemark[1]
Duncan A. Forbes$^{4}$\footnotemark[1]
\newauthor
and Gretchen L. H. Harris$^{5}$\footnotemark[1]
\\
$^{1}$ School of Physical and Geographical Sciences,
       Lennard-Jones Laboratories,
       Keele University, Staffordshire ST5 5BG \\
$^{2}$ Department of Physics and Astronomy,
       University of Western Ontario,
       London, ON N6A 3K7 \ Canada \\
$^{3}$ Department of Physics and Astronomy,
       McMaster University,
       Hamilton, ON L8S 4M1 \ Canada \\
$^{4}$ Centre for Astrophysics and Supercomputing,
       Swinburne University,
       Hawthorn, VIC 3122 \ Australia \\
$^{5}$ Department of Physics and Astronomy,
       University of Waterloo,
       Waterloo, ON N2L 3G1 \ Canada
}

\date{\large MNRAS, {\bf \sl in press}}
 
\pagerange{\pageref{firstpage}--\pageref{lastpage}}
\pubyear{2007}
 
\begin{document}

\maketitle
\label{firstpage}

\begin{abstract}

We present internal surface-brightness profiles, based on HST/ACS
imaging in the $F606W$ bandpass, for 131 globular cluster (GC)
candidates with luminosities $L\simeq 10^4$--$3\times10^6\,L_\odot$ 
in the giant elliptical galaxy NGC 5128. Several structural models
are fit to the profile of each cluster and combined with mass-to-light
ratios from population-synthesis models, to derive a catalogue of
fundamental structural and dynamical parameters parallel in form to
the catalogues recently produced by 
\citeauthor{mcl05} and by \citeauthor{barmby07} for GCs and massive young star
clusters in Local Group galaxies. As part of this, we provide corrected and
extended parameter estimates for another 18 clusters in NGC 5128, which we
observed previously. We show that, like GCs in the Milky Way and some
of its satellites, the majority of globulars in NGC 5128 are well
fit by isotropic \citeauthor{wil75} models, which have intrinsically more
distended envelope structures than the standard \citeauthor{king66} lowered
isothermal spheres. We use our models to predict internal velocity
dispersions for every cluster in our sample. These predictions
agree well in general with the observed dispersions in a small number
of clusters for which spectroscopic data are available. In a
subsequent paper, we use these results to investigate scaling
relations for GCs in NGC 5128.

\end{abstract}


\section{Introduction}
\label{sec:intro}

The spatial structures and internal stellar kinematics of old globular
clusters (GCs) contain information on both their initial conditions and their
dynamical evolution over a Hubble time. An efficient way of extracting this
information is to fit detailed models to the surface-brightness profiles and
(where available) velocity-dispersion data of individual clusters, and then
to look for possible correlations between the physical properties
of large numbers of GCs. This has been done for most of
the $\simeq\! 150$ globulars in the Milky Way, yielding
comprehensive catalogues of their structural and dynamical parameters
\citep{djo93,pry93,har96,mcl05}. Much of this work has traditionally
started from the assumption that individual globulars are well described
by the classic \citet{king66} models of single-mass, isotropic, modified
isothermal spheres, although recently alternative models have also been
employed \citep{mcl05}. 

Explorations of numerous scaling relations and interdependences between
the properties of Galactic GCs \citep[e.g.,][]{djo94} have led to
the definition of a fundamental plane for globulars that is analogous to but
physically distinct from that for early-type galaxies and bulges
\citep{djo95,bur97,mcl00}. Understanding the GC fundamental plane in
full detail is still not a completely solved problem, but important
advances have been made in recent years as it has become possible to measure
the internal properties of GCs in many other galaxies. High-resolution
Hubble Space Telescope (HST) imaging has been used to fit \citet{king66}
and other models to the surface-brightness profiles of scores of globulars in
the Large and Small Magellanic Clouds and the Fornax dwarf spheroidal
\citep{mg03a,mg03b,mg03c,mcl05}, M31 \citep[e.g.,][]{barmby02,barmby07}, M33
\citep{lar02}, and the giant elliptical galaxy NGC 5128 = Centaurus A
(\citealt{hol99,har02}). Internal velocity dispersions and dynamical mass
estimates are also available for smaller but growing numbers of GCs in these
systems \citep{djo97,dub97,lar02,marho04,rej07}.

Here we add to this database with structural measurements of 131 GCs
in NGC 5128. This galaxy is an attractive target for such studies in part
because of its large GC population, estimated by \citet{har06} at
${\cal N}_{\rm GC} \simeq 1500$. It thus contains many objects at the high end
of the star-cluster mass range ($10^6$--$10^7\ M_{\odot}$), which is largely
unprobed in the ten-times smaller GC system of the Milky Way but where it is 
increasingly suggested that the cluster population may encompass a variety of
objects including classic globulars, the compact nuclei of dwarf elliptical
galaxies, and the new class of ultra-compact dwarf galaxies
\citep{hilker99,drink00,hasegan05}. In addition, the proximity of NGC 5128
($D=3.8$~Mpc; see below) makes it possible to resolve
the core radii as well as just the half-light radii of GCs over nearly
their full mass range ($M\ga 10^4\ M_\odot$), and thus to fit them
rigorously with detailed structural models.

This paper is the third in a series of four dealing with HST observations of
GCs in NGC 5128. In \citet[][$\equiv$~\citetalias{har02}]{har02} the Space
Telescope Imaging Spectrograph (STIS) and Wide Field Planetary Camera 2
(WFPC2) aboard 
HST were used to measure surface-brightness profiles for 27 very bright GCs in
NGC 5128. \citet{king66} models were fitted to these profiles to
derive a structural fundamental plane that could be compared directly to that
of the Milky Way globulars.
In \citet[][$\equiv$~\citetalias{har06}]{har06} we published
the first results from a new HST-based survey of GCs in NGC 5128, using the
Advanced Camera for Surveys (ACS) in its Wide Field Channel (WFC) 
to image a total of 131 GC candidates at a resolution of 0\farcs05
(linear resolution $\simeq\! 0.9$ parsec). \citetalias{har06} gives the
full description of the cluster sample along with some rough overall
characteristics of the ensemble of objects. In the present paper, we
derive surface brightness profiles for all of these clusters and fit
each of them with a number of different structural models. The final paper in
this series \citep[][$\equiv$~\citetalias{mcl07}]{mcl07} uses these results
to examine a number of structural correlations for GCs in NGC 5128, which are
then compared to the globulars in the Milky Way and to various other types of
massive clusters. In related work, \citet{barmby07} present the
results of a similar ACS study of GCs in M31 and compare the fundamental
planes of old GCs in that galaxy, the Milky Way, NGC 5128, the Large and Small
Magellanic Clouds, and the Fornax dwarf spheroidal.

In the next Section we describe the steps we have taken to derive surface
brightness profiles for the GC candidates from \citetalias{har06}, to
characterise the point-spread function (PSF) that blurs the very central
regions ($R\la 2$--3~pc) of these profiles, to transform the
surface-brightness data from their native HST filter to the standard $V$
bandpass, and to estimate metallicities for the clusters from separate,
ground-based Washington photometry.

In \S\ref{sec:models}, we apply publicly available population-synthesis models
to estimate individual $V$-band mass-to-light ratios for the clusters, given
their metallicities and assuming various (old) ages. Following this, we
summarise the main properties of each of three structural models (those of
\citealt{king66}, \citealt{wil75}, and \citealt{sersic}) that we have
convolved with the ACS/WFC PSF and fit to every observed surface-brightness
profile.

Section \ref{sec:fits} gives the results of these fits and uses them to infer
a wide range of structural and dynamical parameters, including total
cluster luminosities and masses, effective and core radii and stellar
densities, concentration indices, relaxation times, total binding energies, 
predicted central velocity dispersions, and $\kappa$-space parameters for the
fundamental plane \citep{bbf92}. We present these in tables that are available
in machine-readable format either online\footnotemark
\footnotetext{See http://www.astro.keele.ac.uk/$\sim$dem/clusters.html}
or upon request from the first
author. Note that our measurements of GC luminosities and intrinsic sizes in 
particular supersede the recent estimates of \citet{vdb07}, who based his
numbers on a less detailed analysis of some very basic cluster characteristics
given in \citetalias{har06}.

In \S\ref{sec:fits} we also address the question of whether the standard
\citet{king66} model specifically gives the best possible fit to GC
surface-brightness profiles in NGC 5128. We then extend the range of
physical parameters calculated for the smaller sample of clusters previously
fitted with \citeauthor{king66} models in \citetalias{har02}, and we provide
important corrections to some of the more basic parameters (in particular, the
intrinsic central surface brightnesses) already published in that earlier
work. The results of this re-analysis are tabulated in Appendix
\ref{sec:STIStables}. 

In \S\ref{sec:apvel} we combine our structural modeling and
population-synthesis mass-to-light ratios to predict line-of-sight
velocity dispersions within a series of circular apertures with physical radii
suited to realistic observational set-ups. We compare these predictions with
spectroscopic data from \citet{marho04} and \citet{rej07} for some of our
current cluster sample. Finally, \S\ref{sec:summary} summarises the paper.

Our modeling analysis in this paper is in all respects very similar to that 
undertaken by \citet{mcl05} for a sample of 103 old GCs and 50 young massive
clusters drawn from the Milky Way, the Large and Small Magellanic Clouds, and
the Fornax dwarf spheroidal. The catalogues of structural and dynamical
properties that we produce here for GCs in NGC 5128 are likewise very close
in form and content to those in \citeauthor{mcl05}.
We have recently completed the same type of modeling and produced parallel
catalogues for a further 93 GCs in M31 \citep{barmby07}. As we mentioned above,
\citeauthor{barmby07} combine results to compare the
fundamental planes of the old GCs in all six galaxies. In \citetalias{mcl07}
\citep{mcl07} we directly compare GC structural correlations only between the
Milky Way and NGC 5128, but we also examine how they relate to other kinds of
massive star clusters.  

In all of what follows, we adopt a distance of 3.8 Mpc to
NGC 5128. This value is representative of recent measurements based on the
tip of the red-giant branch [$(m-M)_0 = 27.98 \pm 0.15$], the planetary
nebulae luminosity function [$(m-M)_0 = 27.97 \pm 0.14$],
surface-brightness fluctuations [$(m-M)_0 = 27.78 \pm 0.10$],
Mira variables [$(m-M)_0 = 27.96 \pm 0.11$], and Cepheids
[$(m-M)_0 = 27.67 \pm 0.20$]; see \citet{har99}, \citet{rej04}, and
\citet{ferr07}. The nominal average of these five, reasonably high-precision
distances is $(m-M)_0 = 27.88 \pm 0.06$, or $3.76 \pm 0.11$~Mpc. All these
methods have undergone recent calibration revisions of various kinds
\citep[cf.][]{ferr07} but the net results have been to shift the mean up or
down by amounts at the level of only $0.1$~mag. At a distance of 3.8
Mpc, 1 arcsecond is subtended by 18.4 parsec. One ACS/WFC pixel
(0\farcs05) then corresponds to 0.92~pc.

\section{Data}
\label{sec:data}

The GC sample from \citetalias{har06} consists of 62 previously known clusters
in NGC 5128, and 69 newly discovered candidates. All these objects fall
in 12 target fields imaged in the $F606W$ (``wide $V$'') band on the
ACS/WFC. We observed 16 clusters twice, since they appeared in two overlapping
target fields. These are listed in Table \ref{tab:dual}. We measured two
independent surface-brightness profiles for each of them, so that in all we
have 147 profiles for 131 distinct objects. In \S\ref{subsec:dualcomp} we use
these duplications to assess whether variations in the PSF over the ACS field
of view might have systematically affected our results.

Of the 27 GCs observed with STIS or WFPC2 in \citetalias{har02}, 9
were re-observed with the ACS/WFC for \citetalias{har06} and this paper;
these are C007, C025, C029, C032, C037, C104, C105, G221, and G293.
Our analysis of them here supersedes that in \citetalias{har02}. We
eventually fold the other 18 STIS/WFPC2 clusters into the sample for
correlations work in \citetalias{mcl07}, although with structural
parameters updated as discussed in \S\ref{subsec:STIS} and Appendix
\ref{sec:STIStables} below.

We repeatedly convert between luminosities
and masses by assigning individual $V$-band mass-to-light ratios to all
GCs in our total sample. As we describe in more detail below
(\S\ref{subsec:transmet} and \S\ref{subsec:popsyn}), to do this we first
estimate a metallicity for each cluster from its $(C-T_1)$ colour in the
Washington filter system, using a relation calibrated
against genuinely old GCs. Then, we input this and an assumed old age
(normally 13 Gyr) to a standard population-synthesis code. Spectroscopy
indicates that most of the GCs in NGC 5128 are indeed old, but a younger
(few Gyr) population cannot be ruled out at this stage. If some of the
objects in our sample are young, then the mass-to-light ratio we assign to
them would be slightly high, and all physical parameters deriving from it
would be slightly biased. One way to guard somewhat against this is
not to include exceedingly blue objects with unrealistically low inferred
metallicities (see Table \ref{tab:n5128mtol}) in detailed studies of parameter
correlations and the like.


\begin{table}
\caption{NGC 5128 clusters measured independently on two fields$^{~a}$
\label{tab:dual}}
\begin{tabular}{@{}llcll}
\hline
\multicolumn{1}{c}{Cluster} &
\multicolumn{1}{c}{Fields}  &
~~                          &
\multicolumn{1}{c}{Cluster} &
\multicolumn{1}{c}{Fields}  \\
\multicolumn{1}{c}{(1)} &
\multicolumn{1}{c}{(2)} &
~~                      &
\multicolumn{1}{c}{(1)} &
\multicolumn{1}{c}{(2)}  \\
\hline
AAT118198        & {\bf C018}, C019 & ~~ &    C171   & {\bf C007}, C025 \\
AAT120976        & C007, {\bf C025} & ~~ &    C173   & {\bf C007}, C025 \\
C007             & {\bf C007}, C025 & ~~ &    C176   & C007, {\bf C025} \\
C018             & {\bf C018}, C019 & ~~ &    F1GC20 & {\bf C007}, C025 \\
C025             & C007, {\bf C025} & ~~ &    G221   & {\bf C007}, C025 \\
C104             & {\bf C007}, C025 & ~~ &    G293   & {\bf C007}, C025 \\
C156$^{~b}$      & {\bf C018}, C019 & ~~ &    PFF021 & {\bf C003}, C030 \\
C158             & {\bf C018}, C019 & ~~ &    WHH22  & {\bf C018}, C019 \\
\hline
\end{tabular}

\medskip
$^{a}$ Boldface in column (2) denotes the field in which the cluster in column
(1) is closest to the centre of the chip. Model fits to the intensity profiles
from these images are the ones used in the correlation analyses of
\citetalias{mcl07}.

$^{b}$ Possible star; see Table \ref{tab:badstuff}.

\end{table}


\subsection{Surface-brightness profiles}
\label{subsec:SBprofs}

We have used the STSDAS ELLIPSE task to obtain $F606W$ surface-brightness
profiles for all cluster candidates from \citetalias{har06}. As part of
the same HST program (GO-10260), we also obtained ACS images for a series
of clusters in M31.  These data were reduced and modeled simultaneously with
the present sample and are discussed in \citet{barmby07}. Full details
of the surface-photometry procedures are given in that paper. Here we note that
we forced the isophote ellpticity in ELLIPSE to be identically 0 at all radii.
We thus always have circularly symmetric $I_{F606}(R)$ profiles,
which we then fit with spherical structural models. In \citetalias{har06}, we
showed that the actual ellipticities of the clusters are generally quite
small, averaging $\langle \epsilon \rangle = 0.08$ over our whole 
sample. The assumption $\epsilon \equiv 0$ for the purposes of modeling is
therefore not a significant limitation.

The raw output from ELLIPSE is in terms of counts per second per pixel,
which we convert to ${\rm cts/s}$ per square arcsecond by multiplying by
$400=(1\,{\rm px}/0\farcs05)^2$. Normally, these counts
would then be transformed to $F606W$ surface brightnesses, calibrated on the
VEGAMAG system according to (ACS Handbook)
\begin{equation}
\mu_{F606}/{\rm mag\ arcsec^{-2}} =
        26.398 - 2.5\,\log({\rm cts/s}/\sq\arcsec)\ .
\label{eq:f606calib}
\end{equation}
However, we quickly found that the average, global sky background
that was automatically subtracted from each ACS image during the
multi-drizzling in the data reduction pipeline often underestimated and
sometimes overestimated the local background level around individual
clusters. The latter case in particular led to the occurrence of pixels with
unphysical negative counts. We thus had to work immediately in terms of
linear intensity (which we chose to express right away as
$L_\odot\ {\rm pc}^{-2}$) rather than  
going through the usual logarithmic $\mu_{F606}$ and then converting to linear
quantities later. For
the solar magnitude we adopt $M_{\odot,F606}=4.64$,\footnotemark
\footnotetext{See http://www.ucolick.org/$\sim$cnaw/sun.html}
and combining this with equation (\ref{eq:f606calib}) gives
\begin{equation}
I_{F606}/L_\odot\,{\rm pc}^{-2} \simeq
      0.8427 \times ({\rm cts/s}/\sq\arcsec)\ .
\label{eq:f606int}
\end{equation}

To begin with, we obtained $I_{F606}(R)$ profiles out to $R>10\arcsec$
(more than 180 pc) for all clusters. This limit exceeds
the expected tidal radius for most of them, but such a large
field of view enables us to correct for the inaccurate average sky subtraction
in the multi-drizzling, by fitting the profiles with PSF-convolved structural
models that include a constant background term (allowed to be negative).  
We did have to exclude a number of isophotes from most of the intensity
profiles during this fitting, however.

First, at very large radii the signal from most clusters
is clearly swamped by noise, and thus we restricted all fitting
to radii $R<150$ WFC pixels, corresponding to $R<7\farcs5\simeq140$~pc.

Second, the central pixels in some of
the brighter clusters were saturated. We adopted a saturation limit of
70 cts/s/px and did not fit to any isophotal intensities brighter than
this. This corresponds to a ``good'' data range of about
$I_{F606}\le 2.36\times10^4\ L_\odot\,{\rm pc}^{-2}$, or
$\mu_{F606}\ge 15.28$ mag~arcsec$^{-2}$.

Third, at intermediate clustercentric radii there are some individual
isophotes with ELLIPSE intensities that deviate strongly from those of
immediately neighboring isophotes. To prevent such ``blips'' from skewing
the model fits, we
first ran the ELLIPSE output through a boxcar filter to make a
smoothed cluster profile, and identified points deviating from this by more
than twice their own (internal) uncertainty. Such points were not included in
the error-weighted model fitting of \S\ref{sec:fits}.

Fourth, the ELLIPSE estimates of isophotal intensities at
clustercentric radii $R<2\ {\rm px}=0\farcs1$ are all derived
from the data in the same innermost 13 pixels; but the task nevertheless
outputs brightnesses for 15 radii inside 2 px. Clearly not all of these
are statistically independent. In order to avoid having
such correlations (and excessive weighting of the central regions of the
cluster) bias our fits, we decided to include only the ELLIPSE intensities
reported for the innermost unsaturated isophotal radius, $\equiv R_{\rm min}$
(which was always at least 0.5 px), and then for
$(R_{\rm min}+0.5\ {\rm px})$, $(R_{\rm min}+1.0\ {\rm px})$,
$(R_{\rm min}+2.0\ {\rm px})$, and all $R>2.5\ {\rm px}=0\farcs125$.


\begin{table*}
\begin{minipage}{140mm}
\caption{Irregularities in some NGC 5128 cluster profiles
  \label{tab:badstuff}} 
\begin{tabular}{@{}ll}
\hline
Cluster & Comments \\
\hline
AAT111563    & Bright star nearby. Fits restricted to
                $R<5\farcs5 \simeq107$ pc.    \\
AAT113992    & Profile dips at $R=1\farcs5$ (30px), near image edge.
               Fits restricted to $R<1\arcsec\simeq20$~pc. \\
AAT118198    & Diffraction spike from nearby star. Fits restricted to
               $R<2\arcsec$.      \\
C104 ON C025 & Two nearby stars. Fits restricted to
               $R<5\arcsec \simeq 97$~pc.  \\
C118         & Bright object nearby. Region $0\farcs55<R<1\arcsec$ masked out
               of fits. \\
C134         & Bright star nearby. Fits restricted to $R\la 3\arcsec$. \\
C137         & Bright object nearby. Region $1\farcs2<R<2\farcs2$ masked
               out of fits.  \\
C154         & Cluster C153 at $R\approx2\farcs5$. Model fits restricted
               to $R<1\arcsec\simeq20$~pc. \\
C162         & Profile dips slightly around $R\simeq2\farcs5$, but fits not
               biased. No restrictions. \\
C168         & Next to very bright star. Fits restricted to $R<1\arcsec$,
               which likely misses some cluster. \\
~~~~         & Object excluded from sample for correlation analyses
               in \citetalias{mcl07}. \\
C171         & Stars at $R=3\farcs5$ and $R=6\farcs6$. Fits
               restricted to $R<6\farcs6\simeq130$~pc. \\
C174         & Star at $R=1\farcs6$. Fits restricted to
               $R<1\arcsec\simeq20$~pc.  \\
F1GC34       & Bright object at $R=2\farcs5$, and F1GC14 at $R=3\farcs3$.
               Fits restricted to $R<1\arcsec$. \\
~~~~         & Object excluded from sample for correlation analyses
	       in \citetalias{mcl07}. \\
F2GC14       & In middle of image artifact (bright star ghost?). Fits
               restricted to $R<0\farcs7\simeq13.5$~pc. \\
~~~~         & Object excluded from sample for correlation analyses
	       in \citetalias{mcl07}. \\
G170         & Profile dips due to image edge nearby. Fits restricted to
               $R<2\farcs1\simeq40$~pc. \\
WHH22        & Diffraction spike from nearby star at $R=4\farcs8$ (star also
               in sky annulus). \\
~~~~         & Fits restricted to $R<3\farcs7\simeq70$~pc. \\
\hline
C145         & Very compact, not clearly resolved. Possible star. \\
C152         & Very compact, not clearly resolved. Possible star. \\
C156         & Very compact, not clearly resolved. Possible star. \\
C177         & Very extended; half-light radius
               $R_h \simeq 6$~kpc in a \citet{king66} model fit.
               Possible background galaxy. \\
\hline
\end{tabular}
\end{minipage}
\end{table*}



\begin{table*}
\begin{minipage}{95mm}
\caption{147 $F606W$ intensity profiles for 131 GCs in NGC 5128
\label{tab:N5128sbprofs}}
\begin{tabular}{@{}ccccccl}
\hline
\multicolumn{1}{c}{Name} &
\multicolumn{1}{c}{Detector } &
\multicolumn{1}{c}{Filter} &
\multicolumn{1}{c}{$R$} &
\multicolumn{1}{c}{$I_{F606}$} &
\multicolumn{1}{c}{uncertainty} &
\multicolumn{1}{c}{Flag} \\
  &
  &
  &
\multicolumn{1}{c}{[arcsec]} &
\multicolumn{1}{c}{[$L_\odot\ {\rm pc}^{-2}$]} &
\multicolumn{1}{c}{[$L_\odot\ {\rm pc}^{-2}$]} &
~~  \\
\multicolumn{1}{c}{(1)} &
\multicolumn{1}{c}{(2)} &
\multicolumn{1}{c}{(3)} &
\multicolumn{1}{c}{(4)} &
\multicolumn{1}{c}{(5)} &
\multicolumn{1}{c}{(6)} &
\multicolumn{1}{c}{(7)} \\
\hline
AAT111563 &  WFC & $F606$ & 0.0260 &  1984.157 &  19.537 &    OK  \\
AAT111563 &  WFC & $F606$ & 0.0287 &  1963.700 &  18.380 &   DEP  \\
AAT111563 &  WFC & $F606$ & 0.0315 &  1939.018 &  17.086 &   DEP  \\
AAT111563 &  WFC & $F606$ & 0.0347 &  1907.682 &  15.106 &   DEP  \\
AAT111563 &  WFC & $F606$ & 0.0381 &  1875.142 &  13.808 &   DEP  \\
AAT111563 &  WFC & $F606$ & 0.0420 &  1836.267 &  14.456 &   DEP  \\
AAT111563 &  WFC & $F606$ & 0.0461 &  1792.333 &  15.835 &   DEP  \\
AAT111563 &  WFC & $F606$ & 0.0508 &  1741.324 &  17.563 &   DEP  \\
AAT111563 &  WFC & $F606$ & 0.0558 &  1680.360 &  16.994 &    OK  \\
AAT111563 &  WFC & $F606$ & 0.0614 &  1612.842 &  15.518 &   DEP  \\
AAT111563 &  WFC & $F606$ & 0.0676 &  1538.833 &  14.016 &   DEP  \\
\hline
\end{tabular}

\medskip
  A machine-readable version of the full Table \ref{tab:N5128sbprofs} is
  available online
  (http://www.astro.keele.ac.uk/$\sim$dem/clusters.html)
  or upon request from the first author. 
  Only a short extract from it is shown here, for guidance
  regarding its form and content.
  Note that the reported $F606W$-band intensities are calibrated on the VEGAMAG
  scale, but {\it not} corrected for extinction. In terms of {\it magnitude},
  $A_{F606}=0.308$ for the average foreground $E(B-V)=0.11$ in the direction of
  NGC 5128; see \S\ref{subsec:transmet} for more details.

\end{minipage}
\end{table*}


Finally, we looked at every intensity
profile individually and in a number of cases found irregular features,
which we masked out by hand. These are summarised in Table
\ref{tab:badstuff}.

At the end of Table \ref{tab:badstuff} we also note three GC candidates
(C145, C152, and C156) that are probably foreground stars rather than clusters
in NGC 5128, and one (C177) that is more likely a background galaxy.
We retain these objects in our catalogues of intensity profiles and
model fits, but only for completeness; none is included in any physical
analyses. 

Table \ref{tab:N5128sbprofs} gives our final, calibrated $F606W$ intensity
profiles for the 131 objects in our sample (including the duplicate
profiles for the 16 in Table \ref{tab:dual}). These are not corrected
for extinction, which we discuss below. Note that only the first few lines of
the table are reported here; an ascii file containing the full data can be
obtained from the first author or online.
Most of the columns in this table are self-explanatory (the second and third,
which are always WFC and $F606$, are present only for compatibility with
the analogous table for M31 GCs in \citealt{barmby07}, where the detector and
filter vary from cluster to cluster). The final column gives a flag for every
point, which can take one of four values: ``BAD'' if the radius is beyond our
upper limit of 7\farcs5 or the intensity value is otherwise deemed dubious
according to the third or final points just above; ``SAT'' if the isophotal
intensity above our imposed saturation limit of
$\simeq\!23,600\ L_\odot\ {\rm pc}^{-2}$; ``DEP'' if the radius is inside
$R<2\ {\rm px}=0\farcs1$ and the isophotal intensity is dependent on its
neighbours (as per the fourth point above); or ``OK'' if none of these apply
and the point is used when we fit models. 

\subsection{Point-spread function}
\label{subsec:PSF}


\begin{figure}
\centerline{\hfil
   \includegraphics[width=84mm]{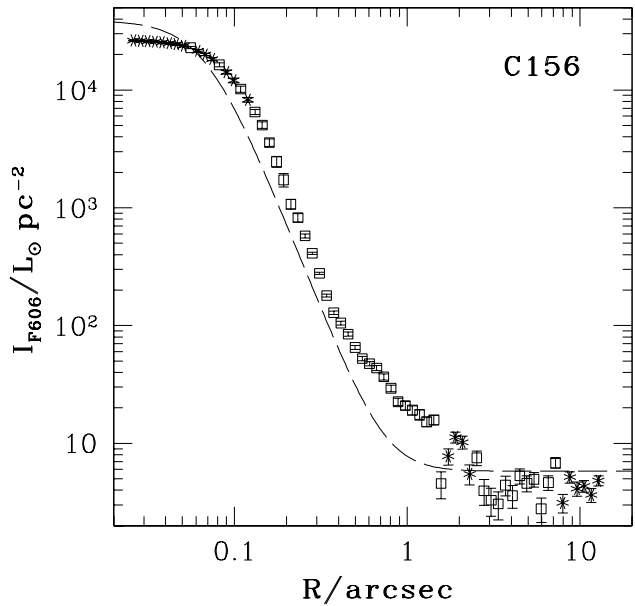}
\hfil}
\caption{
$F606W$ intensity profile, from Table \ref{tab:N5128sbprofs}, of an
object that is not clearly resolved. Open squares represent points
flagged as ``OK'' in the table; asterisks are points flagged as
``SAT'', ``DEP'', or ``BAD''. The dashed
line traces the PSF in equation (\ref{eq:f606psf}), after
adding a constant background and normalising to run through the
innermost unsaturated datapoint.
\label{fig:marginal}
}
\end{figure}


The ACS/WFC has a scale of $0\farcs05 = 0.92$~pc per pixel,
and thus most globular clusters (with typical effective radii
$R_h\sim 3$--4 pc)
are clearly resolved with it. Their apparent core structures, however,
are still strongly influenced by the point-spread function. Rather than
attempt to deconvolve the data, we instead fit structural models after
convolving them with a simple analytic description of the PSF. From
\citet{barmby07}, for the combination of the WFC and $F606W$ filter,
\begin{equation}
I_{\rm PSF}/I_0 = \left[1+\left(R/0\farcs0686\right)^3\right]^{-1.23}\ ,
\label{eq:f606psf}
\end{equation}
which has a full width at half-maximum of ${\rm FWHM}=0\farcs125$, or about
2.5 px. Since this PSF formula is radially symmetric and the models we fit are
intrinsically spherical, the convolved models to be fitted to the data are
also circularly symmetric.

Figure \ref{fig:marginal} illustrates the shape of this PSF and
compares it to the intensity profile from Table \ref{tab:N5128sbprofs}
for C156, one of the three GC candidates noted in Table
\ref{tab:badstuff} as being potential stars.

\subsection{Extinction, transformation to standard $V$, and
cluster metallicities}
\label{subsec:transmet}


\begin{table*}
\begin{minipage}{170mm}
\scriptsize
\caption{Colours and photometric metallicities for 149 GCs in NGC 5128$^{~a}$
\label{tab:n5128colors}}
\begin{tabular}{@{}lcccccclccccc}
\hline
\multicolumn{1}{c}{Name} &
\multicolumn{1}{c}{$T_1$} &
\multicolumn{1}{c}{$(C-T_1)$} &
\multicolumn{1}{c}{$(C-T_1)_0$} &
\multicolumn{1}{c}{$(V-F606)_0$} &
\multicolumn{1}{c}{[Fe/H]} &
~~~ &
\multicolumn{1}{c}{Name} &
\multicolumn{1}{c}{$T_1$} &
\multicolumn{1}{c}{$(C-T_1)$} &
\multicolumn{1}{c}{$(C-T_1)_0$} &
\multicolumn{1}{c}{$(V-F606)_0$} &
\multicolumn{1}{c}{[Fe/H]} \\
\multicolumn{1}{c}{(1)} &
\multicolumn{1}{c}{(2)} &
\multicolumn{1}{c}{(3)} &
\multicolumn{1}{c}{(4)} &
\multicolumn{1}{c}{(5)} &
\multicolumn{1}{c}{(6)} &
~~~ &
\multicolumn{1}{c}{(1)} &
\multicolumn{1}{c}{(2)} &
\multicolumn{1}{c}{(3)} &
\multicolumn{1}{c}{(4)} &
\multicolumn{1}{c}{(5)} &
\multicolumn{1}{c}{(6)} \\
\hline
AAT111563 & $20.049$ & $1.091$ & $0.874\pm0.030$ & $0.091\pm0.050$ &
            $-2.46\pm0.1$ & ~~~ &
C146 & $19.940$ & $1.773$ & $1.556\pm0.030$ & $0.177\pm0.050$ &
            $-0.70\pm0.1$ \\
AAT113992 & $19.818$ & $1.955$ & $1.738\pm0.030$ & $0.200\pm0.050$ &
            $-0.39\pm0.1$ & ~~~ &
C147 & $20.055$ & $1.523$ & $1.306\pm0.030$ & $0.146\pm0.050$ &
            $-1.24\pm0.1$ \\
AAT115339 & $19.561$ & $1.546$ & $1.329\pm0.030$ & $0.148\pm0.050$ &
            $-1.18\pm0.1$ & ~~~ &
C148 & $20.207$ & $1.042$ & $0.825\pm0.030$ & $0.085\pm0.050$ &
            $-2.62\pm0.1$ \\
AAT117287 & $20.450$ & $1.374$ & $1.157\pm0.030$ & $0.127\pm0.050$ &
            $-1.62\pm0.1$ & ~~~ &
C149 & $19.680$ & $1.343$ & $1.126\pm0.030$ & $0.123\pm0.050$ &
            $-1.70\pm0.1$ \\
AAT118198 & $19.031$ & $2.110$ & $1.893\pm0.030$ & $0.220\pm0.050$ &
            $-0.17\pm0.1$ & ~~~ &
C150 & $19.780$ & --- & $1.400\pm0.300$ & $0.157\pm0.063$ &
            $-1.00\pm0.6$ \\
AAT119508 & $19.857$ & $1.832$ & $1.615\pm0.030$ & $0.185\pm0.050$ &
            $-0.59\pm0.1$ & ~~~ &
C151 & $19.952$ & $2.317$ & $2.100\pm0.030$ & $0.246\pm0.050$ &
            $+0.05\pm0.1$ \\
AAT120336 & $19.668$ & $1.809$ & $1.592\pm0.030$ & $0.182\pm0.050$ &
            $-0.63\pm0.1$ & ~~~ &
C152 & $17.820$ & --- & $1.400\pm0.300$ & $0.157\pm0.063$ &
            $-1.00\pm0.6$ \\
AAT120976 & $19.991$ & $1.511$ & $1.294\pm0.030$ & $0.144\pm0.050$ &
            $-1.27\pm0.1$ & ~~~ &
C153 & $18.230$ & --- & $1.400\pm0.300$ & $0.157\pm0.063$ &
            $-1.00\pm0.6$ \\
C002 & $17.937$ & $1.546$ & $1.329\pm0.030$ & --- &
            $-1.18\pm0.1$ & ~~~ &
C154 & $19.580$ & --- & $1.400\pm0.300$ & $0.157\pm0.063$ &
            $-1.00\pm0.6$ \\
C003 & $17.081$ & $1.940$ & $1.723\pm0.030$ & $0.198\pm0.050$ &
            $-0.41\pm0.1$ & ~~~ &
C155 & $21.358$ & $1.627$ & $1.410\pm0.030$ & $0.159\pm0.050$ &
            $-1.00\pm0.1$ \\
C004 & $17.498$ & $1.451$ & $1.234\pm0.030$ & $0.137\pm0.050$ &
            $-1.42\pm0.1$ & ~~~ &
C156 & $17.651$ & $2.106$ & $1.889\pm0.030$ & $0.219\pm0.050$ &
            $-0.18\pm0.1$ \\
C006 & $16.510$ & $1.858$ & $1.641\pm0.030$ & $0.188\pm0.050$ &
            $-0.55\pm0.1$ & ~~~ &
C157 & $19.210$ & --- & $1.400\pm0.300$ & $0.157\pm0.063$ &
            $-1.00\pm0.6$ \\
C007 & $16.644$ & $1.534$ & $1.317\pm0.030$ & $0.147\pm0.050$ &
            $-1.21\pm0.1$ & ~~~ &
C158 & $20.061$ & $1.606$ & $1.389\pm0.030$ & $0.156\pm0.050$ &
            $-1.05\pm0.1$ \\
C011 & $17.197$ & $2.011$ & $1.794\pm0.030$ & --- &
            $-0.30\pm0.1$ & ~~~ &
C159 & $19.929$ & $1.986$ & $1.769\pm0.030$ & $0.204\pm0.050$ &
            $-0.34\pm0.1$ \\
C012 & $17.358$ & $1.984$ & $1.767\pm0.030$ & $0.204\pm0.050$ &
            $-0.34\pm0.1$ & ~~~ &
C160 & $19.990$ & --- & $1.400\pm0.300$ & $0.157\pm0.063$ &
            $-1.00\pm0.6$ \\
C014 & $17.407$ & $1.655$ & $1.438\pm0.030$ & $0.162\pm0.050$ &
            $-0.94\pm0.1$ & ~~~ &
C161 & $19.261$ & $1.953$ & $1.736\pm0.030$ & $0.200\pm0.050$ &
            $-0.39\pm0.1$ \\
C017 & $17.186$ & $1.422$ & $1.205\pm0.030$ & --- &
            $-1.49\pm0.1$ & ~~~ &
C162 & $20.909$ & $1.046$ & $0.829\pm0.030$ & $0.085\pm0.050$ &
            $-2.60\pm0.1$ \\
C018 & $16.891$ & $1.603$ & $1.386\pm0.030$ & $0.156\pm0.050$ &
            $-1.05\pm0.1$ & ~~~ &
C163 & $20.431$ & $1.547$ & $1.330\pm0.030$ & $0.149\pm0.050$ &
            $-1.18\pm0.1$ \\
C019 & $17.554$ & $1.661$ & $1.444\pm0.030$ & $0.163\pm0.050$ &
            $-0.93\pm0.1$ & ~~~ &
C164 & $19.600$ & --- & $1.400\pm0.300$ & $0.157\pm0.063$ &
            $-1.00\pm0.6$ \\
C021 & $17.397$ & $1.576$ & $1.359\pm0.030$ & --- &
            $-1.12\pm0.1$ & ~~~ &
C165 & $18.173$ & $1.931$ & $1.714\pm0.030$ & $0.197\pm0.050$ &
            $-0.42\pm0.1$ \\
C022 & $17.696$ & $1.516$ & $1.299\pm0.030$ & --- &
            $-1.26\pm0.1$ & ~~~ &
C166 & $20.720$ & --- & $1.400\pm0.300$ & $0.157\pm0.063$ &
            $-1.00\pm0.6$ \\
C023 & $16.686$ & $1.904$ & $1.687\pm0.030$ & --- &
            $-0.47\pm0.1$ & ~~~ &
C167 & $20.410$ & --- & $1.400\pm0.300$ & $0.157\pm0.063$ &
            $-1.00\pm0.6$ \\
C025 & $17.965$ & $1.952$ & $1.735\pm0.030$ & $0.200\pm0.050$ &
            $-0.39\pm0.1$ & ~~~ &
C168 & $19.710$ & --- & $1.400\pm0.300$ & $0.157\pm0.063$ &
            $-1.00\pm0.6$ \\
C029 & $17.533$ & $1.924$ & $1.707\pm0.030$ & $0.196\pm0.050$ &
            $-0.44\pm0.1$ & ~~~ &
C169 & $20.389$ & $1.212$ & $0.995\pm0.030$ & $0.106\pm0.050$ &
            $-2.08\pm0.1$ \\
C030 & $16.681$ & $1.789$ & $1.572\pm0.030$ & $0.179\pm0.050$ &
            $-0.67\pm0.1$ & ~~~ &
C170 & $22.161$ & $1.148$ & $0.931\pm0.030$ & $0.098\pm0.050$ &
            $-2.27\pm0.1$ \\
C031 & $17.710$ & $2.023$ & $1.806\pm0.030$ & --- &
            $-0.29\pm0.1$ & ~~~ &
C171 & $20.670$ & $1.577$ & $1.360\pm0.030$ & $0.152\pm0.050$ &
            $-1.11\pm0.1$ \\
C032 & $17.854$ & $2.006$ & $1.789\pm0.030$ & $0.206\pm0.050$ &
            $-0.31\pm0.1$ & ~~~ &
C172 & $20.907$ & $1.242$ & $1.025\pm0.030$ & $0.110\pm0.050$ &
            $-1.99\pm0.1$ \\
C036 & $17.944$ & $1.378$ & $1.161\pm0.030$ & $0.127\pm0.050$ &
            $-1.61\pm0.1$ & ~~~ &
C173 & $21.057$ & $2.092$ & $1.875\pm0.030$ & $0.217\pm0.050$ &
            $-0.19\pm0.1$ \\
C037 & $17.962$ & $1.691$ & $1.474\pm0.030$ & $0.167\pm0.050$ &
            $-0.86\pm0.1$ & ~~~ &
C174 & $21.424$ & $1.812$ & $1.595\pm0.030$ & $0.182\pm0.050$ &
            $-0.63\pm0.1$ \\
C040 & $18.490$ & $1.630$ & $1.413\pm0.030$ & --- &
            $-0.99\pm0.1$ & ~~~ &
C175 & $21.838$ & $1.668$ & $1.451\pm0.030$ & $0.164\pm0.050$ &
            $-0.91\pm0.1$ \\
C041 & $17.969$ & $1.980$ & $1.763\pm0.030$ & --- &
            $-0.35\pm0.1$ & ~~~ &
C176 & $21.297$ & $1.053$ & $0.836\pm0.030$ & $0.086\pm0.050$ &
            $-2.58\pm0.1$ \\
C043 & $18.068$ & $1.521$ & $1.304\pm0.030$ & $0.145\pm0.050$ &
            $-1.24\pm0.1$ & ~~~ &
C177 & $21.087$ & $1.867$ & $1.650\pm0.030$ & $0.189\pm0.050$ &
            $-0.53\pm0.1$ \\
C044 & $18.148$ & $1.441$ & $1.224\pm0.030$ & --- &
            $-1.44\pm0.1$ & ~~~ &
C178 & $21.534$ & $1.481$ & $1.264\pm0.030$ & $0.140\pm0.050$ &
            $-1.34\pm0.1$ \\
C100 & $19.030$ & $1.409$ & $1.192\pm0.030$ & --- &
            $-1.53\pm0.1$ & ~~~ &
C179 & $21.043$ & $1.088$ & $0.871\pm0.030$ & $0.091\pm0.050$ &
            $-2.47\pm0.1$ \\
C101 & $20.077$ & $1.762$ & $1.545\pm0.030$ & --- &
            $-0.72\pm0.1$ & ~~~ &
F1GC14 & $19.671$ & $1.421$ & $1.204\pm0.030$ & $0.133\pm0.050$ &
            $-1.49\pm0.1$ \\
C102 & $21.038$ & $1.673$ & $1.456\pm0.030$ & --- &
            $-0.90\pm0.1$ & ~~~ &
F1GC15 & $19.531$ & $2.224$ & $2.007\pm0.030$ & $0.234\pm0.050$ &
            $-0.04\pm0.1$ \\
C103 & $18.439$ & $1.993$ & $1.776\pm0.030$ & --- &
            $-0.33\pm0.1$ & ~~~ &
F1GC20 & $21.216$ & $1.839$ & $1.622\pm0.030$ & $0.185\pm0.050$ &
            $-0.58\pm0.1$ \\
C104 & $19.403$ & $1.405$ & $1.188\pm0.030$ & $0.131\pm0.050$ &
            $-1.54\pm0.1$ & ~~~ &
F1GC21 & $21.358$ & $2.071$ & $1.854\pm0.030$ & $0.215\pm0.050$ &
            $-0.22\pm0.1$ \\
C105 & $21.758$ & $1.922$ & $1.705\pm0.030$ & $0.196\pm0.050$ &
            $-0.44\pm0.1$ & ~~~ &
F1GC34 & $20.961$ & $1.896$ & $1.679\pm0.030$ & $0.193\pm0.050$ &
            $-0.48\pm0.1$ \\
C106 & $20.657$ & $1.879$ & $1.662\pm0.030$ & --- &
            $-0.51\pm0.1$ & ~~~ &
F2GC14 & $20.824$ & $1.678$ & $1.461\pm0.030$ & $0.165\pm0.050$ &
            $-0.89\pm0.1$ \\
C111 & $21.363$ & $1.172$ & $0.955\pm0.030$ & $0.101\pm0.050$ &
            $-2.20\pm0.1$ & ~~~ &
F2GC18 & $21.147$ & --- & $1.400\pm0.300$ & $0.157\pm0.063$ &
            $-1.00\pm0.6$ \\
C112 & $21.310$ & $1.668$ & $1.451\pm0.030$ & $0.164\pm0.050$ &
            $-0.91\pm0.1$ & ~~~ &
F2GC20 & $21.013$ & $1.870$ & $1.653\pm0.030$ & $0.189\pm0.050$ &
            $-0.53\pm0.1$ \\
C113 & $19.120$ & $1.378$ & $1.161\pm0.030$ & $0.127\pm0.050$ &
            $-1.61\pm0.1$ & ~~~ &
F2GC28 & $21.156$ & $1.701$ & $1.484\pm0.030$ & $0.168\pm0.050$ &
            $-0.84\pm0.1$ \\
C114 & $21.423$ & $2.037$ & $1.820\pm0.030$ & $0.210\pm0.050$ &
            $-0.27\pm0.1$ & ~~~ &
F2GC31 & $20.154$ & $1.395$ & $1.178\pm0.030$ & $0.129\pm0.050$ &
            $-1.56\pm0.1$ \\
C115 & $19.507$ & $1.372$ & $1.155\pm0.030$ & $0.127\pm0.050$ &
            $-1.62\pm0.1$ & ~~~ &
F2GC69 & $19.301$ & $1.813$ & $1.596\pm0.030$ & $0.182\pm0.050$ &
            $-0.63\pm0.1$ \\
C116 & $21.837$ & $1.967$ & $1.750\pm0.030$ & $0.202\pm0.050$ &
            $-0.37\pm0.1$ & ~~~ &
F2GC70 & $20.039$ & $1.253$ & $1.036\pm0.030$ & $0.112\pm0.050$ &
            $-1.96\pm0.1$ \\
C117 & $19.262$ & $2.011$ & $1.794\pm0.030$ & $0.207\pm0.050$ &
            $-0.30\pm0.1$ & ~~~ &
G019 & $18.636$ & $1.422$ & $1.205\pm0.030$ & --- &
            $-1.49\pm0.1$ \\
C118 & $20.672$ & $1.924$ & $1.707\pm0.030$ & $0.196\pm0.050$ &
            $-0.44\pm0.1$ & ~~~ &
G170 & $18.727$ & $1.844$ & $1.627\pm0.030$ & $0.186\pm0.050$ &
            $-0.57\pm0.1$ \\
C119 & $20.669$ & $0.487$ & $0.270\pm0.030$ & $0.015\pm0.050$ &
            $-4.77\pm0.1$ & ~~~ &
G221 & $18.828$ & $1.715$ & $1.498\pm0.030$ & $0.170\pm0.050$ &
            $-0.82\pm0.1$ \\
C120 & $21.433$ & $1.828$ & $1.611\pm0.030$ & $0.184\pm0.050$ &
            $-0.60\pm0.1$ & ~~~ &
G277 & $18.607$ & $1.530$ & $1.313\pm0.030$ & --- &
            $-1.22\pm0.1$ \\
C121 & $22.446$ & $1.080$ & $0.863\pm0.030$ & $0.090\pm0.050$ &
            $-2.49\pm0.1$ & ~~~ &
G284 & $19.412$ & $1.853$ & $1.636\pm0.030$ & $0.187\pm0.050$ &
            $-0.56\pm0.1$ \\
C122 & $22.330$ & --- & $1.400\pm0.300$ & $0.157\pm0.063$ &
            $-1.00\pm0.6$ & ~~~ &
G293 & $18.693$ & $1.349$ & $1.132\pm0.030$ & $0.124\pm0.050$ &
            $-1.69\pm0.1$ \\
C123 & $20.244$ & $1.637$ & $1.420\pm0.030$ & $0.160\pm0.050$ &
            $-0.98\pm0.1$ & ~~~ &
G302 & $18.728$ & $1.450$ & $1.233\pm0.030$ & --- &
            $-1.42\pm0.1$ \\
C124 & $21.574$ & $1.498$ & $1.281\pm0.030$ & $0.142\pm0.050$ &
            $-1.30\pm0.1$ & ~~~ &
K131 & $18.741$ & $2.104$ & $1.887\pm0.030$ & $0.219\pm0.050$ &
            $-0.18\pm0.1$ \\
C125 & $20.664$ & $1.671$ & $1.454\pm0.030$ & $0.164\pm0.050$ &
            $-0.91\pm0.1$ & ~~~ &
PFF011 & $18.553$ & $1.398$ & $1.181\pm0.030$ & $0.130\pm0.050$ &
            $-1.55\pm0.1$ \\
C126 & $22.663$ & $1.354$ & $1.137\pm0.030$ & $0.124\pm0.050$ &
            $-1.67\pm0.1$ & ~~~ &
PFF016 & $19.334$ & $1.853$ & $1.636\pm0.030$ & $0.187\pm0.050$ &
            $-0.56\pm0.1$ \\
C127 & $21.620$ & $1.593$ & $1.376\pm0.030$ & $0.154\pm0.050$ &
            $-1.08\pm0.1$ & ~~~ &
PFF021 & $18.814$ & $1.318$ & $1.101\pm0.030$ & $0.120\pm0.050$ &
            $-1.77\pm0.1$ \\
C128 & $20.947$ & $2.131$ & $1.914\pm0.030$ & $0.222\pm0.050$ &
            $-0.14\pm0.1$ & ~~~ &
PFF023 & $18.971$ & $1.511$ & $1.294\pm0.030$ & $0.144\pm0.050$ &
            $-1.27\pm0.1$ \\
C129 & $20.831$ & $1.757$ & $1.540\pm0.030$ & $0.175\pm0.050$ &
            $-0.73\pm0.1$ & ~~~ &
PFF029 & $19.098$ & $0.621$ & $0.404\pm0.030$ & $0.032\pm0.050$ &
            $-4.20\pm0.1$ \\
C130 & $19.851$ & $1.368$ & $1.151\pm0.030$ & $0.126\pm0.050$ &
            $-1.63\pm0.1$ & ~~~ &
PFF031 & $18.952$ & $1.479$ & $1.262\pm0.030$ & $0.140\pm0.050$ &
            $-1.35\pm0.1$ \\
C131 & $20.202$ & $1.363$ & $1.146\pm0.030$ & $0.125\pm0.050$ &
            $-1.65\pm0.1$ & ~~~ &
PFF034 & $19.384$ & $1.502$ & $1.285\pm0.030$ & $0.143\pm0.050$ &
            $-1.29\pm0.1$ \\
C132 & $18.896$ & $1.480$ & $1.263\pm0.030$ & $0.140\pm0.050$ &
            $-1.34\pm0.1$ & ~~~ &
PFF035 & $19.132$ & $2.000$ & $1.783\pm0.030$ & $0.206\pm0.050$ &
            $-0.32\pm0.1$ \\
C133 & $19.300$ & --- & $1.400\pm0.300$ & $0.157\pm0.063$ &
            $-1.00\pm0.6$ & ~~~ &
PFF041 & $19.007$ & $1.471$ & $1.254\pm0.030$ & $0.139\pm0.050$ &
            $-1.37\pm0.1$ \\
C134 & $20.711$ & $1.735$ & $1.518\pm0.030$ & $0.172\pm0.050$ &
            $-0.78\pm0.1$ & ~~~ &
PFF052 & $19.416$ & $1.442$ & $1.225\pm0.030$ & $0.135\pm0.050$ &
            $-1.44\pm0.1$ \\
C135 & $18.900$ & --- & $1.400\pm0.300$ & $0.157\pm0.063$ &
            $-1.00\pm0.6$ & ~~~ &
PFF059 & $19.479$ & $1.943$ & $1.726\pm0.030$ & $0.199\pm0.050$ &
            $-0.41\pm0.1$ \\
C136 & $21.190$ & $1.295$ & $1.078\pm0.030$ & $0.117\pm0.050$ &
            $-1.84\pm0.1$ & ~~~ &
PFF063 & $19.452$ & $1.176$ & $0.959\pm0.030$ & $0.102\pm0.050$ &
            $-2.19\pm0.1$ \\
C137 & $19.040$ & --- & $1.400\pm0.300$ & $0.157\pm0.063$ &
            $-1.00\pm0.6$ & ~~~ &
PFF066 & $19.436$ & $1.629$ & $1.412\pm0.030$ & $0.159\pm0.050$ &
            $-1.00\pm0.1$ \\
C138 & $19.974$ & $1.555$ & $1.338\pm0.030$ & $0.150\pm0.050$ &
            $-1.16\pm0.1$ & ~~~ &
PFF079 & $19.140$ & $1.484$ & $1.267\pm0.030$ & $0.141\pm0.050$ &
            $-1.33\pm0.1$ \\
C139 & $18.863$ & $1.866$ & $1.649\pm0.030$ & $0.189\pm0.050$ &
            $-0.53\pm0.1$ & ~~~ &
PFF083 & $19.436$ & $1.683$ & $1.466\pm0.030$ & $0.166\pm0.050$ &
            $-0.88\pm0.1$ \\
C140 & $19.830$ & $1.570$ & $1.353\pm0.030$ & $0.152\pm0.050$ &
            $-1.13\pm0.1$ & ~~~ &
R203 & $19.384$ & $1.240$ & $1.023\pm0.030$ & $0.110\pm0.050$ &
            $-2.00\pm0.1$ \\
C141 & $20.945$ & $1.288$ & $1.071\pm0.030$ & $0.116\pm0.050$ &
            $-1.86\pm0.1$ & ~~~ &
R223 & $18.186$ & $1.708$ & $1.491\pm0.030$ & $0.169\pm0.050$ &
            $-0.83\pm0.1$ \\
C142 & $17.640$ & --- & $1.400\pm0.300$ & $0.157\pm0.063$ &
            $-1.00\pm0.6$ & ~~~ &
WHH09 & $18.300$ & $1.979$ & $1.762\pm0.030$ & $0.203\pm0.050$ &
            $-0.35\pm0.1$ \\
C143 & $20.525$ & $1.345$ & $1.128\pm0.030$ & $0.123\pm0.050$ &
            $-1.70\pm0.1$ & ~~~ &
WHH16 & $18.599$ & $2.036$ & $1.819\pm0.030$ & $0.210\pm0.050$ &
            $-0.27\pm0.1$ \\
C144 & $21.959$ & $1.257$ & $1.040\pm0.030$ & $0.112\pm0.050$ &
            $-1.95\pm0.1$ & ~~~ &
WHH22 & $18.032$ & $1.631$ & $1.414\pm0.030$ & $0.159\pm0.050$ &
            $-0.99\pm0.1$ \\
C145 & $17.814$ & $1.510$ & $1.293\pm0.030$ & $0.144\pm0.050$ &
            $-1.27\pm0.1$ & ~~~ &
   &   &   &   &   &   \\
\hline
\end{tabular}

\medskip
$^{a}$~A blank entry in Column (5) indicates a cluster appearing in the
sample of \citet{har02} but not observed by us here, and for which we
therefore do not need a $(V-F606)_0$ colour.

\end{minipage}
\end{table*}


Once we have fitted models to our clusters' brightness profiles, we will want
to correct the inferred intensity/magnitude parameters for extinction, and
to transform the ACS/WFC magnitudes in $F606W$ to standard $V$ for easy
comparison with catalogues of other old GCs. At the same time, we are
interested in predicting observable dynamical properties of the clusters
(e.g., projected velocity dispersions), which will require some estimate of a
mass-to-light ratio to apply to our surface-brightness fits. We use
population-synthesis models to predict $V$-band $M/L$ ratios, and these
require as input an assumed (old) age and an estimate of [Fe/H] for every
cluster.

The effective wavelength of the $F606W$ bandpass is
$\lambda_{\rm eff}=5917$~\AA\ . Using this in the formula developed by
\citet{card89} implies the extinction relation
$A_{F606}\simeq 2.8\times E(B-V)$, in good 
agreement with the conclusions of \citet{siri05}. The foreground reddening in
the direction of NGC 5128 is $E(B-V)=0.11$~mag, which we adopt for all objects
in our sample; thus, $A_{F606}=0.308$~mag in all cases, and the
intensities in Table \ref{tab:N5128sbprofs} all need to be multiplied
by a corrective factor $1.328$. As discussed in \citetalias{har06}, the only
object for which this might seriously be in error is the cluster C150, which
is projected near the central dust lane in NGC 5128.

Transforming the extinction-corrected intensities to standard $V$ is a
two-step process. \citet{siri05} give two transformations from $F606W$ to $V$
magnitude, both including a linear dependence on de-reddened
$(V-R)_0$ colour (see their
Table 22). Over the colour range $0.4\la (V-R)_0\la 0.6$, appropriate to old
globulars with ${\rm [Fe/H]}\la 0$, these two transformations are offset
from each other by about 0.05 mag. We therefore take their average:
$(V-F606)_0=-0.042 + 0.461(V-R)_0$,
with an estimated rms scatter of $\simeq\!0.03$~mag. Reassuringly, we
find that this transformation agrees very well
with the relation between $(V-F606)_0$ and $(V-R)_0$ {\it predicted} for old
globulars in the population-synthesis models of \citet{mar98,mar05}. [The
VEGAMAG $(V-F606)_0$ colours in this model were kindly computed for us by
C.~Maraston.]

However, we have $(C-T_1)$ colours on the Washington photometric
system for most of the clusters in our sample (\citealt{ghar92,ghar04}; see
\citetalias{har06}),
rather than $(V-R)$ indices. \citet{geis96} gives an accurate
transformation between $(V-R)_0$ and $(C-T_1)_0$ (see his Table 4), and
combining this with our $(V-R)_0$--$(V-F606)_0$ relation yields
\begin{equation}
(V-F606)_0 = -0.019 + 0.126 (C-T1)_0 \ ,
\label{eq:ct1v606}
\end{equation}
for which we estimate a precision of about $\pm0.05$~mag. We emphasize again
that this conversion is applied {\it after} correcting our measured
$F606W$ intensities for extinction and de-reddening the $(C-T_1)$ colours using
$E(C-T_1)=1.97 E(B-V)=0.217$~mag \citep{harcan79}.

We also estimate metallicities for our clusters from their $(C-T_1)$
colour indices, using the relation of
\citet{harhar02} after correcting the colours for reddening:
\begin{equation}
{\rm [Fe/H]} = -6.037\left[1-0.82(C-T_1)_0+0.162(C-T_1)_0^2\right] \ .
\label{eq:ct1feh}
\end{equation}
This relation has been
calibrated for classically old GCs, and it could give spurious metallicities
for significantly younger clusters, if any such objects are in our sample.

Table \ref{tab:n5128colors} lists the $(V-F606)_0$ colours and [Fe/H] values
we have estimated in this way for the full ACS sample from
\citetalias{har06}. We have also added the 18 clusters from the sample of
Paper I, which we have not re-observed. The first column of
the table is the cluster ID. The second and third columns are taken directly
from Tables 1 and 2 of Paper II; they are the observed Washington $T_1$
magnitudes and $(C-T_1)$ colours (not corrected for extinction/reddening). The
fourth column is the de-reddened $(C-T_1)_0=(C-T_1)-0.217$. We have adopted a
uniform uncertainty of $\pm0.03$~mag on this colour for most clusters.
Note that there are 16 clusters (all in our subset of newly discovered
objects) for which there is no observed $(C-T_1)$ index. To these 16
objects we have {\it assigned} $(C-T_1)_0=1.4\pm0.3$~mag, which is the
average and dispersion of the measured $(C-T_1)_0$ for the other clusters.
Columns 5 then gives the colour $(V-F606)_0$; it is left blank for the
18 clusters from Paper I, whose surface-brightness profiles we
are not modeling here. Column (6) reports the [Fe/H] inferred from
equations (\ref{eq:ct1v606}) and (\ref{eq:ct1feh}).

There are two objects (C119, PFF029) with colours so blue that
equation (\ref{eq:ct1feh}) implies ${\rm [Fe/H]}<-4$. These are likely to
be much younger objects than the classical $\sim\! 13$-Gyr globular cluster
ages (see Paper II), and their metallicities should not be taken seriously.

\section{Models}
\label{sec:models}

\subsection{Population synthesis models: $M/L$ ratios}
\label{subsec:popsyn}


\begin{table*}
\begin{minipage}{155mm}
\scriptsize
\caption{Model $V$-band mass-to-light ratios for 149 GCs in NGC 5128
         \label{tab:n5128mtol}}
\begin{tabular}{@{}lcccccclccccc}
\hline
\multicolumn{1}{c}{Name} &
\multicolumn{1}{c}{7 Gyr} &
\multicolumn{1}{c}{9 Gyr} &
\multicolumn{1}{c}{11 Gyr} &
\multicolumn{1}{c}{13 Gyr} &
\multicolumn{1}{c}{15 Gyr} &
 ~~~ &
\multicolumn{1}{c}{Name} &
\multicolumn{1}{c}{7 Gyr} &
\multicolumn{1}{c}{9 Gyr} &
\multicolumn{1}{c}{11 Gyr} &
\multicolumn{1}{c}{13 Gyr} &
\multicolumn{1}{c}{15 Gyr} \\
\hline
AAT111563$^{~a}$ & $1.18^{+0.01}_{-0.01}$ &
      $1.46^{+0.01}_{-0.02}$ & $1.69^{+0.01}_{-0.02}$ &
      $1.94^{+0.01}_{-0.02}$ & $2.18^{+0.01}_{-0.02}$ & ~~~ &
C146 & $1.47^{+0.10}_{-0.08}$ & $1.81^{+0.11}_{-0.10}$ &
      $2.12^{+0.12}_{-0.11}$ & $2.44^{+0.14}_{-0.12}$ &
      $2.72^{+0.17}_{-0.15}$ \\
AAT113992 & $1.79^{+0.12}_{-0.12}$ &
      $2.17^{+0.13}_{-0.13}$ & $2.53^{+0.16}_{-0.15}$ &
      $2.91^{+0.20}_{-0.18}$ & $3.30^{+0.22}_{-0.21}$ & ~~~ &
C147 & $1.21^{+0.03}_{-0.02}$ & $1.48^{+0.04}_{-0.03}$ &
      $1.74^{+0.05}_{-0.04}$ & $1.99^{+0.06}_{-0.05}$ &
      $2.21^{+0.06}_{-0.04}$ \\
AAT115339 & $1.23^{+0.03}_{-0.03}$ &
      $1.50^{+0.05}_{-0.04}$ & $1.77^{+0.05}_{-0.04}$ &
      $2.02^{+0.07}_{-0.05}$ & $2.24^{+0.07}_{-0.05}$ & ~~~ &
C148$^{~a}$ & $1.18^{+0.01}_{-0.01}$ &
      $1.46^{+0.01}_{-0.02}$ & $1.69^{+0.01}_{-0.02}$ &
      $1.94^{+0.01}_{-0.02}$ & $2.18^{+0.01}_{-0.02}$ \\
AAT117287 & $1.18^{+0.01}_{-0.01}$ &
      $1.42^{+0.01}_{-0.01}$ & $1.67^{+0.02}_{-0.01}$ &
      $1.89^{+0.02}_{-0.01}$ & $2.13^{+0.01}_{-0.01}$ & ~~~ &
C149 & $1.18^{+0.01}_{-0.01}$ & $1.42^{+0.01}_{-0.01}$ &
      $1.66^{+0.01}_{-0.01}$ & $1.88^{+0.01}_{-0.01}$ &
      $2.13^{+0.01}_{-0.01}$ \\
AAT118198 & $2.04^{+0.12}_{-0.12}$ &
      $2.45^{+0.15}_{-0.13}$ & $2.89^{+0.19}_{-0.17}$ &
      $3.36^{+0.24}_{-0.22}$ & $3.78^{+0.25}_{-0.23}$ & ~~~ &
C150 & $1.29^{+0.50}_{-0.11}$ & $1.58^{+0.58}_{-0.16}$ &
      $1.86^{+0.66}_{-0.20}$ & $2.13^{+0.77}_{-0.25}$ &
      $2.37^{+0.92}_{-0.24}$ \\
AAT119508 & $1.57^{+0.11}_{-0.10}$ &
      $1.92^{+0.13}_{-0.11}$ & $2.25^{+0.14}_{-0.13}$ &
      $2.59^{+0.16}_{-0.14}$ & $2.90^{+0.20}_{-0.17}$ & ~~~ &
C151 & $2.31^{+0.16}_{-0.14}$ & $2.80^{+0.22}_{-0.18}$ &
      $3.33^{+0.26}_{-0.22}$ & $3.91^{+0.30}_{-0.27}$ &
      $4.37^{+0.34}_{-0.30}$ \\
AAT120336 & $1.53^{+0.11}_{-0.09}$ &
      $1.88^{+0.12}_{-0.11}$ & $2.20^{+0.13}_{-0.12}$ &
      $2.53^{+0.15}_{-0.14}$ & $2.83^{+0.19}_{-0.16}$ & ~~~ &
C152 & $1.29^{+0.50}_{-0.11}$ & $1.58^{+0.58}_{-0.16}$ &
      $1.86^{+0.66}_{-0.20}$ & $2.13^{+0.77}_{-0.25}$ &
      $2.37^{+0.92}_{-0.24}$ \\
AAT120976 & $1.21^{+0.03}_{-0.02}$ &
      $1.47^{+0.04}_{-0.03}$ & $1.73^{+0.04}_{-0.03}$ &
      $1.97^{+0.05}_{-0.04}$ & $2.20^{+0.05}_{-0.04}$ & ~~~ &
C153 & $1.29^{+0.50}_{-0.11}$ & $1.58^{+0.58}_{-0.16}$ &
      $1.86^{+0.66}_{-0.20}$ & $2.13^{+0.77}_{-0.25}$ &
      $2.37^{+0.92}_{-0.24}$ \\
C002 & $1.23^{+0.03}_{-0.03}$ &
       $1.50^{+0.05}_{-0.04}$ & $1.76^{+0.05}_{-0.04}$ &
       $2.01^{+0.06}_{-0.05}$ & $2.24^{+0.07}_{-0.05}$ & ~~~ &
C154 & $1.29^{+0.50}_{-0.11}$ & $1.58^{+0.58}_{-0.16}$ &
      $1.86^{+0.66}_{-0.20}$ & $2.13^{+0.77}_{-0.25}$ &
      $2.37^{+0.92}_{-0.24}$ \\
C003 & $1.77^{+0.12}_{-0.12}$ & $2.14^{+0.13}_{-0.13}$ &
      $2.50^{+0.16}_{-0.15}$ & $2.88^{+0.19}_{-0.17}$ &
      $3.26^{+0.22}_{-0.21}$ & ~~~ &
C155 & $1.29^{+0.05}_{-0.04}$ & $1.58^{+0.07}_{-0.05}$ &
      $1.86^{+0.08}_{-0.06}$ & $2.13^{+0.09}_{-0.08}$ &
      $2.37^{+0.10}_{-0.08}$ \\
C004 & $1.19^{+0.02}_{-0.01}$ & $1.44^{+0.02}_{-0.02}$ &
      $1.69^{+0.03}_{-0.02}$ & $1.92^{+0.04}_{-0.03}$ &
      $2.16^{+0.03}_{-0.02}$ & ~~~ &
C156 & $2.03^{+0.12}_{-0.12}$ & $2.44^{+0.15}_{-0.13}$ &
      $2.87^{+0.19}_{-0.17}$ & $3.33^{+0.24}_{-0.22}$ &
      $3.76^{+0.25}_{-0.23}$ \\
C006 & $1.61^{+0.12}_{-0.10}$ & $1.97^{+0.13}_{-0.12}$ &
      $2.30^{+0.15}_{-0.13}$ & $2.65^{+0.17}_{-0.15}$ &
      $2.97^{+0.20}_{-0.18}$ & ~~~ &
C157 & $1.29^{+0.50}_{-0.11}$ & $1.58^{+0.58}_{-0.16}$ &
      $1.86^{+0.66}_{-0.20}$ & $2.13^{+0.77}_{-0.25}$ &
      $2.37^{+0.92}_{-0.24}$ \\
C007 & $1.22^{+0.03}_{-0.02}$ & $1.49^{+0.04}_{-0.03}$ &
      $1.75^{+0.05}_{-0.04}$ & $2.00^{+0.06}_{-0.05}$ &
      $2.23^{+0.06}_{-0.05}$ & ~~~ &
C158 & $1.27^{+0.05}_{-0.04}$ & $1.55^{+0.06}_{-0.05}$ &
      $1.83^{+0.07}_{-0.06}$ & $2.10^{+0.08}_{-0.07}$ &
      $2.33^{+0.09}_{-0.07}$ \\
C011 & $1.89^{+0.12}_{-0.12}$ &
       $2.28^{+0.13}_{-0.13}$ & $2.67^{+0.17}_{-0.16}$ &
       $3.08^{+0.21}_{-0.19}$ & $3.48^{+0.23}_{-0.22}$ & ~~~ &
C159 & $1.85^{+0.12}_{-0.12}$ & $2.23^{+0.13}_{-0.13}$ &
      $2.61^{+0.17}_{-0.16}$ & $3.01^{+0.20}_{-0.19}$ &
      $3.40^{+0.22}_{-0.22}$ \\
C012 & $1.85^{+0.12}_{-0.12}$ & $2.23^{+0.13}_{-0.13}$ &
      $2.61^{+0.17}_{-0.16}$ & $3.01^{+0.20}_{-0.19}$ &
      $3.40^{+0.22}_{-0.22}$ & ~~~ &
C160 & $1.29^{+0.50}_{-0.11}$ & $1.58^{+0.58}_{-0.16}$ &
      $1.86^{+0.66}_{-0.20}$ & $2.13^{+0.77}_{-0.25}$ &
      $2.37^{+0.92}_{-0.24}$ \\
C014 & $1.31^{+0.06}_{-0.05}$ & $1.61^{+0.07}_{-0.06}$ &
      $1.90^{+0.09}_{-0.07}$ & $2.18^{+0.10}_{-0.08}$ &
      $2.42^{+0.11}_{-0.09}$ & ~~~ &
C161 & $1.79^{+0.12}_{-0.12}$ & $2.17^{+0.13}_{-0.13}$ &
      $2.53^{+0.16}_{-0.15}$ & $2.91^{+0.20}_{-0.18}$ &
      $3.30^{+0.22}_{-0.21}$ \\
C017 & $1.19^{+0.01}_{-0.01}$ &
       $1.43^{+0.02}_{-0.01}$ & $1.68^{+0.02}_{-0.02}$ &
       $1.90^{+0.03}_{-0.02}$ & $2.14^{+0.02}_{-0.02}$ & ~~~ &
C162$^{~a}$ & $1.18^{+0.01}_{-0.01}$ &
      $1.46^{+0.01}_{-0.02}$ & $1.69^{+0.01}_{-0.02}$ &
      $1.94^{+0.01}_{-0.02}$ & $2.18^{+0.01}_{-0.02}$ \\
C018 & $1.27^{+0.05}_{-0.04}$ & $1.55^{+0.06}_{-0.05}$ &
      $1.83^{+0.07}_{-0.06}$ & $2.10^{+0.08}_{-0.07}$ &
      $2.33^{+0.09}_{-0.07}$ & ~~~ &
C163 & $1.23^{+0.03}_{-0.03}$ & $1.50^{+0.05}_{-0.04}$ &
      $1.77^{+0.05}_{-0.04}$ & $2.02^{+0.07}_{-0.05}$ &
      $2.24^{+0.07}_{-0.05}$ \\
C019 & $1.32^{+0.06}_{-0.05}$ & $1.62^{+0.08}_{-0.06}$ &
      $1.91^{+0.09}_{-0.07}$ & $2.19^{+0.10}_{-0.09}$ &
      $2.43^{+0.11}_{-0.09}$ & ~~~ &
C164 & $1.29^{+0.50}_{-0.11}$ & $1.58^{+0.58}_{-0.16}$ &
      $1.86^{+0.66}_{-0.20}$ & $2.13^{+0.77}_{-0.25}$ &
      $2.37^{+0.92}_{-0.24}$ \\
C021 & $1.25^{+0.04}_{-0.03}$ &
       $1.52^{+0.05}_{-0.04}$ & $1.80^{+0.06}_{-0.05}$ &
       $2.05^{+0.07}_{-0.06}$ & $2.28^{+0.08}_{-0.06}$ & ~~~ &
C165 & $1.76^{+0.12}_{-0.12}$ & $2.13^{+0.13}_{-0.13}$ &
      $2.49^{+0.16}_{-0.15}$ & $2.86^{+0.19}_{-0.17}$ &
      $3.23^{+0.22}_{-0.21}$ \\
C022 & $1.21^{+0.03}_{-0.02}$ &
       $1.48^{+0.04}_{-0.03}$ & $1.74^{+0.05}_{-0.04}$ &
       $1.98^{+0.06}_{-0.04}$ & $2.21^{+0.05}_{-0.04}$ & ~~~ &
C166 & $1.29^{+0.50}_{-0.11}$ & $1.58^{+0.58}_{-0.16}$ &
      $1.86^{+0.66}_{-0.20}$ & $2.13^{+0.77}_{-0.25}$ &
      $2.37^{+0.92}_{-0.24}$ \\
C023 & $1.70^{+0.12}_{-0.11}$ &
       $2.07^{+0.13}_{-0.13}$ & $2.42^{+0.15}_{-0.14}$ &
       $2.78^{+0.18}_{-0.16}$ & $3.13^{+0.21}_{-0.20}$ & ~~~ &
C167 & $1.29^{+0.50}_{-0.11}$ & $1.58^{+0.58}_{-0.16}$ &
      $1.86^{+0.66}_{-0.20}$ & $2.13^{+0.77}_{-0.25}$ &
      $2.37^{+0.92}_{-0.24}$ \\
C025 & $1.79^{+0.12}_{-0.12}$ & $2.17^{+0.13}_{-0.13}$ &
      $2.53^{+0.16}_{-0.15}$ & $2.91^{+0.20}_{-0.18}$ &
      $3.30^{+0.22}_{-0.21}$ & ~~~ &
C168 & $1.29^{+0.50}_{-0.11}$ & $1.58^{+0.58}_{-0.16}$ &
      $1.86^{+0.66}_{-0.20}$ & $2.13^{+0.77}_{-0.25}$ &
      $2.37^{+0.92}_{-0.24}$ \\
C029 & $1.74^{+0.12}_{-0.12}$ & $2.11^{+0.13}_{-0.13}$ &
      $2.46^{+0.16}_{-0.15}$ & $2.83^{+0.19}_{-0.17}$ &
      $3.19^{+0.22}_{-0.21}$ & ~~~ &
C169 & $1.18^{+0.01}_{-0.01}$ & $1.44^{+0.01}_{-0.01}$ &
      $1.68^{+0.01}_{-0.01}$ & $1.91^{+0.02}_{-0.02}$ &
      $2.16^{+0.02}_{-0.02}$ \\
C030 & $1.50^{+0.10}_{-0.09}$ & $1.84^{+0.12}_{-0.10}$ &
      $2.15^{+0.13}_{-0.11}$ & $2.48^{+0.15}_{-0.13}$ &
      $2.76^{+0.18}_{-0.15}$ & ~~~ &
C170$^{~a}$ & $1.18^{+0.01}_{-0.01}$ &
      $1.46^{+0.01}_{-0.02}$ & $1.69^{+0.01}_{-0.02}$ &
      $1.94^{+0.01}_{-0.02}$ & $2.18^{+0.01}_{-0.02}$ \\
C031 & $1.91^{+0.12}_{-0.12}$ &
       $2.30^{+0.13}_{-0.13}$ & $2.70^{+0.17}_{-0.16}$ &
       $3.11^{+0.21}_{-0.20}$ & $3.52^{+0.23}_{-0.22}$ & ~~~ &
C171 & $1.25^{+0.04}_{-0.03}$ & $1.53^{+0.05}_{-0.04}$ &
      $1.80^{+0.06}_{-0.05}$ & $2.06^{+0.08}_{-0.06}$ &
      $2.28^{+0.08}_{-0.06}$ \\
C032 & $1.88^{+0.12}_{-0.12}$ & $2.27^{+0.13}_{-0.13}$ &
      $2.66^{+0.17}_{-0.16}$ & $3.06^{+0.21}_{-0.19}$ &
      $3.47^{+0.23}_{-0.22}$ & ~~~ &
C172 & $1.18^{+0.01}_{-0.01}$ & $1.44^{+0.01}_{-0.01}$ &
      $1.67^{+0.01}_{-0.01}$ & $1.90^{+0.02}_{-0.02}$ &
      $2.15^{+0.02}_{-0.01}$ \\
C036 & $1.18^{+0.01}_{-0.01}$ & $1.42^{+0.01}_{-0.01}$ &
      $1.67^{+0.02}_{-0.01}$ & $1.89^{+0.02}_{-0.01}$ &
      $2.13^{+0.01}_{-0.01}$ & ~~~ &
C173 & $2.01^{+0.12}_{-0.12}$ & $2.43^{+0.14}_{-0.13}$ &
      $2.85^{+0.19}_{-0.17}$ & $3.31^{+0.24}_{-0.21}$ &
      $3.74^{+0.25}_{-0.23}$ \\
C037 & $1.36^{+0.07}_{-0.06}$ & $1.67^{+0.09}_{-0.07}$ &
      $1.97^{+0.10}_{-0.08}$ & $2.26^{+0.11}_{-0.10}$ &
      $2.50^{+0.13}_{-0.11}$ & ~~~ &
C174 & $1.53^{+0.11}_{-0.09}$ & $1.88^{+0.12}_{-0.11}$ &
      $2.20^{+0.13}_{-0.12}$ & $2.53^{+0.15}_{-0.14}$ &
      $2.83^{+0.19}_{-0.16}$ \\
C040 & $1.29^{+0.05}_{-0.04}$ &
       $1.58^{+0.07}_{-0.06}$ & $1.87^{+0.08}_{-0.06}$ &
       $2.14^{+0.09}_{-0.08}$ & $2.37^{+0.10}_{-0.08}$ & ~~~ &
C175 & $1.33^{+0.07}_{-0.05}$ & $1.63^{+0.08}_{-0.07}$ &
      $1.93^{+0.09}_{-0.08}$ & $2.21^{+0.10}_{-0.09}$ &
      $2.45^{+0.12}_{-0.10}$ \\
C041 & $1.84^{+0.12}_{-0.12}$ &
       $2.22^{+0.13}_{-0.13}$ & $2.60^{+0.16}_{-0.16}$ &
       $2.99^{+0.20}_{-0.19}$ & $3.39^{+0.22}_{-0.22}$ & ~~~ &
C176$^{~a}$ & $1.18^{+0.01}_{-0.01}$ &
      $1.46^{+0.01}_{-0.02}$ & $1.69^{+0.01}_{-0.02}$ &
      $1.94^{+0.01}_{-0.02}$ & $2.18^{+0.01}_{-0.02}$ \\
C043 & $1.21^{+0.03}_{-0.02}$ & $1.48^{+0.04}_{-0.03}$ &
      $1.74^{+0.05}_{-0.04}$ & $1.99^{+0.06}_{-0.05}$ &
      $2.21^{+0.06}_{-0.04}$ & ~~~ &
C177 & $1.64^{+0.12}_{-0.11}$ & $1.99^{+0.13}_{-0.12}$ &
      $2.33^{+0.15}_{-0.13}$ & $2.68^{+0.17}_{-0.15}$ &
      $3.01^{+0.21}_{-0.19}$ \\
C044 & $1.19^{+0.02}_{-0.01}$ &
       $1.44^{+0.02}_{-0.02}$ & $1.69^{+0.03}_{-0.02}$ &
       $1.91^{+0.03}_{-0.02}$ & $2.15^{+0.03}_{-0.02}$ & ~~~ &
C178 & $1.20^{+0.02}_{-0.02}$ & $1.46^{+0.03}_{-0.02}$ &
      $1.71^{+0.04}_{-0.03}$ & $1.94^{+0.05}_{-0.04}$ &
      $2.18^{+0.04}_{-0.03}$ \\
C100 & $1.18^{+0.01}_{-0.01}$ &
       $1.43^{+0.02}_{-0.01}$ & $1.67^{+0.02}_{-0.01}$ &
       $1.90^{+0.03}_{-0.02}$ & $2.14^{+0.02}_{-0.01}$ & ~~~ &
C179$^{~a}$ & $1.18^{+0.01}_{-0.01}$ &
      $1.46^{+0.01}_{-0.02}$ & $1.69^{+0.01}_{-0.02}$ &
      $1.94^{+0.01}_{-0.02}$ & $2.18^{+0.01}_{-0.02}$ \\
C101 & $1.45^{+0.09}_{-0.08}$ &
       $1.78^{+0.11}_{-0.09}$ & $2.10^{+0.12}_{-0.10}$ &
       $2.41^{+0.14}_{-0.12}$ & $2.68^{+0.17}_{-0.14}$ & ~~~ &
F1GC14 & $1.19^{+0.01}_{-0.01}$ &
      $1.43^{+0.02}_{-0.01}$ & $1.68^{+0.02}_{-0.02}$ &
      $1.90^{+0.03}_{-0.02}$ & $2.14^{+0.03}_{-0.02}$ \\
C102 & $1.33^{+0.07}_{-0.05}$ &
       $1.64^{+0.08}_{-0.07}$ & $1.93^{+0.09}_{-0.08}$ &
       $2.22^{+0.11}_{-0.09}$ & $2.46^{+0.12}_{-0.10}$ & ~~~ &
F1GC15 & $2.19^{+0.14}_{-0.13}$ &
      $2.64^{+0.18}_{-0.15}$ & $3.14^{+0.23}_{-0.20}$ &
      $3.67^{+0.28}_{-0.25}$ & $4.11^{+0.30}_{-0.26}$ \\
C103 & $1.86^{+0.12}_{-0.12}$ &
       $2.25^{+0.13}_{-0.13}$ & $2.63^{+0.17}_{-0.16}$ &
       $3.03^{+0.21}_{-0.19}$ & $3.43^{+0.22}_{-0.22}$ & ~~~ &
F1GC20 & $1.58^{+0.11}_{-0.10}$ &
      $1.93^{+0.13}_{-0.11}$ & $2.26^{+0.14}_{-0.13}$ &
      $2.60^{+0.16}_{-0.14}$ & $2.92^{+0.20}_{-0.18}$ \\
C104 & $1.18^{+0.01}_{-0.01}$ & $1.43^{+0.02}_{-0.01}$ &
      $1.67^{+0.02}_{-0.01}$ & $1.90^{+0.03}_{-0.02}$ &
      $2.14^{+0.02}_{-0.01}$ & ~~~ &
F1GC21 & $1.98^{+0.12}_{-0.12}$ &
      $2.39^{+0.14}_{-0.13}$ & $2.80^{+0.18}_{-0.17}$ &
      $3.25^{+0.23}_{-0.21}$ & $3.67^{+0.24}_{-0.23}$ \\
C105 & $1.74^{+0.12}_{-0.12}$ & $2.11^{+0.13}_{-0.13}$ &
      $2.46^{+0.16}_{-0.15}$ & $2.83^{+0.19}_{-0.17}$ &
      $3.19^{+0.22}_{-0.21}$ & ~~~ &
F1GC34 & $1.69^{+0.12}_{-0.11}$ &
      $2.06^{+0.13}_{-0.13}$ & $2.40^{+0.15}_{-0.14}$ &
      $2.76^{+0.18}_{-0.16}$ & $3.11^{+0.21}_{-0.20}$ \\
C106 & $1.66^{+0.12}_{-0.11}$ &
       $2.02^{+0.13}_{-0.12}$ & $2.36^{+0.15}_{-0.14}$ &
       $2.71^{+0.17}_{-0.16}$ & $3.05^{+0.21}_{-0.19}$ & ~~~ &
F2GC14 & $1.34^{+0.07}_{-0.05}$ &
      $1.65^{+0.08}_{-0.07}$ & $1.94^{+0.09}_{-0.08}$ &
      $2.23^{+0.11}_{-0.09}$ & $2.47^{+0.12}_{-0.10}$ \\
C111 & $1.18^{+0.01}_{-0.01}$ & $1.45^{+0.01}_{-0.02}$ &
      $1.69^{+0.01}_{-0.02}$ & $1.93^{+0.01}_{-0.02}$ &
      $2.17^{+0.01}_{-0.02}$ & ~~~ &
F2GC18 & $1.29^{+0.50}_{-0.11}$ &
      $1.58^{+0.58}_{-0.16}$ & $1.86^{+0.66}_{-0.20}$ &
      $2.13^{+0.77}_{-0.25}$ & $2.37^{+0.92}_{-0.24}$ \\
C112 & $1.33^{+0.07}_{-0.05}$ & $1.63^{+0.08}_{-0.07}$ &
      $1.93^{+0.09}_{-0.08}$ & $2.21^{+0.10}_{-0.09}$ &
      $2.45^{+0.12}_{-0.10}$ & ~~~ &
F2GC20 & $1.64^{+0.12}_{-0.11}$ &
      $1.99^{+0.13}_{-0.12}$ & $2.33^{+0.15}_{-0.13}$ &
      $2.68^{+0.17}_{-0.15}$ & $3.01^{+0.21}_{-0.19}$ \\
C113 & $1.18^{+0.01}_{-0.01}$ & $1.42^{+0.01}_{-0.01}$ &
      $1.67^{+0.02}_{-0.01}$ & $1.89^{+0.02}_{-0.01}$ &
      $2.13^{+0.01}_{-0.01}$ & ~~~ &
F2GC28 & $1.37^{+0.07}_{-0.06}$ &
      $1.68^{+0.09}_{-0.07}$ & $1.98^{+0.10}_{-0.09}$ &
      $2.28^{+0.12}_{-0.10}$ & $2.53^{+0.14}_{-0.11}$ \\
C114 & $1.93^{+0.12}_{-0.12}$ & $2.32^{+0.13}_{-0.13}$ &
      $2.72^{+0.17}_{-0.16}$ & $3.14^{+0.22}_{-0.20}$ &
      $3.56^{+0.23}_{-0.22}$ & ~~~ &
F2GC31 & $1.18^{+0.01}_{-0.01}$ &
      $1.43^{+0.01}_{-0.01}$ & $1.67^{+0.02}_{-0.01}$ &
      $1.89^{+0.02}_{-0.01}$ & $2.14^{+0.02}_{-0.01}$ \\
C115 & $1.18^{+0.01}_{-0.01}$ & $1.42^{+0.01}_{-0.01}$ &
      $1.67^{+0.02}_{-0.01}$ & $1.89^{+0.02}_{-0.01}$ &
      $2.13^{+0.01}_{-0.01}$ & ~~~ &
F2GC69 & $1.53^{+0.11}_{-0.09}$ &
      $1.88^{+0.12}_{-0.11}$ & $2.20^{+0.13}_{-0.12}$ &
      $2.53^{+0.15}_{-0.14}$ & $2.83^{+0.19}_{-0.16}$ \\
C116 & $1.81^{+0.12}_{-0.12}$ & $2.20^{+0.13}_{-0.13}$ &
      $2.56^{+0.16}_{-0.15}$ & $2.95^{+0.20}_{-0.18}$ &
      $3.34^{+0.22}_{-0.21}$ & ~~~ &
F2GC70 & $1.18^{+0.01}_{-0.01}$ &
      $1.43^{+0.01}_{-0.01}$ & $1.67^{+0.01}_{-0.01}$ &
      $1.90^{+0.02}_{-0.01}$ & $2.14^{+0.02}_{-0.01}$ \\
C117 & $1.89^{+0.12}_{-0.12}$ & $2.29^{+0.13}_{-0.13}$ &
      $2.67^{+0.17}_{-0.16}$ & $3.08^{+0.21}_{-0.19}$ &
      $3.49^{+0.23}_{-0.22}$ & ~~~ &
G019 & $1.19^{+0.01}_{-0.01}$ &
       $1.43^{+0.02}_{-0.01}$ & $1.68^{+0.02}_{-0.02}$ &
       $1.90^{+0.03}_{-0.02}$ & $2.14^{+0.02}_{-0.02}$ \\
C118 & $1.74^{+0.12}_{-0.12}$ & $2.11^{+0.13}_{-0.13}$ &
      $2.46^{+0.16}_{-0.15}$ & $2.83^{+0.19}_{-0.17}$ &
      $3.19^{+0.22}_{-0.21}$ & ~~~ &
G170 & $1.59^{+0.11}_{-0.10}$ & $1.95^{+0.13}_{-0.12}$ &
      $2.28^{+0.14}_{-0.13}$ & $2.62^{+0.16}_{-0.15}$ &
      $2.94^{+0.20}_{-0.18}$ \\
C119$^{~a}$ & $1.18^{+0.01}_{-0.01}$ &
      $1.46^{+0.01}_{-0.02}$ & $1.69^{+0.01}_{-0.02}$ &
      $1.94^{+0.01}_{-0.02}$ & $2.18^{+0.01}_{-0.02}$ & ~~~ &
G221 & $1.38^{+0.08}_{-0.06}$ & $1.70^{+0.09}_{-0.08}$ &
      $2.00^{+0.10}_{-0.09}$ & $2.30^{+0.12}_{-0.10}$ &
      $2.55^{+0.14}_{-0.12}$ \\
C120 & $1.56^{+0.11}_{-0.10}$ & $1.91^{+0.13}_{-0.11}$ &
      $2.24^{+0.14}_{-0.12}$ & $2.57^{+0.16}_{-0.14}$ &
      $2.88^{+0.19}_{-0.17}$ & ~~~ &
G277 & $1.22^{+0.03}_{-0.02}$ &
       $1.49^{+0.04}_{-0.03}$ & $1.75^{+0.05}_{-0.04}$ &
       $1.99^{+0.06}_{-0.05}$ & $2.22^{+0.06}_{-0.04}$ \\
C121$^{~a}$ & $1.18^{+0.01}_{-0.01}$ &
      $1.46^{+0.01}_{-0.02}$ & $1.69^{+0.01}_{-0.02}$ &
      $1.94^{+0.01}_{-0.02}$ & $2.18^{+0.01}_{-0.02}$ & ~~~ &
G284 & $1.60^{+0.11}_{-0.10}$ & $1.96^{+0.13}_{-0.12}$ &
      $2.29^{+0.14}_{-0.13}$ & $2.63^{+0.17}_{-0.15}$ &
      $2.95^{+0.20}_{-0.18}$ \\
C122 & $1.29^{+0.50}_{-0.11}$ & $1.58^{+0.58}_{-0.16}$ &
      $1.86^{+0.66}_{-0.20}$ & $2.13^{+0.77}_{-0.25}$ &
      $2.37^{+0.92}_{-0.24}$ & ~~~ &
G293 & $1.18^{+0.01}_{-0.01}$ & $1.42^{+0.01}_{-0.01}$ &
      $1.66^{+0.01}_{-0.01}$ & $1.88^{+0.01}_{-0.01}$ &
      $2.13^{+0.01}_{-0.01}$ \\
C123 & $1.30^{+0.06}_{-0.04}$ & $1.59^{+0.07}_{-0.06}$ &
      $1.88^{+0.08}_{-0.07}$ & $2.15^{+0.09}_{-0.08}$ &
      $2.38^{+0.10}_{-0.08}$ & ~~~ &
G302 & $1.19^{+0.02}_{-0.01}$ &
       $1.44^{+0.02}_{-0.02}$ & $1.69^{+0.03}_{-0.02}$ &
       $1.92^{+0.04}_{-0.03}$ & $2.16^{+0.03}_{-0.02}$ \\
C124 & $1.20^{+0.02}_{-0.02}$ & $1.46^{+0.03}_{-0.03}$ &
      $1.72^{+0.04}_{-0.03}$ & $1.96^{+0.05}_{-0.04}$ &
      $2.19^{+0.05}_{-0.03}$ & ~~~ &
K131 & $2.03^{+0.12}_{-0.12}$ & $2.44^{+0.15}_{-0.13}$ &
      $2.87^{+0.19}_{-0.17}$ & $3.33^{+0.24}_{-0.22}$ &
      $3.76^{+0.25}_{-0.23}$ \\
C125 & $1.33^{+0.07}_{-0.05}$ & $1.63^{+0.08}_{-0.07}$ &
      $1.93^{+0.09}_{-0.08}$ & $2.21^{+0.10}_{-0.09}$ &
      $2.45^{+0.12}_{-0.10}$ & ~~~ &
PFF011 & $1.18^{+0.01}_{-0.01}$ &
      $1.43^{+0.01}_{-0.01}$ & $1.67^{+0.02}_{-0.01}$ &
      $1.89^{+0.02}_{-0.02}$ & $2.14^{+0.02}_{-0.01}$ \\
C126 & $1.18^{+0.01}_{-0.01}$ & $1.42^{+0.01}_{-0.01}$ &
      $1.66^{+0.01}_{-0.01}$ & $1.88^{+0.01}_{-0.01}$ &
      $2.13^{+0.01}_{-0.01}$ & ~~~ &
PFF016 & $1.60^{+0.11}_{-0.10}$ &
      $1.96^{+0.13}_{-0.12}$ & $2.29^{+0.14}_{-0.13}$ &
      $2.63^{+0.17}_{-0.15}$ & $2.95^{+0.20}_{-0.18}$ \\
C127 & $1.26^{+0.04}_{-0.03}$ & $1.54^{+0.06}_{-0.05}$ &
      $1.81^{+0.07}_{-0.05}$ & $2.08^{+0.08}_{-0.07}$ &
      $2.30^{+0.08}_{-0.07}$ & ~~~ &
PFF021 & $1.18^{+0.01}_{-0.01}$ &
      $1.42^{+0.01}_{-0.01}$ & $1.66^{+0.01}_{-0.01}$ &
      $1.88^{+0.01}_{-0.01}$ & $2.13^{+0.01}_{-0.01}$ \\
C128 & $2.07^{+0.13}_{-0.12}$ & $2.49^{+0.15}_{-0.14}$ &
      $2.94^{+0.20}_{-0.18}$ & $3.42^{+0.25}_{-0.22}$ &
      $3.85^{+0.26}_{-0.24}$ & ~~~ &
PFF023 & $1.21^{+0.03}_{-0.02}$ &
      $1.47^{+0.04}_{-0.03}$ & $1.73^{+0.04}_{-0.03}$ &
      $1.97^{+0.05}_{-0.04}$ & $2.20^{+0.05}_{-0.04}$ \\
C129 & $1.45^{+0.09}_{-0.08}$ & $1.78^{+0.11}_{-0.09}$ &
      $2.09^{+0.12}_{-0.10}$ & $2.40^{+0.14}_{-0.12}$ &
      $2.67^{+0.16}_{-0.14}$ & ~~~ &
PFF029$^{~a}$ & $1.18^{+0.01}_{-0.01}$ &
      $1.46^{+0.01}_{-0.02}$ & $1.69^{+0.01}_{-0.02}$ &
      $1.94^{+0.01}_{-0.02}$ & $2.18^{+0.01}_{-0.02}$ \\
C130 & $1.18^{+0.01}_{-0.01}$ & $1.42^{+0.01}_{-0.01}$ &
      $1.66^{+0.01}_{-0.01}$ & $1.89^{+0.02}_{-0.01}$ &
      $2.13^{+0.01}_{-0.01}$ & ~~~ &
PFF031 & $1.20^{+0.02}_{-0.02}$ &
      $1.45^{+0.03}_{-0.02}$ & $1.71^{+0.04}_{-0.03}$ &
      $1.94^{+0.04}_{-0.03}$ & $2.17^{+0.04}_{-0.03}$ \\
C131 & $1.18^{+0.01}_{-0.01}$ & $1.42^{+0.01}_{-0.01}$ &
      $1.66^{+0.01}_{-0.01}$ & $1.88^{+0.02}_{-0.01}$ &
      $2.13^{+0.01}_{-0.01}$ & ~~~ &
PFF034 & $1.21^{+0.03}_{-0.02}$ &
      $1.47^{+0.03}_{-0.03}$ & $1.73^{+0.04}_{-0.03}$ &
      $1.96^{+0.05}_{-0.04}$ & $2.19^{+0.05}_{-0.04}$ \\
C132 & $1.20^{+0.02}_{-0.02}$ & $1.46^{+0.03}_{-0.02}$ &
      $1.71^{+0.04}_{-0.03}$ & $1.94^{+0.05}_{-0.04}$ &
      $2.18^{+0.04}_{-0.03}$ & ~~~ &
PFF035 & $1.87^{+0.12}_{-0.12}$ &
      $2.26^{+0.13}_{-0.13}$ & $2.64^{+0.17}_{-0.16}$ &
      $3.05^{+0.21}_{-0.19}$ & $3.45^{+0.23}_{-0.22}$ \\
C133 & $1.29^{+0.50}_{-0.11}$ & $1.58^{+0.58}_{-0.16}$ &
      $1.86^{+0.66}_{-0.20}$ & $2.13^{+0.77}_{-0.25}$ &
      $2.37^{+0.92}_{-0.24}$ & ~~~ &
PFF041 & $1.20^{+0.02}_{-0.01}$ &
      $1.45^{+0.03}_{-0.02}$ & $1.70^{+0.03}_{-0.03}$ &
      $1.93^{+0.04}_{-0.03}$ & $2.17^{+0.04}_{-0.03}$ \\
C134 & $1.41^{+0.08}_{-0.07}$ & $1.73^{+0.10}_{-0.08}$ &
      $2.04^{+0.11}_{-0.09}$ & $2.34^{+0.13}_{-0.11}$ &
      $2.60^{+0.15}_{-0.13}$ & ~~~ &
PFF052 & $1.19^{+0.02}_{-0.01}$ &
      $1.44^{+0.02}_{-0.02}$ & $1.69^{+0.03}_{-0.02}$ &
      $1.92^{+0.04}_{-0.03}$ & $2.15^{+0.03}_{-0.02}$ \\
C135 & $1.29^{+0.50}_{-0.11}$ & $1.58^{+0.58}_{-0.16}$ &
      $1.86^{+0.66}_{-0.20}$ & $2.13^{+0.77}_{-0.25}$ &
      $2.37^{+0.92}_{-0.24}$ & ~~~ &
PFF059 & $1.77^{+0.12}_{-0.12}$ &
      $2.14^{+0.13}_{-0.13}$ & $2.50^{+0.16}_{-0.15}$ &
      $2.88^{+0.19}_{-0.17}$ & $3.26^{+0.22}_{-0.21}$ \\
C136 & $1.18^{+0.01}_{-0.01}$ & $1.43^{+0.01}_{-0.01}$ &
      $1.66^{+0.01}_{-0.01}$ & $1.88^{+0.01}_{-0.01}$ &
      $2.13^{+0.01}_{-0.01}$ & ~~~ &
PFF063 & $1.18^{+0.01}_{-0.01}$ &
      $1.45^{+0.01}_{-0.02}$ & $1.69^{+0.01}_{-0.02}$ &
      $1.93^{+0.02}_{-0.02}$ & $2.17^{+0.01}_{-0.02}$ \\
C137 & $1.29^{+0.50}_{-0.11}$ & $1.58^{+0.58}_{-0.16}$ &
      $1.86^{+0.66}_{-0.20}$ & $2.13^{+0.77}_{-0.25}$ &
      $2.37^{+0.92}_{-0.24}$ & ~~~ &
PFF066 & $1.29^{+0.05}_{-0.04}$ &
      $1.58^{+0.07}_{-0.05}$ & $1.86^{+0.08}_{-0.06}$ &
      $2.13^{+0.09}_{-0.08}$ & $2.37^{+0.10}_{-0.08}$ \\
C138 & $1.23^{+0.04}_{-0.03}$ & $1.51^{+0.05}_{-0.04}$ &
      $1.77^{+0.06}_{-0.05}$ & $2.03^{+0.07}_{-0.06}$ &
      $2.25^{+0.07}_{-0.05}$ & ~~~ &
PFF079 & $1.20^{+0.02}_{-0.02}$ &
      $1.46^{+0.03}_{-0.02}$ & $1.71^{+0.04}_{-0.03}$ &
      $1.95^{+0.05}_{-0.04}$ & $2.18^{+0.04}_{-0.03}$ \\
C139 & $1.64^{+0.12}_{-0.11}$ & $1.99^{+0.13}_{-0.12}$ &
      $2.33^{+0.15}_{-0.13}$ & $2.68^{+0.17}_{-0.15}$ &
      $3.01^{+0.21}_{-0.19}$ & ~~~ &
PFF083 & $1.35^{+0.07}_{-0.06}$ &
      $1.65^{+0.08}_{-0.07}$ & $1.95^{+0.09}_{-0.08}$ &
      $2.24^{+0.11}_{-0.09}$ & $2.48^{+0.13}_{-0.10}$ \\
C140 & $1.24^{+0.04}_{-0.03}$ & $1.52^{+0.05}_{-0.04}$ &
      $1.79^{+0.06}_{-0.05}$ & $2.04^{+0.07}_{-0.06}$ &
      $2.27^{+0.07}_{-0.06}$ & ~~~ &
R203 & $1.18^{+0.01}_{-0.01}$ & $1.44^{+0.01}_{-0.01}$ &
      $1.67^{+0.01}_{-0.01}$ & $1.90^{+0.02}_{-0.02}$ &
      $2.15^{+0.02}_{-0.01}$ \\
C141 & $1.18^{+0.01}_{-0.01}$ & $1.43^{+0.01}_{-0.01}$ &
      $1.66^{+0.01}_{-0.01}$ & $1.89^{+0.01}_{-0.01}$ &
      $2.14^{+0.01}_{-0.01}$ & ~~~ &
R223 & $1.38^{+0.08}_{-0.06}$ & $1.69^{+0.09}_{-0.08}$ &
      $1.99^{+0.10}_{-0.09}$ & $2.29^{+0.12}_{-0.10}$ &
      $2.54^{+0.14}_{-0.11}$ \\
C142 & $1.29^{+0.50}_{-0.11}$ & $1.58^{+0.58}_{-0.16}$ &
      $1.86^{+0.66}_{-0.20}$ & $2.13^{+0.77}_{-0.25}$ &
      $2.37^{+0.92}_{-0.24}$ & ~~~ &
WHH09 & $1.84^{+0.12}_{-0.12}$ &
      $2.22^{+0.13}_{-0.13}$ & $2.60^{+0.16}_{-0.16}$ &
      $2.99^{+0.20}_{-0.18}$ & $3.38^{+0.22}_{-0.22}$ \\
C143 & $1.18^{+0.01}_{-0.01}$ & $1.42^{+0.01}_{-0.01}$ &
      $1.66^{+0.01}_{-0.01}$ & $1.88^{+0.01}_{-0.01}$ &
      $2.13^{+0.01}_{-0.01}$ & ~~~ &
WHH16 & $1.93^{+0.12}_{-0.12}$ &
      $2.32^{+0.13}_{-0.13}$ & $2.72^{+0.17}_{-0.16}$ &
      $3.14^{+0.22}_{-0.20}$ & $3.56^{+0.23}_{-0.22}$ \\
C144 & $1.18^{+0.01}_{-0.01}$ & $1.43^{+0.01}_{-0.01}$ &
      $1.67^{+0.01}_{-0.01}$ & $1.89^{+0.02}_{-0.01}$ &
      $2.14^{+0.02}_{-0.01}$ & ~~~ &
WHH22 & $1.29^{+0.05}_{-0.04}$ &
      $1.58^{+0.07}_{-0.06}$ & $1.87^{+0.08}_{-0.06}$ &
      $2.14^{+0.09}_{-0.08}$ & $2.37^{+0.10}_{-0.08}$ \\
C145 & $1.21^{+0.03}_{-0.02}$ & $1.47^{+0.04}_{-0.03}$ &
      $1.73^{+0.04}_{-0.03}$ & $1.97^{+0.05}_{-0.04}$ &
      $2.20^{+0.05}_{-0.04}$ & ~~~ &
  &  &  &  &  &  \\
\hline
\end{tabular}

$^{a}$~Denotes clusters inferred on the basis of $(C-T_1)_0$ colour to have
${\rm [Fe/H]}<-2.25$ in Table \ref{tab:n5128colors}. Mass-to-light ratios are
computed for them assuming ${\rm [Fe/H]}=-2.25$ instead.

\end{minipage}
\end{table*}



\begin{figure}
\centerline{\hfil
   \includegraphics[width=84mm]{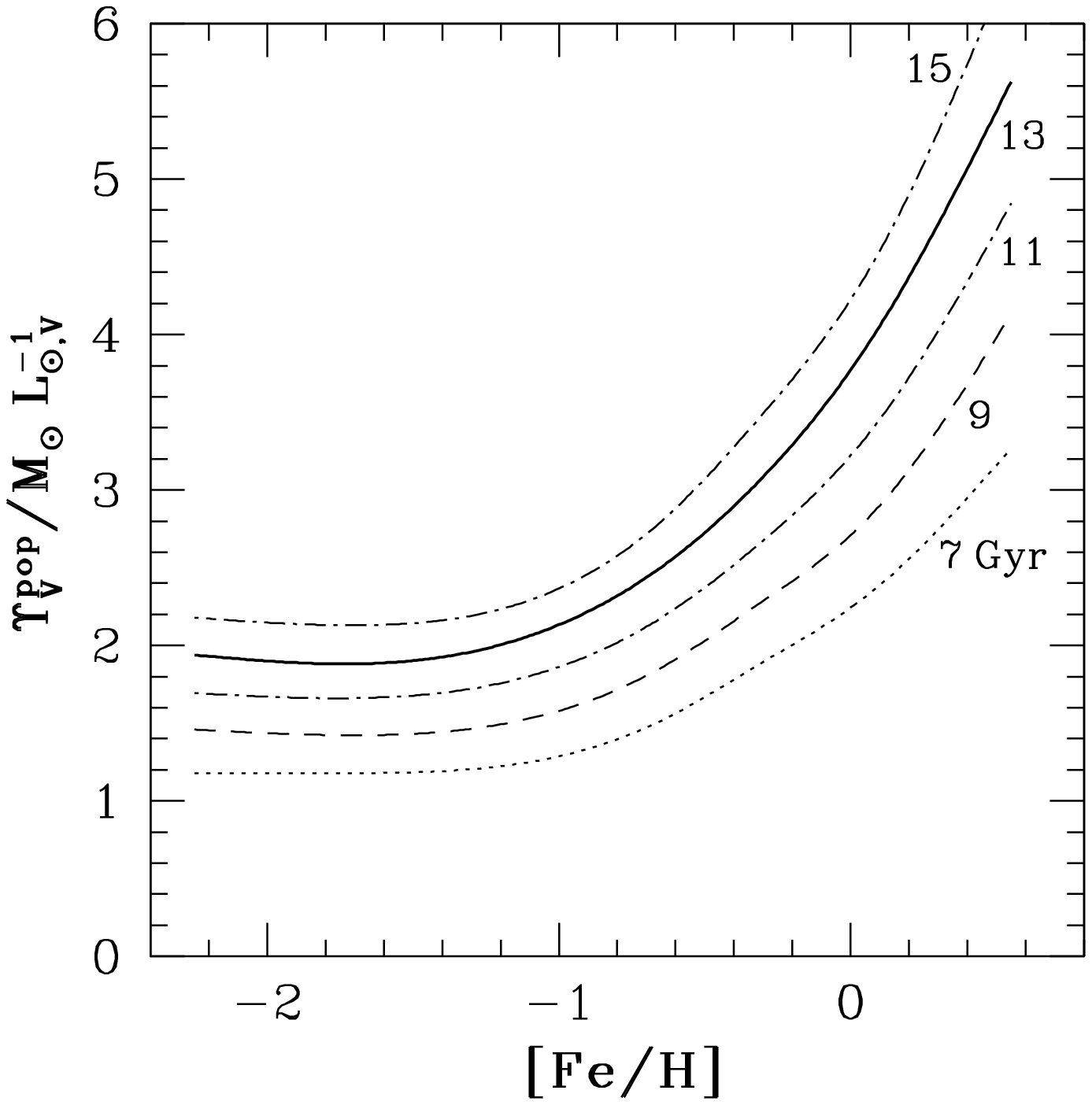}
\hfil}
\caption{
$V$-band mass-to-light ratio vs.~metallicity, from the population-synthesis
models of \citet{bc03} for single-burst clusters with the stellar IMF of
\citet{chab03}. Curves are for cluster ages of 7, 9, 11, 13, and 15 Gyr.
Note that all of our clusters have ${\rm [Fe/H]}\la 0$, and thus
$\Upsilon_V^{\rm pop}<4\ M_\odot\,L_{\odot,V}^{-1}$ for an age of 13 Gyr.
A typical value is closer to
$\Upsilon_V^{\rm pop}\simeq 2.1\ M_\odot\,L_{\odot,V}^{-1}$
for an average $\langle{\rm [Fe/H]}\rangle\simeq-1.0$ in NGC 5128.
\label{fig:modmtol}
}
\end{figure}


Velocity dispersions have been measured for 20 of the GCs in the combined
sample of this paper and \citetalias{har02} \citep{marho04,rej07}.
These can be used in conjunction with surface-brightness modeling to estimate
dynamical mass-to-light ratios. However, since this is possible for only
for a small minority of our clusters, we have chosen instead to use
population-synthesis 
models to {\it predict} mass-to-light ratios for the full sample, and then to
use these to produce (for example) predicted velocity dispersions that can be
compared to current and future spectroscopic data. Since we have
already derived the requisite corrections to put our photometry on the
standard $V$ scale, we need only calculate the mass-to-light ratios in this
bandpass.

This approach is also taken by \citet{mcl05} in their study of 153 globular
and young massive clusters in the Milky Way, the Large and Small Magellanic
Clouds, and the Fornax dwarf spheroidal. \citeauthor{mcl05} discuss the
differences in population-synthesis mass-to-light ratios,
$\Upsilon_V^{\rm pop}$, obtained using
different models (\citealt{bc03} versus \citealt{frv97}) and for a number of
different assumed stellar IMFs. They ultimately work with the results
from the code of \citet{bc03} using the (disk-star) IMF of \citet{chab03} to
produce catalogues of dynamical properties for all of their clusters.
We thus do the same here.

Figure \ref{fig:modmtol} illustrates model curves of $\Upsilon_V^{\rm pop}$
versus cluster metallicity, for ages from 7 to 15 Gyr. Given an age, and a
metallicity from Table \ref{tab:n5128colors}, we interpolate on the curves
in Figure \ref{fig:modmtol} to obtain a model $\Upsilon_V^{\rm pop}$ for any 
cluster in our sample. We eventually assume an age of 13 Gyr for all GCs, in
\S\ref{sec:fits} and subsequent analyses.


\begin{figure}
\centerline{\hfil
   \includegraphics[width=84mm]{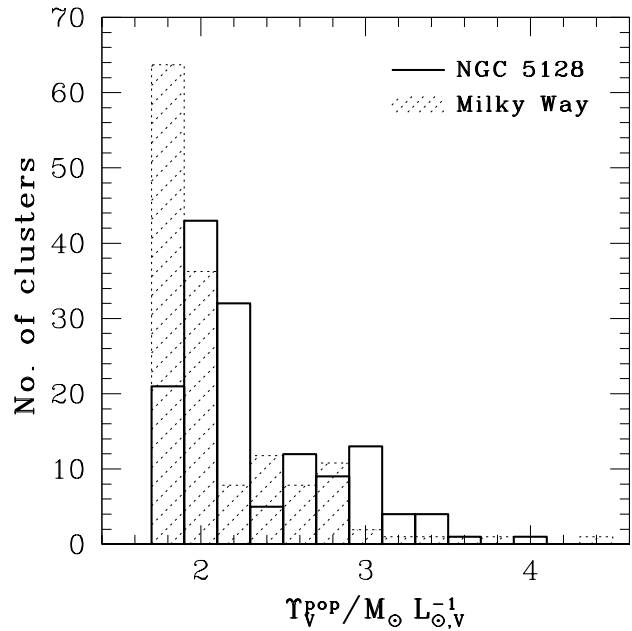}
\hfil}
\caption{
Histogram of population-synthesis $V$-band mass-to-light ratios for 145
GCs in NGC 5128 (excluding C145, C152, C156, and C177; see Table
\ref{tab:badstuff}), for an assumed age of 13 Gyr; and the same distribution
for 148 Galactic globulars (normalized to a total population of 145),
from \citet{mcl05}.
\label{fig:n5128mtol}
}
\end{figure}


Table \ref{tab:n5128mtol} reports $\Upsilon_V^{\rm pop}$ values
for our ACS sample plus the 18 additional GCs from \citetalias{har02}. The
first column lists the cluster name, and subsequent columns give the model
mass-to-light ratios for assumed ages of 7, 9, 11, 13, and 15 Gyr.
As was just suggested, in the
majority of cases we have assumed the cluster metallicity in Column
(6) of Table \ref{tab:n5128colors} to evaluate $\Upsilon_V^{\rm pop}$.
However, there are 9 clusters in the table
with apparent ${\rm [Fe/H]}<-2.25$, below the minimum for which the
\citeauthor{bc03} models are defined. We have reset these to have
${\rm [Fe/H]}=-2.25$ so we can still obtain some estimate of
$\Upsilon_V^{\rm pop}$; but note that such low metallicities follow
from very blue $(C-T_1)_0$ colours, which may be indicating that these
objects are much younger than 13 Gyr (C119 and PFF029, which we
mentioned at the end of \S\ref{subsec:transmet}, are in this group).

Figure \ref{fig:n5128mtol} shows a histogram of $\Upsilon_V^{\rm pop}$ for
GCs in NGC 5128 assuming a common age of 13 Gyr, and compares it to the
distribution of mass-to-light ratios for 148 Galactic
globulars, as calculated by \citet{mcl05}. Note that the distribution in NGC
5128 peaks at a slightly higher $\Upsilon_V^{\rm pop}$ and is rather broader
than that in the Milky Way. This just reflects the different averages and
dispersions of the GC metallicity distributions in the two galaxies.

\subsection{Structural models}
\label{subsec:structure}

We fit a number of  models to the density profile of each cluster observed
with ACS/WFC.

First is the usual \citet{king66} single-mass, isotropic, modified isothermal
sphere, which is defined by the stellar distribution function (phase-space
density)
\begin{equation}
f(E)\propto\left\{
\begin{array}{ll}
\exp[-E/\sigma_0^2] -1 &,\ E<0 \\
0 &,\ E\ge 0\ ,
\end{array}
\right.
\label{eq:king66}
\end{equation}
where $E$ is the stellar energy. Under certain conditions, this
formula roughly approximates a steady-state solution of the Fokker-Planck
equation \citep[e.g.,][]{king65}. It is the standard model that is routinely
fit to GC surface-brightness profiles.

Second is a further modification of a single-mass, isotropic isothermal
sphere, based on a model originally introduced by \citet{wil75} for elliptical
galaxies. It has recently been fitted to globular and young massive clusters
in the Milky Way and its satellites by \citet{mcl05}, and to GCs in M31 by
\citet{barmby07}. This model is again defined by a phase-space distribution
function: 
\begin{equation}
f(E)\propto\left\{
\begin{array}{ll}
\exp[-E/\sigma_0^2] -1 + E/\sigma_0^2 &,\ E<0 \\
0 &,\ E\ge 0\ .
\end{array}
\right.
\label{eq:wilson}
\end{equation}
\citet{wil75} originally included a multiplicative term in the
distribution function depending on the angular momentum $J_z$, in order to
create axisymmetric model galaxies. We omit this term to make
spherical and isotropic cluster models, but we still refer to equation
(\ref{eq:wilson}) as Wilson's model. The connection with the \citet{king66}
model in equation (\ref{eq:king66}) is clear: the extra $+ E/\sigma_0^2$ in
the first line of equation (\ref{eq:wilson}) removes the
linear term from a Taylor series expansion of the Maxwellian
$\exp(-E/\sigma_0^2)$ near the zero-energy (tidal) boundary of the
cluster. Its effect is to make clusters that are spatially more extended than
in isotropic \citet{king66} models. To see this, note that adding
$E/\sigma_0^2$ to the exponential term causes a much more significant decrease
in the phase-space density of tightly bound stars with $E\ll 0$ than it does
in the density of loosely bound stars with $E$ near 0, which have large
orbital apocentres.

As \citet{mcl05} discuss in detail, extended \citeauthor{wil75}-type cluster
models fit the majority of old and young
massive clusters in the Local Group at least as well as, and often
significantly better than, \citeauthor{king66} models.  The difference between
the King and Wilson models when compared with real cluster data tends to show
up most prominently at large radii approaching the nominal tidal radius.
At small and intermediate radii (within the core, and out to slightly beyond
the half-mass radius), the two model profiles are quite similar.  It is worth
noting \citep[see also][]{mcl05} that in the earlier era of measured data
that almost never extended to large radii,
it would not have been possible to decide between these models.  
The much larger amounts of cluster profile data now available, which
extend more reliably out to 
the cluster tidal radii and even beyond, make it possible to start
discriminating 
betwen models more accurately.  In such cases, we find that the Wilson
model treats the 
entire cluster profile more accurately without the need to invoke
arbitrary amounts of 
``extra-tidal light'' beyond the formal King profile boundaries.

Spherically symmetric, dimensionless density and velocity-dispersion profiles
are obtained for \citet{king66} and \citeauthor{wil75} models by appropriate
integrals of $f(E)$ over all velocities in the first case, and over all
spatial radii in the second. These are then integrated along the line
of sight to produce normalized surface densities $I/I_0$ and
velocity dispersions $\sigma_{\rm p}/\sigma_0$ as functions of a dimensionless
projected clustercentric radius $R/r_0$, for comparison with observations.
Here $r_0$ is a scalelength associated with, but not equivalent to, the
observed half-power radius $R_c$, as discussed below. The shapes of
both profiles  
are fully specified by the value of a dimensionless central potential,
$W_0 \equiv -\phi(0)/\sigma_0^2 > 0$. In principle $W_0$ can take
on any real value between 0 and $\infty$, with the latter limit corresponding
in both models to a non-singular isothermal sphere of infinite extent.
$W_0$ bears a one-to-one relationship with the more intuitive
concentration parameter: $c\equiv \log\,(r_t/r_0)$, where $r_t$ is
the tidal radius of the model cluster $[\rho(r_t)=0]$. As suggested above,
however, the tidal radii of \citeauthor{wil75} models are generally larger
than those of otherwise similar \citet{king66} models, so that the same
cluster data will almost always return different $c$ values for the two models;
this parameter, which is frequently mentioned in discussions of GC properties,
is a highly model-dependent quantity. See \citet{mcl05} for further discussion
of this point.


\begin{table*}
\begin{minipage}{170mm}
\scriptsize
\caption{Basic parameters of fits to 147 profiles of 131 GCs in NGC 5128
\label{tab:n5128fits}}
\begin{tabular}{@{}lcccclrrrccrr}
\hline
\multicolumn{1}{c}{Name} &
\multicolumn{1}{c}{Detector} &
\multicolumn{1}{c}{$A_{F606}$} &
\multicolumn{1}{c}{$(V-F606)_0$} &
\multicolumn{1}{c}{$N_{\rm pts}$} &
\multicolumn{1}{c}{Model} &
\multicolumn{1}{c}{$\chi_{\rm min}^2$} &
\multicolumn{1}{c}{$I_{\rm bkg}$} &
\multicolumn{1}{c}{$W_0$} &
\multicolumn{1}{c}{$c/n$} &
\multicolumn{1}{c}{$\mu_0$ (F606W)} &
\multicolumn{1}{c}{$\log\,r_0$} &
\multicolumn{1}{c}{$\log\,r_0$} \\
    &
    &
\multicolumn{1}{c}{[mag]} &
\multicolumn{1}{c}{[mag]} &
    &
    &
    &
\multicolumn{1}{c}{[$L_\odot\,{\rm pc}^{-2}$]} &
    &
    &
\multicolumn{1}{c}{[mag arcsec$^{-2}$]} &
\multicolumn{1}{c}{[arcsec]} &
\multicolumn{1}{c}{[pc]}    \\
\multicolumn{1}{c}{(1)}  &
\multicolumn{1}{c}{(2)}  &
\multicolumn{1}{c}{(3)}  &
\multicolumn{1}{c}{(4)}  &
\multicolumn{1}{c}{(5)}  &
\multicolumn{1}{c}{(6)}  &
\multicolumn{1}{c}{(7)}  &
\multicolumn{1}{c}{(8)}  &
\multicolumn{1}{c}{(9)}  &
\multicolumn{1}{c}{(10)} &
\multicolumn{1}{c}{(11)} &
\multicolumn{1}{c}{(12)} &
\multicolumn{1}{c}{(13)} \\
\hline
   AAT111563  & WFC/F606   & $0.308$  & $0.091\pm0.050$      & $42$       &
                K66   & $42.97$  & $2.56\pm0.15$  & $6.70^{+0.10}_{-0.20}$  &
               $1.44^{+0.03}_{-0.06}$  & $16.64^{+0.06}_{-0.03}$  &
               $-1.206^{+0.025}_{-0.014}$  & $0.059^{+0.025}_{-0.014}$ \\
          ~~        & ~~        & ~~        & ~~        & ~~         &
                W   & $42.87$  & $2.25\pm0.25$  & $6.20^{+0.20}_{-0.20}$  &
               $1.89^{+0.10}_{-0.09}$  & $16.76^{+0.04}_{-0.04}$  &
               $-1.129^{+0.019}_{-0.020}$  & $0.136^{+0.019}_{-0.020}$ \\
          ~~        & ~~        & ~~        & ~~        & ~~        &
                S   & $61.87$  & $2.57\pm0.17$       & ---  &
               $1.92^{+0.13}_{-0.12}$  & $15.27^{+0.20}_{-0.21}$  &
               $-2.140^{+0.131}_{-0.145}$  & $-0.875^{+0.131}_{-0.145}$ \\
   AAT113992  & WFC/F606   & $0.308$  & $0.200\pm0.050$      & $26$       &
                K66   & $10.19$  & $-80.13\pm6.35$  & $6.60^{+0.30}_{-0.20}$
                &
               $1.41^{+0.09}_{-0.05}$  & $16.96^{+0.02}_{-0.02}$  &
               $-1.096^{+0.010}_{-0.014}$  & $0.170^{+0.010}_{-0.014}$ \\
          ~~        & ~~        & ~~        & ~~        & ~~         &
                W   & $15.15$  & $-90.92\pm8.88$  & $7.10^{+0.40}_{-0.40}$  &
               $2.46^{+0.33}_{-0.29}$  & $16.97^{+0.03}_{-0.02}$  &
               $-1.082^{+0.018}_{-0.016}$  & $0.183^{+0.018}_{-0.016}$ \\
          ~~        & ~~        & ~~        & ~~        & ~~        &
                S   & $11.51$  & $-73.95\pm3.30$       & ---  &
               $1.54^{+0.08}_{-0.08}$  & $16.13^{+0.10}_{-0.10}$  &
               $-1.660^{+0.071}_{-0.072}$  & $-0.395^{+0.071}_{-0.072}$ \\
   AAT115339  & WFC/F606   & $0.308$  & $0.148\pm0.050$      & $35$       &
                K66   & $47.87$  & $-42.66\pm0.41$  & $13.70^{+1.10}_{-1.00}$
                &
               $3.08^{+0.23}_{-0.20}$  & $11.08^{+0.43}_{-0.51}$  &
               $-3.018^{+0.191}_{-0.214}$  & $-1.752^{+0.191}_{-0.214}$ \\
          ~~        & ~~        & ~~        & ~~        & ~~         &
                W  & $116.80$  & $-45.30\pm2.73$  & $6.40^{+0.40}_{-0.40}$  &
               $2.00^{+0.25}_{-0.20}$  & $15.78^{+0.10}_{-0.11}$  &
               $-1.255^{+0.043}_{-0.047}$  & $0.010^{+0.043}_{-0.047}$ \\
          ~~        & ~~        & ~~        & ~~        & ~~        &
                S   & $38.38$  & $-43.78\pm0.92$       & ---  &
               $2.15^{+0.15}_{-0.15}$  & $13.92^{+0.26}_{-0.26}$  &
               $-2.511^{+0.173}_{-0.177}$  & $-1.246^{+0.173}_{-0.177}$ \\
\hline
\end{tabular}

\medskip
  A machine-readable version of the full Table \ref{tab:n5128fits} is
  available online
  (http://www.astro.keele.ac.uk/$\sim$dem/clusters.html)
  or upon request from the first author.
  Only a short extract from it is shown here, for guidance
  regarding its form and content.

\end{minipage}
\end{table*}


The third model we fit to our data is defined by directly parametrizing the
observable surface-density profile. It is the \citet{sersic} or $R^{1/n}$
model, which, like the \citeauthor{wil75} model, was
also originally designed for application to galaxies. We write this slightly
differently than is usually done:
\begin{equation}
I(R)=I_0\ \exp\left[-\ln(2)\times(R/r_0)^{1/n}\right] \ ,
\label{eq:sersic}
\end{equation}
in which $n>0$, and $r_0$ is the projected radius at which the surface density
falls to half its central value $I_0$. Often \citeauthor{sersic} models are
defined in terms of the projected half-light (or effective) radius---which we
denote by $R_h$---such that the exponentiated term in brackets in equation
(\ref{eq:sersic}) is $[ -b_n (R/R_h)^{1/n} ]$, with
$b_n=\ln(2)\times(R_h/r_0)^{1/n}$ a function of $n$ that we compute
numerically. Note that \citeauthor{sersic} models have a formally infinite
spatial extent. However, the density profile falls steeply at very large $R$,
and the models thus have finite total luminosities and 
well-defined $R_h$ for any $n$. One slight complication is that the
{\it de-projected} density profiles $j(r)$ corresponding to the observable
$I(R)$ are weakly divergent in the central limit $r\rightarrow 0$ for
$n\ge 1$. We return to this point just below.

As we have already discussed, setting $W_0$ at some value for \citet{king66}
and \citeauthor{wil75} models leads to the full definition of
three-dimensional and projected density and velocity-dispersion profiles, the
shapes of which are fixed by $W_0$ or, equivalently, a
value of $c=\log(r_t/r_0)$. The analogous shape parameter for
the \citeauthor{sersic} models is the index $n$ in equation
(\ref{eq:sersic}). Even though there is no longer a connection between the
shape parameter and any tidal radius in this case, there is still a one-to-one
correspondence between $n$ and a dimensionless central potential
$W_0=-\phi(0)/\sigma_0^2$, which is always finite. This works in the sense
that a higher $n$ implies density profiles that fall off more slowly with
increasing radius, which in turn implies a deeper central potential, or higher
$W_0$.  

Given a value for $n$, we numerically deproject equation (\ref{eq:sersic})
to compute the volume luminosity density $j(r)$, which is then used to solve
the spherical Jeans equation \citep{bt87}---assuming unit mass-to-light ratio
and an isotropic velocity ellipsoid---to obtain a normalized
velocity-dispersion profile. This can be re-projected along the line of sight
for comparison with data. The complete structural and dynamical details of any
\citeauthor{sersic} model cluster are then known in full, just as they are for
\citet{king66} and \citeauthor{wil75} models.

Equations (\ref{eq:king66}) and (\ref{eq:wilson}) differ formally from
equation (\ref{eq:sersic}): the first two include a velocity scale parameter 
$\sigma_0$, while the last uses an explicit radial scale $r_0$. In the
formulation of his model, \citet{king66} defined a radial scale associated
with $\sigma_0$: $r_0^2\equiv 9\sigma_0^2/(4\pi G \rho_0)$,
where $\rho_0$ is the central mass density of the model and the numerical
coefficients are chosen so as to make $I(r_0) \approx I_0/2$ in most models.
We adopt the same definition for our single-mass, isotropic \citeauthor{wil75}
models.  For both of these, we therefore have
\begin{equation}
\sigma_0^2\equiv \frac{4\pi G \rho_0 r_0^2}{9}
     \qquad {\rm (King\ and\ Wilson\ models)}\ .
\label{eq:sigma0}
\end{equation}
For the \citeauthor{sersic} model, however, as we
mentioned above it happens that $\rho_0\rightarrow\infty$ when $n\ge 1$.
Thus, for any $n$ in these models we
define a velocity scale in terms of the central {\it surface} mass density,
$\Sigma_0=\Upsilon I_0$, and the core radius $r_0$:
\begin{equation}
\sigma_0^2 \equiv \frac{2\pi G \Sigma_0 r_0}{9}
     \qquad {\rm (Sersic\ models\ only)} \ .
\label{eq:sersigma0}
\end{equation}
The factor of two in the numerator here is chosen for
maximum compatibility with equation (\ref{eq:sigma0}) for the other models,
which tend to have $\Sigma_0\approx 2 \rho_0 r_0$
for the $W_0$ or $c$ values of most real clusters.

Although the model scale $r_0$ in \citet{king66} and \citeauthor{wil75} models
is generally close to an observational core (projected half-power) radius,
it is important to recognize that the two are not identical in principle; an
equivalence holds only for \citeauthor{sersic} models, by virtue of our
definition of it in equation (\ref{eq:sersic}). Similarly, the
model velocity scale $\sigma_0$ is {\it never} equal to the velocity
dispersion (projected or unprojected) at the centres of clusters. The
connections between the model parameters and these observable quantities are
straightforward to derive for any member of our model families, but this can
only be done numerically.

We additionally fit all of our cluster profiles with the analytical
parametrization of $I(R)$ introduced by \citet{king62}, and with ``power-law''
models in which $I(R) \propto R^{1-\gamma}$ at large projected radii and
$I(R) \rightarrow {\rm constant}$ at small $R$. Other authors have also fitted
these models to many old GCs and young massive clusters, but here we
have found that doing so yields no substantively new information beyond
what can be learned from \citet{king66}, \citeauthor{wil75}, and
\citeauthor{sersic} fits. We therefore do not discuss these other alternatives
any further, except to show explicitly (in \S\ref{subsec:chicomp} below)
that they never outperform \citeauthor{wil75} fits to the NGC 5128 cluster
profiles in any case.

\section{Fits}
\label{sec:fits}

The models described in \S\ref{subsec:structure} are fitted
to our data after first being convolved with the ACS/WFC
PSF for the $F606W$ filter. Given a value for the scale radius $r_0$
discussed in \S\ref{subsec:structure}, and some specified shape parameter,
we compute a dimensionless model profile
$\widetilde{I}_{\rm mod}\equiv I_{\rm mod}/I_0$ and then perform
the convolution
\begin{equation}
\widetilde{I}_{\rm mod}^{*} (R | r_0) =
\int\!\!\!\int_{-\infty}^{\infty}
               \widetilde{I}_{\rm mod}(R^\prime/r_0) \times
               \widetilde{I}_{\rm PSF}
                    \left[(x-x^\prime),(y-y^\prime)\right]
       \ dx^\prime \, dy^\prime\ ,
\label{eq:convol}
\end{equation}
in which $R^2=x^2+y^2$; $R^{\prime\,2}=x^{\prime\,2}+y^{\prime\,2}$; and
$\widetilde{I}_{\rm PSF}$ is the PSF profile normalized to unit total
luminosity. We use the circularly symmetric function in equation
(\ref{eq:f606psf}) to approximate $\widetilde{I}_{\rm PSF}$. We also allow
for a non-zero background $I_{\rm bkg}$, and so ultimately minimize the
standard
\begin{equation}
\chi^2  =
  \sum_i{
 \frac{\left[I_{\rm obs}(R_i)
             - I_0\times \widetilde{I}_{\rm mod}^{*}(R_i | r_0)
             - I_{\rm bkg}\right]^2}
      {\sigma_i^2}
        }
\label{eq:chi2}
\end{equation}
for the measured intensity profile and uncertainties of any cluster in
Table \ref{tab:N5128sbprofs}.

In practice, we identify the best-fit member
of a model family by first computing unconvolved model profiles
for a large number of fixed values of the appropriate shape parameter ($W_0$
for \citealt{king66} and \citeauthor{wil75} models; $n$ for
\citeauthor{sersic} models). Given any one model in such a pre-set sequence,
we convolve it with the PSF and then vary $r_0$ (with a new convolution
required for every change), $I_0$, and $I_{\rm bkg}$ until $\chi^2$ is
minimized for that particular value of $W_0$ or $n$. We do this in turn for
every model in the grid to identify the single best fit, with the smallest
$\chi^2$, from the chosen family. This procedure further allows us to
estimate uncertainties for all fitted and derived model
parameters, from the range of their values over all models for which $\chi^2$
is within a specified distance of the absolute minimum (e.g.,
$\delta\chi^2\le 1$ for 68\% confidence intervals). 

\subsection{Main model parameters and example fits}
\label{subsec:profs}

Table \ref{tab:n5128fits} summarises the basic ingredients of all model fits
to our full ACS cluster sample, including the duplicate profiles of the
obects listed in Table \ref{tab:dual}. Only a sample of Table
\ref{tab:n5128fits} is shown here; it is available in full, in
machine-readable format, online or upon request from the first author.

The first column of Table \ref{tab:n5128fits} gives the cluster name. The
second column 
reports the detector/filter combination from which we derived our observed
density profile. This is always WFC/$F606$ here, and is written out only for
compatibility with analogous tables reporting our parallel work on M31
clusters in \citet{barmby07}. The third column of the table lists the
$F606W$-band extinction [a constant $A_{F606}=2.8 E(B-V)=0.308$ mag for all
clusters in NGC 5128]; the fourth column is the colour term $(V-F606)_0$ to
transform photometry from the native bandpass of the data to the standard $V$
scale (from Table \ref{tab:n5128colors}); and the fifth column records the
number of points in the intensity profile that are flagged as OK in Table
\ref{tab:N5128sbprofs} above, and thus were used to constrain our model fits.

The subsequent columns in Table \ref{tab:n5128fits} cover three lines for each
cluster, one line for each type of model fit. Each line records:

\medskip
Column (6): identification of the model being fitted.

Column (7): the minimum $\chi^2$ ({\it not} divided by the number of
degrees of freedom) obtained for the best fit in that class of model.

Column (8): the best-fit background intensity in the $F606W$ bandpass.

Column (9): the dimensionless central potential $W_0$ of the best-fitting
model (for \citealt{king66} and \citeauthor{wil75} models only).

Column (10): the concentration $c\equiv\log(r_t/r_0)$ for
\citet{king66} and \citeauthor{wil75} models, or the index $n$ of the
best \citeauthor{sersic} fit.

Column (11): the extinction-corrected central surface
brightness {\it in the $F606W$ bandpass}; the intrinsic $V$-band
central surface brightness follows from adding the colour in column (4).

Column (12): the logarithm of the best-fit scale radius $r_0$ in arcsec
(see \S\ref{subsec:structure}).

Column (13): the logarithm of $r_0$
in units of pc (obtained from the angular scale assuming $D=3.8$ Mpc for NGC
5128).

\medskip
To estimate the errorbars on these parameters (and those on all the
derived quantities discussed in \S\ref{subsec:derived} below), we first
rescaled the $\chi^2$ for all fitted models in any one family, by a common
factor chosen to make the global minimum $\chi_{\rm min}^2 = (N_{\rm pts}-4)$,
where $N_{\rm pts}$ is the number of points used in the model 
fitting. Under this re-scaling, the global minimum $\chi^2$ per
degree of freedom is exactly one. We then found the minimum and maximum values
of any chosen parameter in all models with a re-scaled
$\chi^2 \le \chi_{\rm min}^2+4$. Normally, if $\chi^2$ were not re-scaled,
this would give a 95\% confidence interval on the parameter. Here it is
just a well-defined, quantitative way to assign reasonable uncertainties to
any model quantity for any cluster.

\begin{figure*}
\centerline{\hfil
   \includegraphics[width=160mm]{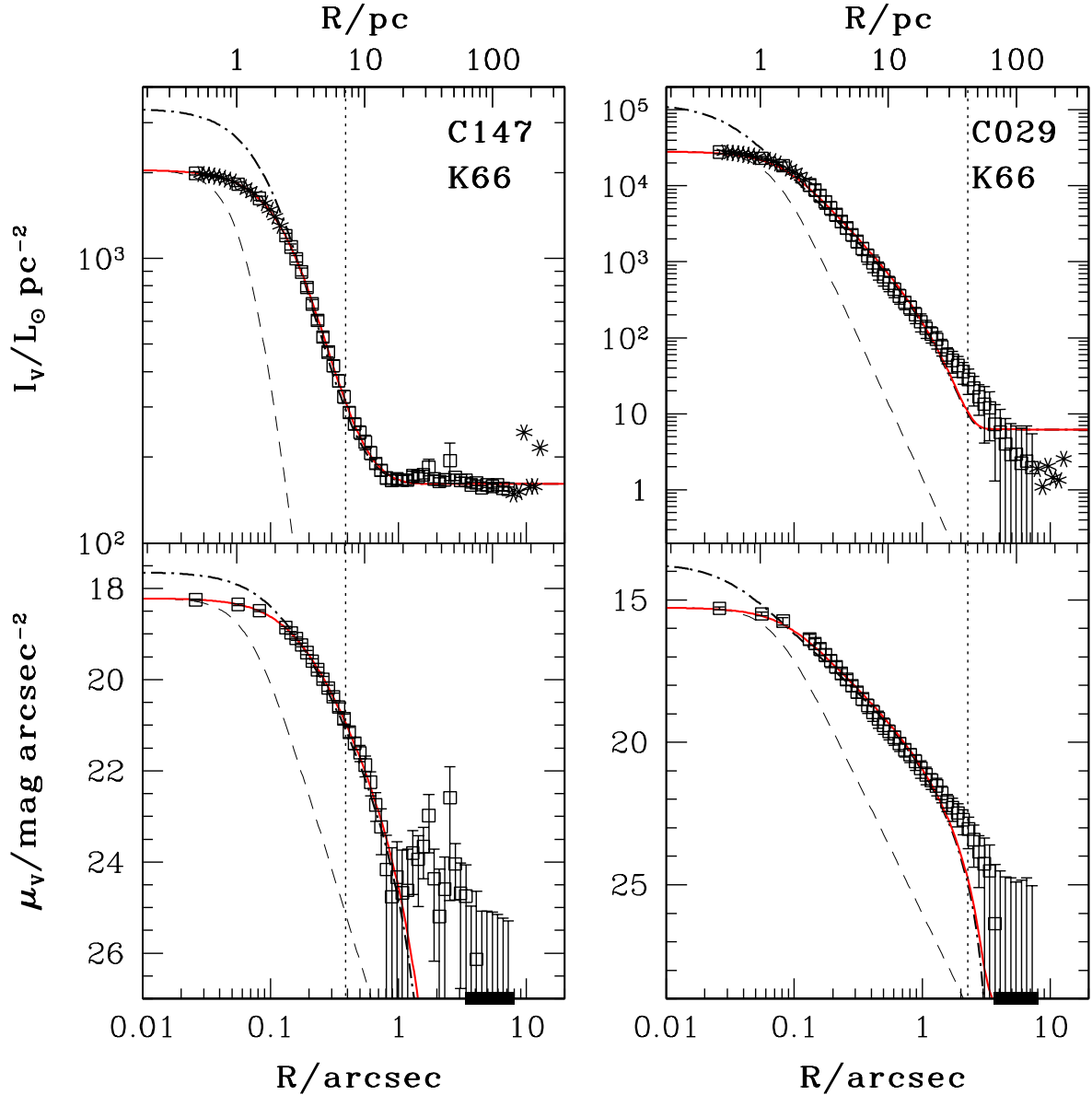}
\hfil}
\caption{
Examples of \citet{king66} model fits to two clusters in our sample: the
fairly average C147 (left panels), and the very bright C029 (right panels),
which has a large halo not well described by this standard model
\citepalias[cf.][]{har02}. Dashed curves trace the PSF intensity profile; bold
dash-dot curves, the unconvolved best-fit model (with background added in the
upper panels but not in the lower ones); and solid (red) curves, the
PSF-convolved best fits. In each panel the lower $x$-axis is marked in
arcsec and the upper $x$-axis in parsec.
Vertical dotted lines mark the radius where the
intrinsic cluster intensity is equal to the best-fit background. See text for
more details.
\label{fig:exfit}
}
\end{figure*}

Figure \ref{fig:exfit} shows the best-fit \citet{king66} models for the
GCs C147 and C029. The first of these is of average brightness and size
($M_V=-7.6$ and effective radius $R_h=3.8$~pc from the model fit). The second
is much brighter and obviously more extended (fitted $M_V=-10.2$ and
$R_h=5.1$~pc). C029 was previously observed with STIS as part of the
sample of \citetalias{har02}, where it was identified as one of several large
clusters with extended haloes showing significant excess power beyond the
nominal \citet{king66} tidal radius. This is also obvious in our new
analysis.


The top panels of Figure \ref{fig:exfit} show the intensity versus radius
profiles output from ELLIPSE for each cluster, after our conversion
from $F606W$ 
to extinction-corrected $V$, described in \S\ref{sec:data}, but before
subtracting a constant background. The lower panels of the figure show the
intensity profiles after 
subtracting the fitted backgrounds and converting to surface brightness,
$\mu_V=26.402-2.5\,\log(I_V/L_\odot\,{\rm pc}^{-2})$. In all panels, the
lower $x$-axis marks the projected radius $R$ in arcsec and the upper $x$-axis
gives the equivalent scale in pc.
The dashed curves falling steeply towards large $R$ show the
(arbitrarily normalized) shape of the PSF in equation (\ref{eq:f606psf}). The
bold, dot-dash curves are the intrinsic (unconvolved) best-fit \citet{king66}
models, with an added background in the upper panels but not in the lower
panels. The solid (red) lines are the PSF-convolved models. Open squares
are data points that were included in the least-squares model fitting,
i.e., those flagged as OK in Table \ref{tab:N5128sbprofs}.
Asterisks in the upper panels are points that were not used to
constrain the fits (flagged as BAD, SAT, or DEP in Table
\ref{tab:N5128sbprofs}); they have been omitted altogether from the lower
plots. The dotted vertical lines in all four panels are at the radius in
each cluster where the intrinsic model intensity is equal to the fitted
background level. For radii larger than this, the observed intensities are at
least 50\% background according to the modeling. Points with intensities below
the subtracted backgrounds are represented in the lower panels by solid points
placed on the lower $x$-axes, with errorbars extending upwards.

The plots of the average C147 in Figure \ref{fig:exfit} are typical of the
results for most of the clusters in our sample, in the sense that our
derived intensity profiles extend to large enough radius that a constant
background level has been reached. In this majority of cases, our procedure
of fitting for the background at the same time as the intrinsic model
parameters is very robust, with the same $I_{\rm bkg}$ estimated assuming any
one of the three intrinsic cluster models discussed in
\S\ref{subsec:structure}. C147 is further typical of many moderate- and
low-brightness objects with $L\la 10^5\,L_\odot$, in that the local ``sky''
level dominates any cluster signal outside of just a few intrinsic effective
radii. Deviations of the data from the best-fit models in any of our five
families occur mostly in this region of noise, and thus there is little
difference in the overall quality of fit (and no large changes in the derived
cluster parameters) from one type of model to another. 

By contrast, the plots of C029 in Figure \ref{fig:exfit} show an interesting
phenomenon that is seen in most of the very bright clusters in our
sample. As we also mentioned above, it is clear here that a \citet{king66}
model is simply not a good description of the data. One way of expressing this
is to note the systematic excess of measured intensity over the best-fit model
at radii $R\ga 1\arcsec\simeq18$~pc, or about 3.5 fitted half-light radii. In
order to minimize $\chi^2$ in this case, the fitted background intensity is
spuriously high, coming in at roughly the average level of the outermost
datapoints but failing utterly to reflect the clear, continual decreasing
density profile of the real cluster. A stronger way of stating this
``problem'' is that any \citet{king66} model is too concave to match
the data for this cluster: the model does not have the shape of the observed
surface-brightness profile even at intermediate radii, outside of the
PSF-blurred core but inside of where the fitted background level is reached.

Either way, on the basis of this failure of the
standard \citet{king66} models, \citetalias{har02} identified C029 and 5 other
clusters similar to it as showing possible evidence for ``extratidal light.''
We have now found several other examples like this, but we have also fit them 
with alternate structural models that have intrinsically more extended haloes
by construction. We generally find that the \citeauthor{wil75} models for
such objects return lower fitted background levels, different global cluster 
parameters such as $R_h$ and $L_{\rm tot}$, and often drastically smaller
$\chi^2$ values. It would appear that the ``excess'' light implied by
\citet{king66} model fits to some large clusters is likely a symptom of 
generic shortcomings in the model itself---the theoretical basis for it is
weak in the low-density and unrelaxed farthest reaches of cluster
envelopes---rather than the signature of genuine tidal debris. \citet{mcl05}
reached a similar conclusion after comparing \citet{king66} and
\citeauthor{wil75} fits to more than 150 old and young massive clusters in the
Milky Way and some of its satellites. 


\begin{figure*}
\centerline{\hfil
   \includegraphics[width=175mm,height=215mm]{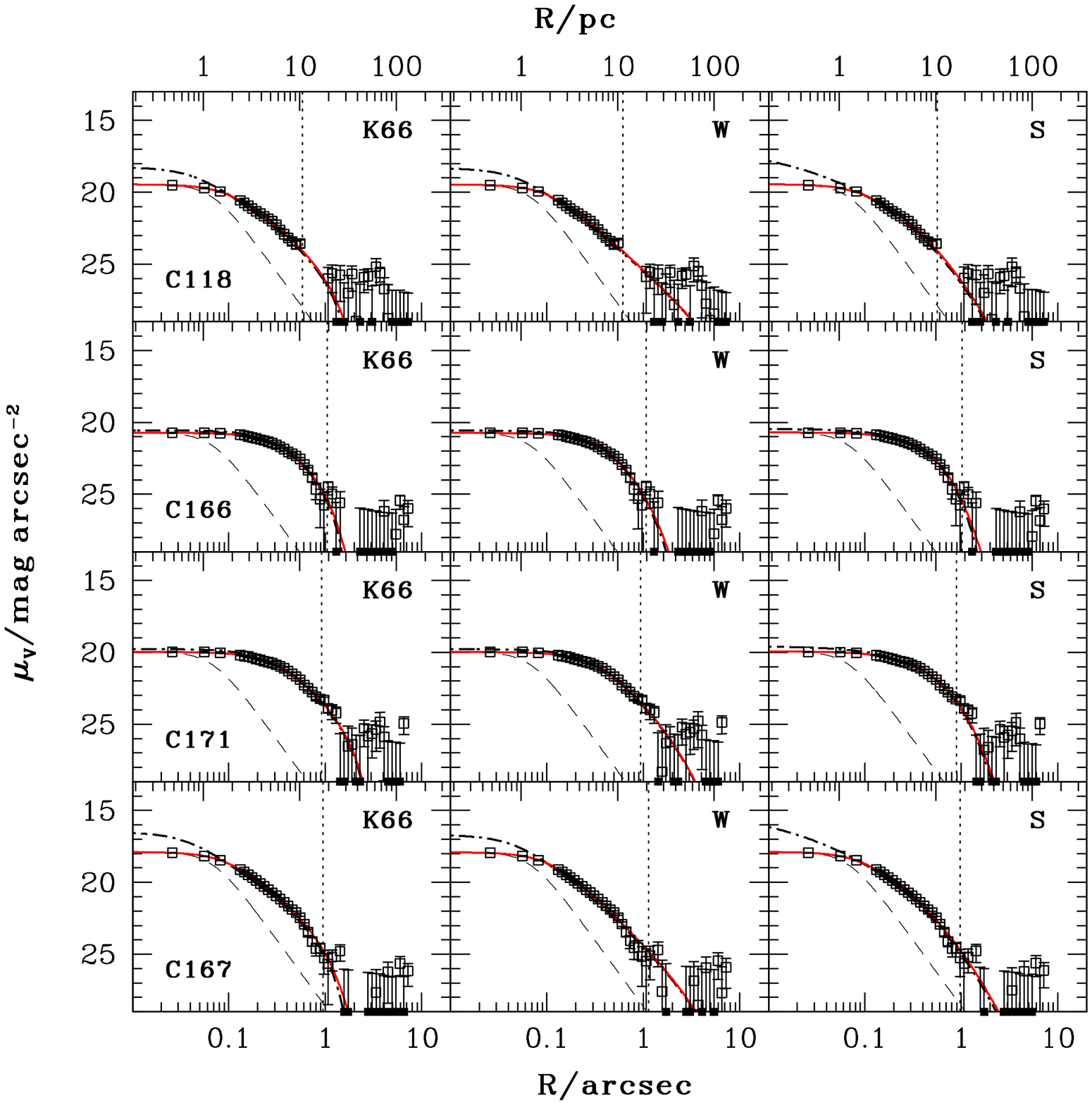}
\hfil}
\caption{
Examples of \citet{king66}, \citeauthor{wil75}, and \citeauthor{sersic} model
fits to 8 globular clusters in NGC 5128. Results are shown in the same
background-subtracted format as the lower panels of Figure \ref{fig:exfit},
with all points and curves having the same meaning here as there.
Clustercentric distance is marked in arcsec along the lower
$x$-axis in each panel, and in parsec along the upper $x$-axis.
Clusters are presented in order of increasing brightness from top to bottom:
$M_V=-6.0$ for C118, $M_V=-6.5$ for C166, $M_V=-7.4$ for C171,
$M_V=-7.4$ for C167, $M_V=-8.7$ for C113, $M_V=-8.8$ for F1GC15,
$M_V=-10.2$ for C029, and $M_V=-11.2$ for C007
(all magnitudes from \citealt{king66} model fits and assuming $D=3.8$~Mpc).
\label{fig:morefits}
}
\end{figure*}

\begin{figure*}
\centerline{\hfil
   \includegraphics[width=175mm,height=215mm]{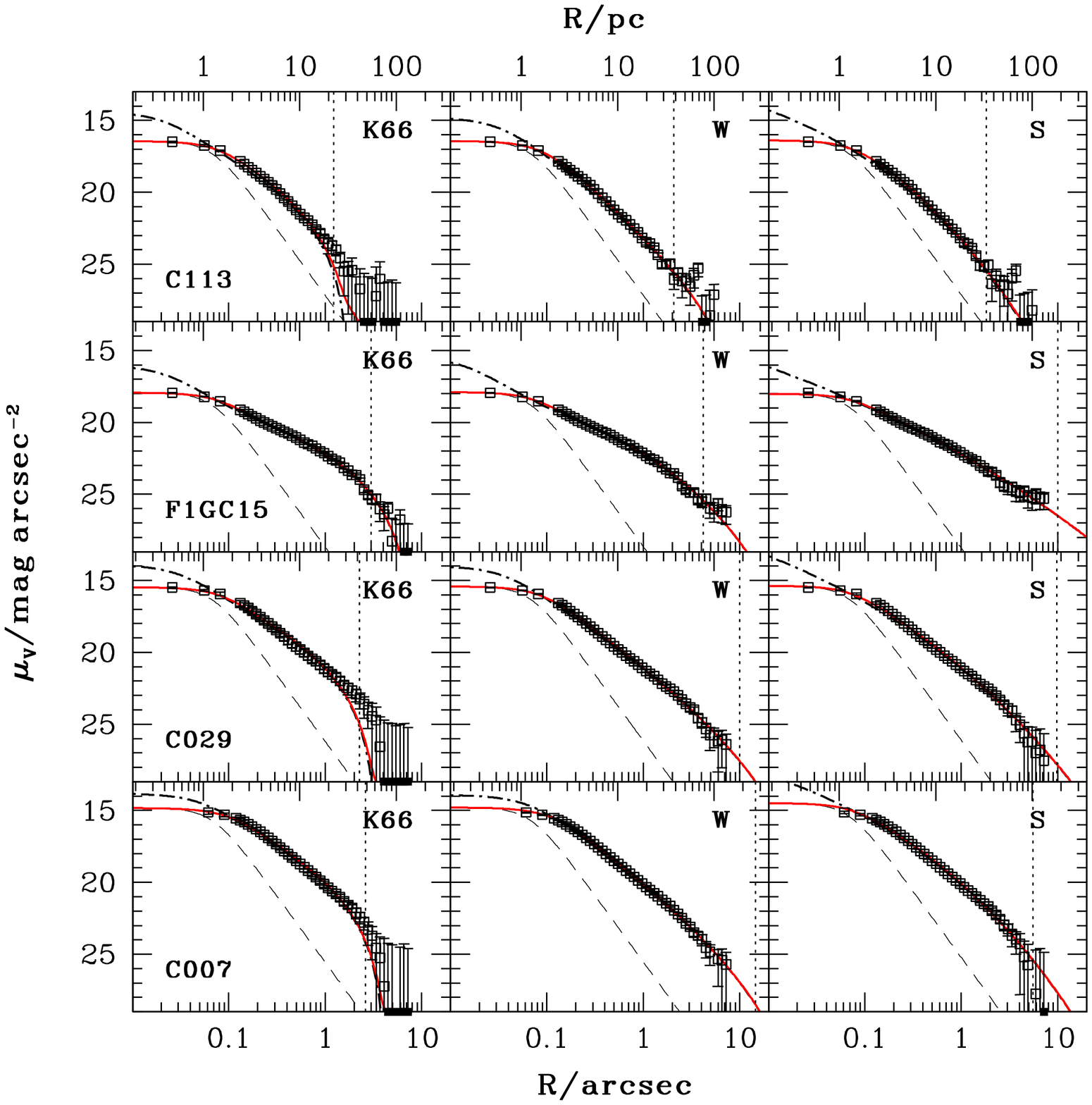}
\hfil}
\contcaption{
Examples of \citet{king66}, \citeauthor{wil75}, and \citeauthor{sersic} model
fits to 8 globular clusters in NGC 5128. Results are shown in the same
background-subtracted format as the lower panels of Figure \ref{fig:exfit},
with all points and curves having the same meaning here as there.
Clustercentric distance is marked in arcsec along the lower
$x$-axis in each panel, and in parsec along the upper $x$-axis.
Clusters are presented in order of increasing brightness from top to bottom:
$M_V=-6.0$ for C118, $M_V=-6.5$ for C166, $M_V=-7.4$ for C171,
$M_V=-7.4$ for C167, $M_V=-8.7$ for C113, $M_V=-8.8$ for F1GC15,
$M_V=-10.2$ for C029, and $M_V=-11.2$ for C007
(all magnitudes from \citealt{king66} model fits and assuming $D=3.8$~Mpc).
}
\end{figure*}



\begin{table*}
\begin{minipage}{161mm}
\scriptsize
\caption{Derived structural and photometric parameters from 147 profiles
              of 131 GCs in NGC 5128
\label{tab:n5128phot}}
\begin{tabular}{@{}lclcrccccccc}
\hline
\multicolumn{1}{c}{Name} &
\multicolumn{1}{c}{Detector} &
\multicolumn{1}{c}{Model} &
\multicolumn{1}{c}{$\log\,r_{\rm tid}$} &
\multicolumn{1}{c}{$\log\,R_c$} &
\multicolumn{1}{c}{$\log\,R_h$} &
\multicolumn{1}{c}{$\log\,(R_h/R_c)$  } &
\multicolumn{1}{c}{$\log\,I_0$ } &
\multicolumn{1}{c}{$\log\,j_0$} &
\multicolumn{1}{c}{$\log\,L_{V}$ } &
\multicolumn{1}{c}{$V_{\rm tot}$} &
\multicolumn{1}{c}{$\log\,I_h$} \\
 &
 &
 &
\multicolumn{1}{c}{[pc]} &
\multicolumn{1}{c}{[pc]} &
\multicolumn{1}{c}{[pc]} &
 &
\multicolumn{1}{c}{[$L_{\odot, V}\,{\rm pc}^{-2}$]} &
\multicolumn{1}{c}{[$L_{\odot, V}\,{\rm pc}^{-3}$]} &
\multicolumn{1}{c}{[$L_{\odot, V}$]} &
\multicolumn{1}{c}{[mag]} &
\multicolumn{1}{c}{[$L_{\odot, V}\,{\rm pc}^{-2}$]}    \\
\multicolumn{1}{c}{(1) } &
\multicolumn{1}{c}{(2) } &
\multicolumn{1}{c}{(3) } &
\multicolumn{1}{c}{(4) } &
\multicolumn{1}{c}{(5) } &
\multicolumn{1}{c}{(6) } &
\multicolumn{1}{c}{(7) } &
\multicolumn{1}{c}{(8) } &
\multicolumn{1}{c}{(9) } &
\multicolumn{1}{c}{(10)} &
\multicolumn{1}{c}{(11)} &
\multicolumn{1}{c}{(12)} \\
\hline
   AAT111563  & WFC/F606       & K66  & $1.50^{+0.01}_{-0.03}$  &
                $0.035^{+0.022}_{-0.012}$  & $0.468^{+0.005}_{-0.010}$  &
                $0.433^{+0.017}_{-0.032}$  & $3.87^{+0.02}_{-0.03}$  &
                $3.53^{+0.03}_{-0.05}$  & $5.05^{+0.02}_{-0.02}$  &
                $20.10^{+0.05}_{-0.05}$  & $3.32^{+0.02}_{-0.02}$ \\
          ~~        & ~~         & W  & $2.03^{+0.08}_{-0.08}$  &
                $0.089^{+0.015}_{-0.016}$  & $0.486^{+0.017}_{-0.014}$  &
                $0.396^{+0.033}_{-0.029}$  & $3.82^{+0.03}_{-0.03}$  &
                $3.42^{+0.04}_{-0.04}$  & $5.07^{+0.02}_{-0.02}$  &
                $20.05^{+0.06}_{-0.06}$  & $3.30^{+0.03}_{-0.03}$ \\
          ~~        & ~~         & S  & $\infty$  &
                $-0.875^{+0.131}_{-0.145}$  & $0.478^{+0.010}_{-0.009}$  &
                $1.353^{+0.155}_{-0.140}$  & $4.42^{+0.09}_{-0.08}$  &
                $4.23^{+0.21}_{-0.19}$  & $5.05^{+0.02}_{-0.02}$  &
                $20.10^{+0.05}_{-0.05}$  & $3.30^{+0.02}_{-0.02}$ \\
   AAT113992  & WFC/F606       & K66  & $1.58^{+0.07}_{-0.04}$  &
                $0.144^{+0.008}_{-0.011}$  & $0.560^{+0.041}_{-0.023}$  &
                $0.416^{+0.051}_{-0.030}$  & $3.70^{+0.02}_{-0.02}$  &
                $3.25^{+0.03}_{-0.02}$  & $5.08^{+0.04}_{-0.03}$  &
                $20.02^{+0.07}_{-0.09}$  & $3.16^{+0.03}_{-0.05}$ \\
          ~~        & ~~         & W  & $2.64^{+0.31}_{-0.27}$  &
                $0.151^{+0.012}_{-0.011}$  & $0.749^{+0.144}_{-0.097}$  &
                $0.599^{+0.155}_{-0.109}$  & $3.69^{+0.02}_{-0.02}$  &
                $3.24^{+0.03}_{-0.03}$  & $5.21^{+0.08}_{-0.06}$  &
                $19.69^{+0.16}_{-0.21}$  & $2.92^{+0.13}_{-0.21}$ \\
          ~~        & ~~         & S  & $\infty$  &
                $-0.395^{+0.071}_{-0.072}$  & $0.528^{+0.016}_{-0.015}$  &
                $0.923^{+0.088}_{-0.086}$  & $4.03^{+0.04}_{-0.04}$  &
                $3.45^{+0.10}_{-0.09}$  & $5.05^{+0.02}_{-0.02}$  &
                $20.11^{+0.06}_{-0.06}$  & $3.19^{+0.03}_{-0.03}$ \\
   AAT115339  & WFC/F606       & K66  & $1.33^{+0.02}_{-0.01}$  &
                $-1.752^{+0.190}_{-0.214}$  & $0.376^{+0.003}_{-0.000}$  &
                $2.127^{+0.216}_{-0.190}$  & $6.07^{+0.20}_{-0.17}$  &
                $7.52^{+0.42}_{-0.36}$  & $5.19^{+0.02}_{-0.02}$  &
	                $19.76^{+0.05}_{-0.05}$  & $3.64^{+0.02}_{-0.02}$ \\
          ~~        & ~~         & W  & $2.01^{+0.20}_{-0.16}$  &
                $-0.033^{+0.035}_{-0.040}$  & $0.397^{+0.044}_{-0.028}$  &
                $0.430^{+0.084}_{-0.062}$  & $4.19^{+0.05}_{-0.04}$  &
                $3.92^{+0.09}_{-0.08}$  & $5.22^{+0.03}_{-0.03}$  &
                $19.67^{+0.07}_{-0.08}$  & $3.63^{+0.04}_{-0.06}$ \\
          ~~        & ~~         & S  & $\infty$  &
                $-1.246^{+0.173}_{-0.177}$  & $0.384^{+0.009}_{-0.009}$  &
                $1.630^{+0.186}_{-0.182}$  & $4.93^{+0.11}_{-0.11}$  &
                $5.08^{+0.26}_{-0.25}$  & $5.20^{+0.02}_{-0.02}$  &
                $19.72^{+0.05}_{-0.05}$  & $3.64^{+0.02}_{-0.02}$ \\
\hline
\end{tabular}

\medskip
  A machine-readable version of the full Table \ref{tab:n5128phot} is
  available online
  (http://www.astro.keele.ac.uk/$\sim$dem/clusters.html)
  or upon request from the first author.
  Only a short extract from it is shown here, for guidance
  regarding its form and content.

\end{minipage}
\end{table*}



\begin{table*}
\begin{minipage}{173mm}
\scriptsize
\caption{Derived dynamical parameters from 147 profiles of 131 GCs
              in NGC 5128
\label{tab:n5128mass}}
\begin{tabular}{@{}lcclrrrrrrrrr}
\hline
\multicolumn{1}{c}{Name} &
\multicolumn{1}{c}{Detector} &
\multicolumn{1}{c}{$\Upsilon_V^{\rm pop}$} &
\multicolumn{1}{c}{Model} &
\multicolumn{1}{c}{$\log\,M_{\rm tot}$} &
\multicolumn{1}{c}{$\log\,E_b$} &
\multicolumn{1}{c}{$\log\,\Sigma_0$} &
\multicolumn{1}{c}{$\log\,\rho_0$ } &
\multicolumn{1}{c}{$\log\,\Sigma_h$} &
\multicolumn{1}{c}{$\log\,\sigma_{{\rm p},0}$} &
\multicolumn{1}{c}{$\log\,v_{{\rm esc},0}$} &
\multicolumn{1}{c}{$\log\,t_{\rm rh}$} &
\multicolumn{1}{c}{$\log\,f_0$} \\
 &
 &
\multicolumn{1}{c}{[$M_\odot\,L_{\odot,V}^{-1}$]} &
 &
\multicolumn{1}{c}{[$M_\odot$]} &
\multicolumn{1}{c}{[erg]} &
\multicolumn{1}{c}{[$M_\odot\,{\rm pc}^{-2}$]} &
\multicolumn{1}{c}{[$M_\odot\,{\rm pc}^{-3}$]} &
\multicolumn{1}{c}{[$M_\odot\,{\rm pc}^{-2}$]} &
\multicolumn{1}{c}{[km~s$^{-1}$]} &
\multicolumn{1}{c}{[km~s$^{-1}$]} &
\multicolumn{1}{c}{[yr]} &
~~   \\
\multicolumn{1}{c}{(1)}  &
\multicolumn{1}{c}{(2)}  &
\multicolumn{1}{c}{(3)}  &
\multicolumn{1}{c}{(4)}  &
\multicolumn{1}{c}{(5)}  &
\multicolumn{1}{c}{(6)}  &
\multicolumn{1}{c}{(7)}  &
\multicolumn{1}{c}{(8)}  &
\multicolumn{1}{c}{(9)}  &
\multicolumn{1}{c}{(10)} &
\multicolumn{1}{c}{(11)} &
\multicolumn{1}{c}{(12)} &
\multicolumn{1}{c}{(13)} \\
\hline
   AAT111563  & WFC/F606  & $1.939^{+0.239}_{-0.240}$       & K66  &
                $5.34^{+0.05}_{-0.06}$  & $50.62^{+0.08}_{-0.09}$  &
                $4.16^{+0.06}_{-0.07}$  & $3.82^{+0.06}_{-0.08}$  &
                $3.60^{+0.06}_{-0.06}$  & $0.836^{+0.027}_{-0.030}$  &
                $1.421^{+0.027}_{-0.031}$  & $8.92^{+0.03}_{-0.03}$  &
                $0.079^{+0.038}_{-0.056}$ \\
          ~~        & ~~        & ~~         & W  &
                $5.36^{+0.06}_{-0.06}$  & $50.62^{+0.08}_{-0.09}$  &
                $4.11^{+0.06}_{-0.06}$  & $3.71^{+0.06}_{-0.07}$  &
                $3.59^{+0.06}_{-0.07}$  & $0.839^{+0.027}_{-0.030}$  &
                $1.427^{+0.027}_{-0.031}$  & $8.96^{+0.04}_{-0.04}$  &
                $-0.049^{+0.045}_{-0.045}$ \\
          ~~        & ~~        & ~~         & S  &
                $5.34^{+0.05}_{-0.06}$  & $50.62^{+0.08}_{-0.09}$  &
                $4.70^{+0.10}_{-0.10}$  & $4.52^{+0.21}_{-0.19}$  &
                $3.58^{+0.06}_{-0.06}$  & $0.797^{+0.028}_{-0.032}$  &
                $1.470^{+0.028}_{-0.031}$  & $8.94^{+0.03}_{-0.03}$  &
                $1.134^{+0.287}_{-0.255}$ \\
   AAT113992  & WFC/F606  & $2.914^{+0.431}_{-0.423}$       & K66  &
                $5.55^{+0.07}_{-0.07}$  & $50.94^{+0.09}_{-0.10}$  &
                $4.16^{+0.06}_{-0.07}$  & $3.71^{+0.07}_{-0.07}$  &
                $3.63^{+0.07}_{-0.09}$  & $0.893^{+0.032}_{-0.036}$  &
                $1.476^{+0.032}_{-0.036}$  & $9.16^{+0.08}_{-0.05}$  &
                $-0.197^{+0.042}_{-0.040}$ \\
          ~~        & ~~        & ~~         & W  &
                $5.68^{+0.10}_{-0.09}$  & $51.01^{+0.10}_{-0.11}$  &
                $4.16^{+0.06}_{-0.07}$  & $3.70^{+0.07}_{-0.07}$  &
                $3.38^{+0.15}_{-0.22}$  & $0.895^{+0.032}_{-0.036}$  &
                $1.499^{+0.032}_{-0.036}$  & $9.50^{+0.25}_{-0.17}$  &
                $-0.215^{+0.043}_{-0.046}$ \\
          ~~        & ~~        & ~~         & S  &
                $5.51^{+0.06}_{-0.07}$  & $50.92^{+0.09}_{-0.10}$  &
                $4.49^{+0.07}_{-0.08}$  & $3.91^{+0.10}_{-0.10}$  &
                $3.66^{+0.07}_{-0.07}$  & $0.877^{+0.032}_{-0.036}$  &
                $1.507^{+0.032}_{-0.036}$  & $9.10^{+0.04}_{-0.05}$  &
                $0.158^{+0.136}_{-0.133}$ \\
   AAT115339  & WFC/F606  & $2.015^{+0.238}_{-0.235}$       & K66  &
                $5.49^{+0.05}_{-0.06}$  & $51.08^{+0.07}_{-0.08}$  &
                $6.37^{+0.21}_{-0.18}$  & $7.82^{+0.42}_{-0.37}$  &
                $3.94^{+0.05}_{-0.06}$  & $1.047^{+0.028}_{-0.029}$  &
                $1.767^{+0.028}_{-0.030}$  & $8.85^{+0.03}_{-0.03}$  &
                $3.482^{+0.435}_{-0.391}$ \\
          ~~        & ~~        & ~~         & W  &
                $5.53^{+0.06}_{-0.06}$  & $51.05^{+0.07}_{-0.08}$  &
                $4.49^{+0.07}_{-0.07}$  & $4.22^{+0.10}_{-0.09}$  &
                $3.94^{+0.06}_{-0.08}$  & $0.972^{+0.026}_{-0.029}$  &
                $1.563^{+0.028}_{-0.030}$  & $8.89^{+0.08}_{-0.06}$  &
                $0.067^{+0.088}_{-0.080}$ \\
          ~~        & ~~        & ~~         & S  &
                $5.51^{+0.05}_{-0.06}$  & $51.05^{+0.07}_{-0.08}$  &
                $5.24^{+0.12}_{-0.12}$  & $5.38^{+0.26}_{-0.25}$  &
                $3.94^{+0.05}_{-0.06}$  & $0.913^{+0.029}_{-0.031}$  &
                $1.616^{+0.027}_{-0.030}$  & $8.87^{+0.03}_{-0.03}$  &
                $1.734^{+0.355}_{-0.343}$ \\
\hline
\end{tabular}

\medskip
  A machine-readable version of the full Table \ref{tab:n5128mass} is
  available online
  (http://www.astro.keele.ac.uk/$\sim$dem/clusters.html)
  or upon request from the first author.
  Only a short extract from it is shown here, for guidance
  regarding its form and content.

\end{minipage}
\end{table*}



\begin{table*}
\begin{minipage}{89mm}
\scriptsize
\caption{Galactocentric radii and $\kappa$-space parameters from 147
         profiles of 131 GCs in NGC 5128  \label{tab:n5128kappa}}
\begin{tabular}{@{}lcrlrrr}
\hline
\multicolumn{1}{c}{Name} &
\multicolumn{1}{c}{Detector} &
\multicolumn{1}{c}{$R_{\rm gc}$} &
\multicolumn{1}{c}{Model} &
\multicolumn{1}{c}{$\kappa_{m,1}$} &
\multicolumn{1}{c}{$\kappa_{m,2}$} &
\multicolumn{1}{c}{$\kappa_{m,3}$} \\
   &
   &
\multicolumn{1}{c}{[kpc]} &
   &
   &
~~ \\
\multicolumn{1}{c}{(1)} &
\multicolumn{1}{c}{(2)} &
\multicolumn{1}{c}{(3)} &
\multicolumn{1}{c}{(4)} &
\multicolumn{1}{c}{(5)} &
\multicolumn{1}{c}{(6)} &
\multicolumn{1}{c}{(7)} \\
\hline
   AAT111563  & WFC/F606     & 11.93       & K66  &
                $-0.608^{+0.039}_{-0.044}$  & $4.660^{+0.068}_{-0.075}$  &
                $0.346^{+0.002}_{-0.003}$ \\
          ~~        & ~~        & ~~         & W  &
                $-0.591^{+0.040}_{-0.044}$  & $4.642^{+0.070}_{-0.079}$  &
                $0.349^{+0.004}_{-0.003}$ \\
          ~~        & ~~        & ~~         & S  &
                $-0.657^{+0.039}_{-0.044}$  & $4.607^{+0.070}_{-0.078}$  &
                $0.306^{+0.008}_{-0.009}$ \\
   AAT113992  & WFC/F606      & 4.26       & K66  &
                $-0.462^{+0.052}_{-0.052}$  & $4.689^{+0.084}_{-0.105}$  &
                $0.345^{+0.005}_{-0.002}$ \\
          ~~        & ~~        & ~~         & W  &
                $-0.326^{+0.110}_{-0.084}$  & $4.410^{+0.168}_{-0.246}$  &
                $0.380^{+0.035}_{-0.019}$ \\
          ~~        & ~~        & ~~         & S  &
                $-0.507^{+0.045}_{-0.050}$  & $4.714^{+0.081}_{-0.091}$  &
                $0.328^{+0.003}_{-0.004}$ \\
   AAT115339  & WFC/F606      & 3.82       & K66  &
                $-0.375^{+0.039}_{-0.041}$  & $5.147^{+0.065}_{-0.071}$  &
                $0.447^{+0.011}_{-0.006}$ \\
          ~~        & ~~        & ~~         & W  &
                $-0.467^{+0.050}_{-0.046}$  & $5.070^{+0.075}_{-0.097}$  &
                $0.352^{+0.012}_{-0.006}$ \\
          ~~        & ~~        & ~~         & S  &
                $-0.558^{+0.038}_{-0.042}$  & $5.032^{+0.068}_{-0.074}$  &
                $0.289^{+0.011}_{-0.012}$ \\
\hline
\end{tabular}

\medskip
  A machine-readable version of the full Table \ref{tab:n5128kappa} is
  available online
  (http://www.astro.keele.ac.uk/$\sim$dem/clusters.html)
  or upon request from the first author.
  Only a short extract from it is shown here, for guidance
  regarding its form and content.

\end{minipage}
\end{table*}


Figure \ref{fig:morefits} compares the fits of \citet{king66},
\citeauthor{wil75}, and \citeauthor{sersic} models to the
background-subtracted, $V$-band surface-brightness profiles of eight more
globulars in NGC 5128, displayed in order of increasing total cluster
brightness; C029 appears in the second-last row of this figure. The curves and
points in every plot have the same meaning as in the lower panels of Figure
\ref{fig:exfit}. 

The first cluster shown in Figure \ref{fig:morefits}, C118, is one of the
faintest in our sample ($M_V=-6.0$, or $L_V\simeq2.1\times10^4\,L_\odot$
according to the \citealt{king66} model fit shown) and has an effective
(or half-light) radius of $R_h\simeq 3$--4 pc---typical of most known globular
clusters in any galaxy---depending slightly on the model fit. The next
cluster, C166, is similarly faint ($M_V=-6.5$, or
$L_V\simeq3.4\times10^4\,L_\odot$) but significantly more extended:
$R_h\simeq6.5$~pc for any of the models fit. This conclusion is clearly not
influenced by the PSF; in fact, these plots demonstrate that almost all of our
cluster candidates are very well resolved indeed.

The next two clusters, C171 and C167, both have the same brightness
(fitted $M_V\simeq-7.4$ or $L_V\simeq7.8\times10^4\,L_\odot$, essentially at
the expected peak of the GC luminosity function), but the first is
significantly more diffuse than the second: $R_h\simeq8$~pc
versus $R_h\simeq3$~pc. Evidently, there is substantial scatter
in any size-mass relation that one might try to explore for average and faint
globulars in NGC 5128. This is reminiscent of the well-known situation in the
Milky Way GC system, and we return to the point in our analysis of
structural correlations in \citetalias{mcl07}.

As we discussed in connection with C147 above, the three model fits to the
globulars in this first half of Figure \ref{fig:morefits} are all very similar:
the estimates of the sky level $I_{\rm bkg}$ are all about the same, as are
the total fitted luminosities, effective radii, and other cluster parameters.
The minimum $\chi^2$ for the different models are also very similar for any
one cluster, and there is no sign that the extended \citeauthor{wil75} or
\citeauthor{sersic} haloes are systematically either preferred 
over or bettered by the standard \citet{king66} structure. It is worth noting
that the \citeauthor{sersic} model fits can return significantly smaller
core radii and brighter intrinsic central surface brightnesses than the
other models; but at some level this simply points to a limitation of the
data, since the unavoidable PSF convolution effectively erases the
intrinsically cuspy inner structures of $n>1$ \citeauthor{sersic} models,
which might be disallowed by much higher-resolution data.

In the second half of Figure \ref{fig:morefits}, 
the clusters C113 and F1GC15 are both taken from
the bright side of the GC luminosity function in NGC 5128 ($M_V=-8.7$
and $M_V=-8.8$), and once again one is much larger than the other
($R_h=2.5$~pc against $R_h=16.5$~pc in the most conservative, \citealt{king66}
model fits). Now, however, the quality of fit differs
significantly between the different models. In C113, the minimum
$\chi^2$ is nearly 7 times smaller for the \citeauthor{wil75} model
versus \citet{king66}, and just over 5 times smaller for
\citeauthor{sersic} versus \citet{king66}. For F1GC15, the
\citeauthor{wil75} fit has a $\chi^2$ about 20\% larger than the
\citet{king66} fit (so the latter is formally somewhat better, but only
marginally so), while the \citeauthor{sersic} model is effectively ruled
out with a minimum $\chi^2$ more than 6 times larger than the
\citet{king66} model.

The last two clusters illustrated here are C029---confirming that this object 
is much better fit by a \citeauthor{wil75} model than a \citet{king66}
model, and also showing that the former does better than a \citeauthor{sersic}
model---and C007, which is the brightest GC in our sample ($M_V=-11.2$, or
$L_V=2.6\times10^6\,L_\odot$ from a \citealt{king66} model fit).
It too is better fit by a \citeauthor{wil75} model than by either of
the other two.

We note again that, because the extended-halo structure of all the bright
clusters in Figure \ref{fig:morefits} is poorly fit by the relatively compact
\citet{king66} models, the fitted background intensity is noticeably, but
spuriously, higher in the leftmost panels than in the other columns.

We will return to the question of which model best fits the majority of GCs
in NGC 5128, in \S\ref{subsec:chicomp} below. First, we present a number of
physical cluster properties derived within the framework of each of the model
fits summarized in Table \ref{tab:n5128fits}. 

\subsection{Derived quantities}
\label{subsec:derived}

Table \ref{tab:n5128phot} contains a number of other cluster properties
derived from the basic fit parameters given in Table
\ref{tab:n5128fits}. Columns following the GC name, detector/filter
combination, and fitted model are:

\medskip
Column (4): $\log\,r_t=c+\log\,r_0$, the model tidal radius in pc, which is
   always infinite for \citeauthor{sersic} models.

Column (5): $\log\,R_c$, the projected core radius of the model fitting a
   cluster, in units of pc. This is defined by $I(R_c)=I_0/2$, or 
   $\mu(R_c)\simeq\mu_0+0.753$ and is not necessarily the same
   as the radial scale $r_0$ in Table \ref{tab:n5128fits}, except for
   \citeauthor{sersic} models.

Column (6): $\log R_h$, the projected half-light, or effective,
   radius of a model. Half the total cluster luminosity is projected within
   $R_h$. It is related to $r_0$ by one-to-one functions of $W_0$ (or $c$)
   or $n$ and is reported here in units of pc.

Column (7): $\log\,(R_h/R_c)$, a measure of cluster concentration that is
   somewhat more model-independent than $c$ or $n$, in that it is generically
   well defined in observational terms and its physical meaning is always the
   same (cf.~our earlier discussion in \S\ref{subsec:structure}). We consider
   it a more suitable quantity to use when intercomparing the overall
   properties of clusters that may not all be fit by the same kind of model.

Column (8): $\log\,I_0$, the best-fit central ($R=0$)
   luminosity surface density in the $V$ band, in units of
   $L_{\odot,V}\,{\rm pc}^{-2}$. This is obtained from the fitted central
   surface brightness in Column (11) of Table \ref{tab:n5128fits}, by first
   applying the ``$V$-colour'' correction in Column (4) of that table to
   obtain the central $\mu_{V,0}$, and then using the definition
   $\log\,I_V=0.4\,(26.402-\mu_V)$, where the zeropoint corresponds to a solar
   absolute magnitude of $M_{\odot,V}=+4.83$. 

Column (9): $\log\,j_0$, the $V$-band luminosity volume
   density ($L_{\odot,V}\,{\rm pc}^{-3}$) at $r=0$ for \citet{king66} and
   \citeauthor{wil75} models but at the three-dimensional radius $r=r_0$ for
   \citeauthor{sersic} models. In the first two cases, $j_0={\cal J}I_0/r_0$
   where ${\cal J}$ is a smooth, model-dependent function of $W_0$ or $c$,
   which we have calculated numerically. For \citeauthor{sersic} models,
   as we discussed in \S\ref{subsec:structure}, the unprojected
   density is infinite as $r\rightarrow 0$ when $n\ge 1$, and thus we only
   quote $j_0\equiv j(r_0)$ for any fitted $n$. This finite quantity is
   related to the quotient $I_0/r_0$ by a well-defined function of $n$, which
   we have again computed numerically.

Column (10): $\log\,L_V$, the total integrated model
   luminosity in the $V$ band. It is related to the product $I_0r_0^2$ by
   model-dependent functions of $W_0$ or $c$, or $n$. 

Column (11):
   $V_{\rm tot}=4.83-2.5\,\log\,(L_V/L_\odot)+ 5\,\log\,(D/10\,{\rm pc})$,
   the total, extinction-corrected apparent $V$-band magnitude of a
   model cluster, assuming $D=3.8$ Mpc.

Column (12): $\log I_h \equiv \log\,(L_V/2\pi R_h^2)$, the $V$-band
   luminosity surface density averaged over the half-light or effective
   radius, in units of $L_{\odot,V}\,{\rm pc}^{-2}$. The $V$ surface
   brightness averaged over $R_h$ is
   $\langle \mu_V\rangle_h \equiv
        26.402-2.5\,\log\,(I_h/L_{\odot,V}\,{\rm pc}^{-2})$.

\medskip
The uncertainties on all of these derived parameters have been
estimated from the $\chi^2$
re-scaling procedure described after Table \ref{tab:n5128fits}.
If the distance $D$ to NGC 5128 is different from our adopted 3.8 Mpc, the
quantities in Table \ref{tab:n5128phot} will change according to:
$r_{\rm tid}$, $R_c$, and $R_h$ all $\propto D$; $I_0$ and $I_h$ independent
of $D$; $j_0\propto D^{-1}$; and $L_V\propto D^2$.

Table \ref{tab:n5128mass} next lists a number of cluster properties derived
from the structural parameters already given plus a mass-to-light ratio. 
The first two columns of this table contain the name of each cluster and the
combination of detector/filter for our observations of it, as usual. Column
(3) lists the $V$-band mass-to-light ratio that we have adopted for 
each object from the analysis in \S\ref{subsec:popsyn}, assuming for
definiteness an age of 13 Gyr for all clusters. The errorbars on
$\Upsilon_V^{\rm pop}$ in Table \ref{tab:n5128mass} are larger than those in
the 13-Gyr column of Table \ref{tab:n5128mtol}, as we allow now for a $\pm
2$-Gyr uncertainty in age on top of the previously tabulated uncertainties in
[Fe/H]. The remaining entries in Table \ref{tab:n5128mass} are, for the best
fit of each model to every cluster:

\medskip
Column (5): $\log\,M_{\rm tot}=\log\,\Upsilon_V^{\rm pop}+\log\,L_V$, the
   integrated model mass in solar units, with $\log\,L_V$ taken from Column
   (10) of Table \ref{tab:n5128phot}.

Column (6): $\log\,E_b$, the integrated binding energy in ergs, defined
   through $E_b\equiv -(1/2)\int_{0}^{r_t} 4\pi r^2 \rho \phi\,dr$. Here the
   minus sign makes $E_b$ positive for gravitationally bound objects, and
   $\phi(r)$ is the potential generated (through Poisson's equation) by the
   model mass distribution $\rho(r)$. $E_b$ can be written in terms of the
   fitted central luminosity density $j_0$; scale radius $r_0$; a
   model-dependent function of $W_0$ or $c$, or $n$; and
   $\Upsilon_V^{\rm pop}$. A more detailed outline of this procedure for
   \citet{king66} models may be found in \citet{mcl00}, which we have followed
   closely to evaluate $E_b$ for our other model fits as well.

Column (7): $\log\,\Sigma_0=\log\,\Upsilon_V^{\rm pop}+\log\,I_0$, the central
   surface mass density in the model, in units of
   $M_\odot\,{\rm pc}^{-2}$.  

Column (8): $\log\,\rho_0=\log\,\Upsilon_V^{\rm pop}+\log\,j_0$, the central
   volume density in units of $M_\odot\,{\rm pc}^{-3}$ (except for
   \citeauthor{sersic} models, where $\rho_0$ is the density at the
   three-dimensional radius $r_0$). 

Column (9): $\log\,\Sigma_h=\log\,\Upsilon_V^{\rm pop}+\log\,I_h$, the
   surface mass density averaged over the inner effective radius $R_h$,
   in units of $M_\odot\,{\rm pc}^{-2}$.

Column (10): $\log\,\sigma_{{\rm p},0}$, the predicted line-of-sight velocity
   dispersion at the cluster centre for \citet{king66} and \citeauthor{wil75}
   models, but at $r_0=R_c$ for \citeauthor{sersic} models, in km~s$^{-1}$.
   The solution of Poisson's and Jeans' equations for any model yields a
   dimensionless $\sigma_{{\rm p},0}/\sigma_0$, and with $\sigma_0$ given by
   the fitted $r_0$ and $\rho_0 = \Upsilon_V^{\rm pop} j_0$ through equation
   (\ref{eq:sigma0}) or (\ref{eq:sersigma0}), the predicted observable
   dispersion follows immediately. 

Column (11): $\log\,v_{{\rm esc},0}$, the predicted central ``escape''
   velocity in km~s$^{-1}$. A star moving out from the centre of a cluster
   with speed $v_{{\rm esc},0}$ will just come to rest at infinity. In
   general, then,
   \[
       v_{{\rm esc},0}^2/\sigma_0^2 =
                2\left[W_0+GM_{\rm tot}/r_t\sigma_0^2\right]\ .
   \]
   A (finite) dimensionless $W_0$ is associated with any $n>0$ for 
   \citeauthor{sersic} models by solving Poisson's equation with
   $\phi\rightarrow 0$ in the limit $r\rightarrow \infty$. The second term on
   the right-hand side of the definition of $v_{{\rm esc},0}$ vanishes for
   \citeauthor{sersic} models, in which $r_t\rightarrow\infty$.

Column (12): $\log\,t_{\rm rh}$, the two-body relaxation time at the model
   projected half-mass radius, in years. This is estimated as
   \[
       t_{\rm rh} = \frac{2.06\times10^6\ {\rm yr}}
                      {\ln \left( 0.4M_{\rm tot}/m_\star \right)}
                    \frac{M_{\rm tot}^{1/2} R_h^{3/2}}{m_\star}
   \]
   \citep[eq.~8-72]{bt87}, if $m_\star$ (the average stellar mass in a cluster)
   and $M_{\rm tot}$ are both in solar units and $R_h$ is in pc. We have
   evaluated this timescale assuming an average $m_\star=0.5\,M_\odot$ in all
   clusters.

Column (13):
$\log f_0\equiv \log\,\left[\rho_0/(2\pi \sigma_c^2)^{3/2}\right]$,
a measure of the model's ``central'' phase-space density in units of
$M_\odot\,{\rm pc}^{-3}\,({\rm km\ s^{-1}})^{-3}$.
In this expression, $\sigma_c$ refers to the central one-dimensional velocity
dispersion {\it not} projected along the line of sight. The ratio
$\sigma_c/\sigma_0$ is obtained for \citet{king66} and \citeauthor{wil75}
models from the solution of the Poisson and  Jeans equations for given $W_0$
or $c$, and $\sigma_0$ is known from equation
(\ref{eq:sigma0}). This procedure breaks down for \citeauthor{sersic} 
models in general, since the unprojected $\rho \rightarrow \infty$ and
$\sigma\rightarrow 0$ as $r\rightarrow 0$ for $n\ge 1$. As before, then, for 
these models we instead calculate $f_0$ at the three-dimensional radius
$r=r_0$. In any case, with the central relaxation time $t_{\rm rc}$
of a cluster defined as in equation (8-71) of \citet{bt87},
taking an average stellar mass of $m_\star=0.5\,M_\odot$ and a typical
Coulomb logarithm $\ln\Lambda\approx12$ leads to the
approximate relation $\log\,(t_{\rm rc}/{\rm yr})\simeq 8.28-\log\,f_0$.

\medskip
The uncertainties in these quantities are estimated from their variations
around the minimum of $\chi^2$ on the model grids we fit, as above, but now
combined in quadrature with the uncertainties in the population-synthesis
model $\Upsilon_V^{\rm pop}$. 
Note that if $\Upsilon_V^{\rm pop}$ is changed to any other value for any
cluster, or if any distance other than $D=3.8$~Mpc is adopted for NGC 5128,
the properties in Table \ref{tab:n5128mass} scale as
$M_{\rm tot}\propto \Upsilon\,D^2$;
$E_b\propto \Upsilon^2\,D^3$;
$\Sigma_0\propto \Upsilon\,D^0$ and $\Sigma_h\propto \Upsilon\,D^0$;
$\rho_0\propto \Upsilon\,D^{-1}$;
$\sigma_{{\rm p},0}\propto \Upsilon^{1/2}\,D^{1/2}$ and
$v_{{\rm esc},0}\propto \Upsilon^{1/2}\,D^{1/2}$;
$t_{\rm rh}\propto \Upsilon^{1/2}\,D^{5/2}$;
and $f_0\propto \Upsilon^{-1/2}\,D^{-5/2}$.

Table \ref{tab:n5128kappa} provides the last few parameters
required to construct the fundamental plane of globular clusters in NGC 5128,
under any of the equivalent formulations of it in the literature.
The first of these remaining parameters is the
projected galactocentric distance $R_{\rm gc}$, in kpc, of each
cluster. This is listed in Column (3) of Table \ref{tab:n5128kappa}, following
the cluster name and detector/filter combination. It is obtained from the
angular $R_{\rm gc}$ given in Tables 1 and 2 of \citetalias{har06}, assuming
$D=3.8$~Mpc as always.

The last three columns of Table \ref{tab:n5128kappa} give analogues of the
so-called $\kappa$ parameters introduced by \citet{bbf92}, who define an
orthonormal ``coordinate'' system based on the three main luminosity
observables of galaxies. Similarly to them, but in terms of mass-based
quantities instead, we define 
\begin{equation}
\begin{array}{rcl}
\kappa_{m,1} & \equiv & (\log\,\sigma_{{\rm p},0}^2 + \log\,R_h)/\sqrt{2} \\
\kappa_{m,2} & \equiv & (\log\,\sigma_{{\rm p},0}^2 + 2\,\log\Sigma_h
                      - \log\,R_h)/\sqrt{6}    \\
\kappa_{m,3} & \equiv & (\log\,\sigma_{{\rm p},0}^2 - \log\,\Sigma_h
                      - \log\,R_h)/\sqrt{3} \\
\end{array}
\label{eq:kspace}
\end{equation}
which differ from the $\kappa$'s in \citeauthor{bbf92}
\citep[and][]{bur97} only in our use of the average mass surface density
$\Sigma_h$ rather than the luminosity surface density
$I_h=\Sigma_h/\Upsilon_V$. The main reason for doing this is to
facilitate the comparison of fundamental planes for young and old star
clusters, by removing the influence of age on $I_h$ for two objects of similar
mass and size. It similarly avoids the influence that any metallicity
dependence in $\Upsilon_V$ has on a luminosity-based fundamental plane.

In equation (\ref{eq:kspace}),
$\kappa_{m,1} \leftrightarrow \log\,(\sigma_{{\rm p},0}^2 R_h)$
is related to the total mass of a system, and
$\kappa_{m,3} \leftrightarrow \log\,(\sigma_{{\rm p},0}^2 R_h/M_{\rm tot})$
contains the exact details of this relationship---that is, information
on the internal density profile $\rho(r)$. In fact, the mass-based
$\kappa_{m,3}$ can be viewed as a replacement for \citeauthor{king66}- or
\citeauthor{wil75}-model concentrations $c$ or \citeauthor{sersic}-model
indices $n$, or any other model-specific shape parameter. As such, any trends
involving $\kappa_{m,3}$ are directly of relevance to questions concerning
cluster (non)homology. The definition of
$\kappa_{m,2} \leftrightarrow \log\,(\Sigma_h^3)$ is chosen to
make the three $\kappa_{m}$ axes space mutually orthogonal in
the parameter space of $(\sigma_{{\rm p},0}^2, R_h, \Sigma_h)$.

In calculating $\kappa_{m,1}$, $\kappa_{m,2}$, and $\kappa_{m,3}$ for Table
\ref{tab:n5128kappa}, we have used the $\sigma_{{\rm p},0}$ predicted
in Table \ref{tab:n5128mass} (Column 7) by our adoption of
population-synthesis mass-to-light ratios. Similarly, $\Sigma_h$ is taken from
Column (9) of Table \ref{tab:n5128mass}. As a result, our values for
$\kappa_{m,3}$ are independent of $\Upsilon_V^{\rm pop}$. $R_h$ is taken from
Table \ref{tab:n5128phot} but put in units of kpc rather than pc, for
better compatibility with the original, galaxy-oriented definitions of
\citet{bbf92}.

\subsection{Consistency checks}
\label{subsec:dualcomp}


\begin{figure}
\centerline{\hfil
   \includegraphics[width=82mm]{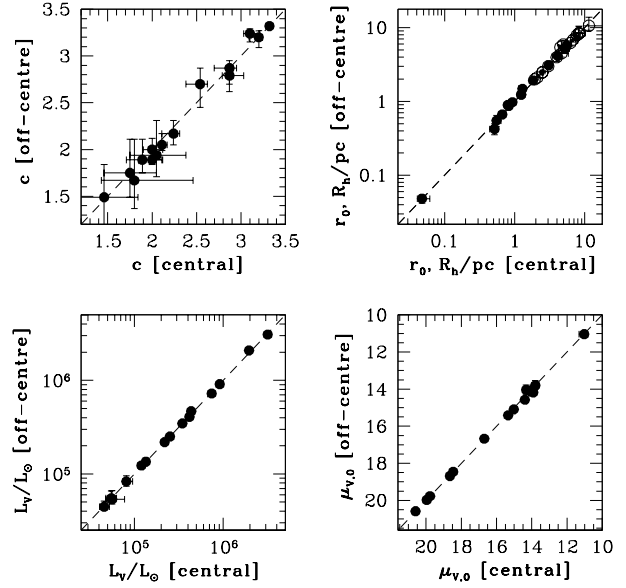}
\hfil}
\caption{
Comparison of concentration parameters, intrinsic radii, total luminosities,
and intrinsic central surface brightnesses for fits of isotropic \citet{wil75}
models to each of the two independent intensity profiles measured for 15 of
the GC candidates listed in Table \ref{tab:dual} (not including the possible
star, C156). In the upper-right panel, filled symbols refer to estimates
of the model scale radii $r_0$ for the clusters, while open circles refer to
the effective (projected half-light) radii $R_h$.
\label{fig:dual}
}
\end{figure}


There are two ways in which we can assess the internal consistency of the
various model fits we have performed.

First, Figure \ref{fig:dual} compares some of the main parameters obtained
from \citeauthor{wil75} fits to the 15 clusters in Table \ref{tab:dual},
for which we have two independent surface-brightness profiles measured
on two different main ACS fields.
(We have excluded the object C156 from this comparison;
cf.~Table \ref{tab:badstuff}.)

The $x$-axis of every panel in Figure \ref{fig:dual} marks the parameter
value measured for the clusters in the survey target fields where they lie
closest to the chip centre (the fields highlighted in bold in Table
\ref{tab:dual}). The $y$-axes then measure the parameter values found in the
ACS field where the clusters lie further from the centre.
There is evidently no substantial or systematic scatter
around the line of equality drawn in each case. For these \citeauthor{wil75}
fits we find that the average and rms scatter in the difference of fitted
concentrations is (in the sense [off-centre] minus [central])
$\langle \delta c \rangle = -0.02\pm0.08$,
as compared to an rms errorbar of $\pm\, 0.13$ on the individual $c$
values. For the scale and half-light radii,
$\langle \delta(\log\,r_0) \rangle = \langle \delta(\log\,R_h) \rangle
= 0.01\pm0.03$~dex,
versus an rms fit uncertainty of about $\pm\, 0.04$~dex in both cases.
Finally, we have
$\langle \delta(\log\,L_V) \rangle=0\pm0.01$~dex and
$\langle \delta \mu_{V,0} \rangle=0.03\pm0.11$~mag~arcsec$^{-2}$, against
rms errorbars of $\pm\, 0.03$~dex and $\pm\, 0.09$~mag~arcsec$^{-2}$.
Nearly identical results hold for comparisons 
of fitted \citet{king66} and \citeauthor{sersic} model parameters.

We conclude that any PSF variations over the ACS/WFC chip have not introduced
any systematic errors into our model fitting, which assumed a single
PSF shape in all cases. For all subsequent analyses, we include the clusters
with double measurements by adopting the fit parameters obtained from
the image on which the cluster is closest to the chip centre.


\begin{figure}
\centerline{\hfil
   \includegraphics[width=82mm]{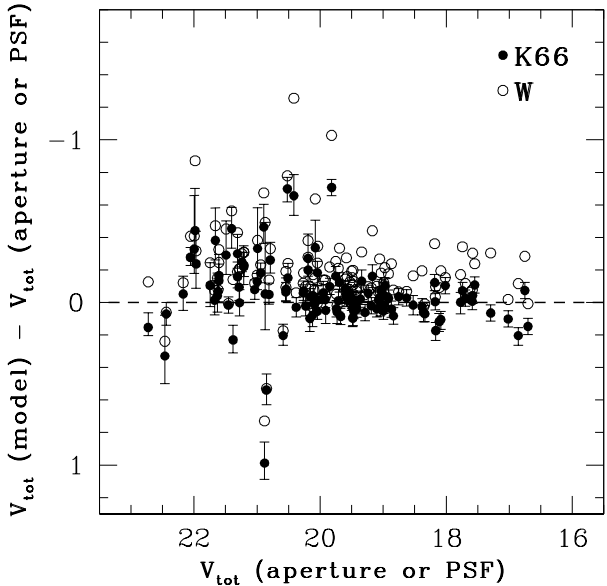}
\hfil}
\caption{
Comparison of total \citet{king66} and \citeauthor{wil75} model $V$ magnitudes
versus aperture or PSF estimates (extinction-corrected) from
\citetalias{har06} \citep[cf.][]{ghar04}, for 124 reliable GC candidates.
\label{fig:modmags}
}
\end{figure}


Second, Figure \ref{fig:modmags} shows the total apparent $V$-band magnitudes
inferred from our \citet{king66} and \citeauthor{wil75} model fits
(from Column [11] of Table \ref{tab:n5128colors}) to
the $V$-band equivalents of Washington $T_1$ magnitudes estimated from
ground-based aperture or PSF photometry and tabulated in \citetalias{har06}
(numbers taken originally from \citealt{ghar04}). 
To transform $T_1$ to $V$, we first corrected for
extinction according to $A_{T1} = 2.54 \times E(B-V) = 0.279$~mag
for all clusters \citep{harcan79}, and then used the relationship
$V_{\rm tot}=T_{1,0}+0.052+0.256\times(C-T_1)_0$ from \citetalias{har02} (see
also \citealt{geis96}) given the de-reddened Washington 
colours in \citetalias{har06} and our Table \ref{tab:n5128colors} above.
We have excluded from this comparison the clusters
C168, F1GC34, and F2GC14, for the reasons given in Table \ref{tab:badstuff};
the objects C145, C152, C156, which are likely foreground stars; and
C177, which is probably a background galaxy.

For bright clusters with $V_{\rm tot} < 20$ ($M_V < -7.9$), our model
luminosities compare rather well in general to the aperture or PSF magnitudes;
with
$\delta V_{\rm tot} \equiv [V_{\rm tot}({\rm model}) -
                  V_{\rm tot}({\rm aperture\ or\ PSF})]$,
we have
$\langle \delta V_{\rm tot} \rangle = 0\pm0.13$~mag (rms) for the
\citet{king66} 
model fits, and $\langle \delta V_{\rm tot} \rangle =
-0.14\pm0.18$~mag (rms) for \citeauthor{wil75} fits. Both results are
quite acceptable, considering 
(1) the number of steps involved in
transforming either our models or the Washington photometry of
\citetalias{har06} and \citet{ghar04} to standard $V$ magnitudes; and
(2) the fact that the aperture or PSF magnitudes are from analyses of
ground-based data and limited to clustercentric radii of only
$R\la 2\arcsec$--$3\arcsec$ (about 35--55 pc), which can easily miss up to
$\sim\!0.2$~mag of cluster light (see Figure \ref{fig:morefits} above, and
also the discussion in \citealt{ghar04}). The 0.15-mag offset between the
average 
\citeauthor{wil75} and \citet{king66} model magnitudes is a direct result
of the fact that the latter models are generally too compact to fit the
brightest clusters well and thus underestimate the true luminosity of the
cluster haloes.

Fainter clusters with $V_{\rm tot} \ga 20$ according to the aperture or PSF
measurements have $\delta V_{\rm tot}$ offsets that are on average
$\simeq\! 0.1$~mag brighter than for the brighter GCs, and
that have a larger rms scatter about the average. This is due to our
ability here (unlike in \citealt{ghar04}) to resolve all of these faint
clusters and robustly to estimate and correct for the local background
intensity around each of them. We thus view our present measurements of
$V_{\rm tot}$ from model fits to resolved surface-brightness profiles as
being {\it more} reliable than the previously published numbers.

A comparison between our \citeauthor{sersic}-model $V_{\rm tot}$ and the
ground-based estimates gives results that are nearly indistinguishable from
those shown here for the \citet{king66} fits.

\subsection{Quality of fit for different models}
\label{subsec:chicomp}


\begin{figure*}
\centerline{\hfil
   \includegraphics[width=175mm]{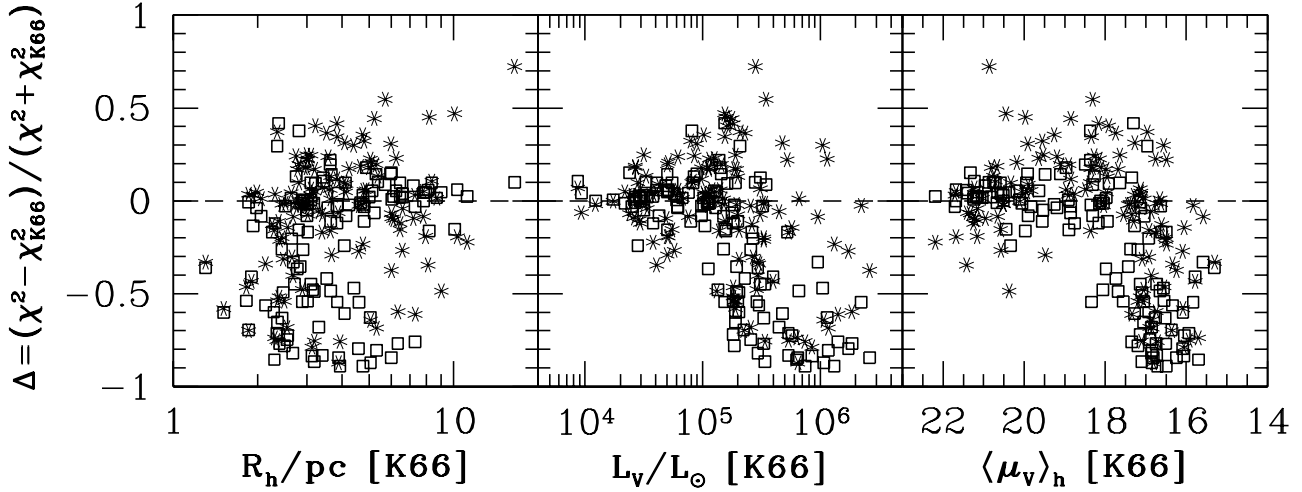}
\hfil}
\caption{
Relative quality-of-fit, $\Delta$ (eq.~[\ref{eq:delchi}]), for isotropic
\citeauthor{wil75} and \citeauthor{sersic} models (open squares and asterisks)
versus \citet{king66} models, for 124 GCs in NGC 5128.
(Objects C168, F1GC34, F2GC14, C145, C152, C156, and C177 are excluded from
the starting sample of 131 cluster candidates, for the reasons given in Table
\ref{tab:badstuff}.) The right-hand panel shows that \citeauthor{wil75} models,
with their relatively more extended haloes, tend to give better fits than
\citet{king66} models for high surface-brightness GCs, which have
envelopes that are empirically better defined against the background sky out
to large clustercentric radii. 
\label{fig:fitcomp}
}
\end{figure*}



\begin{figure*}
\centerline{\hfil
   \includegraphics[width=85mm]{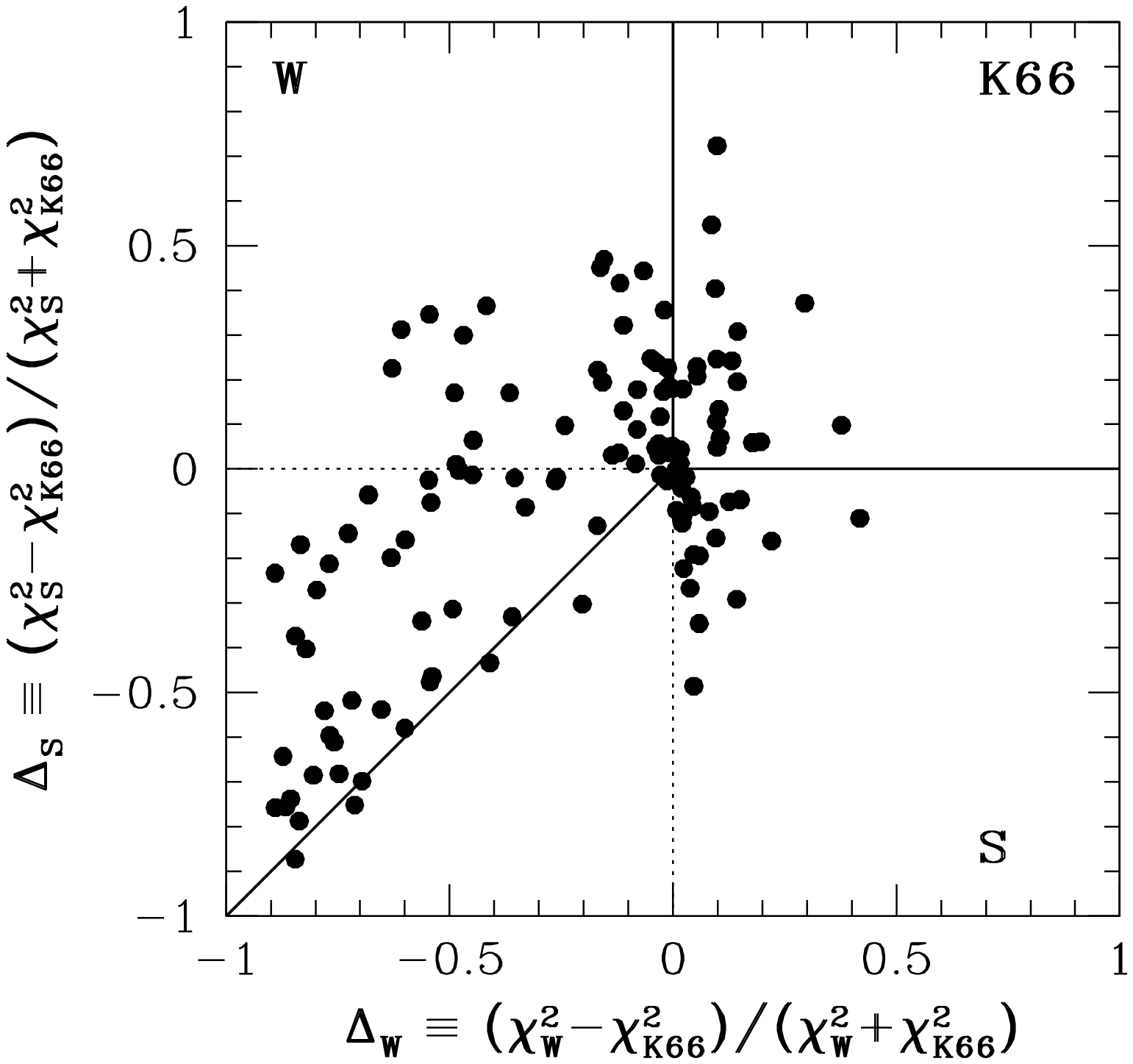}
   \hfill
   \includegraphics[width=85mm]{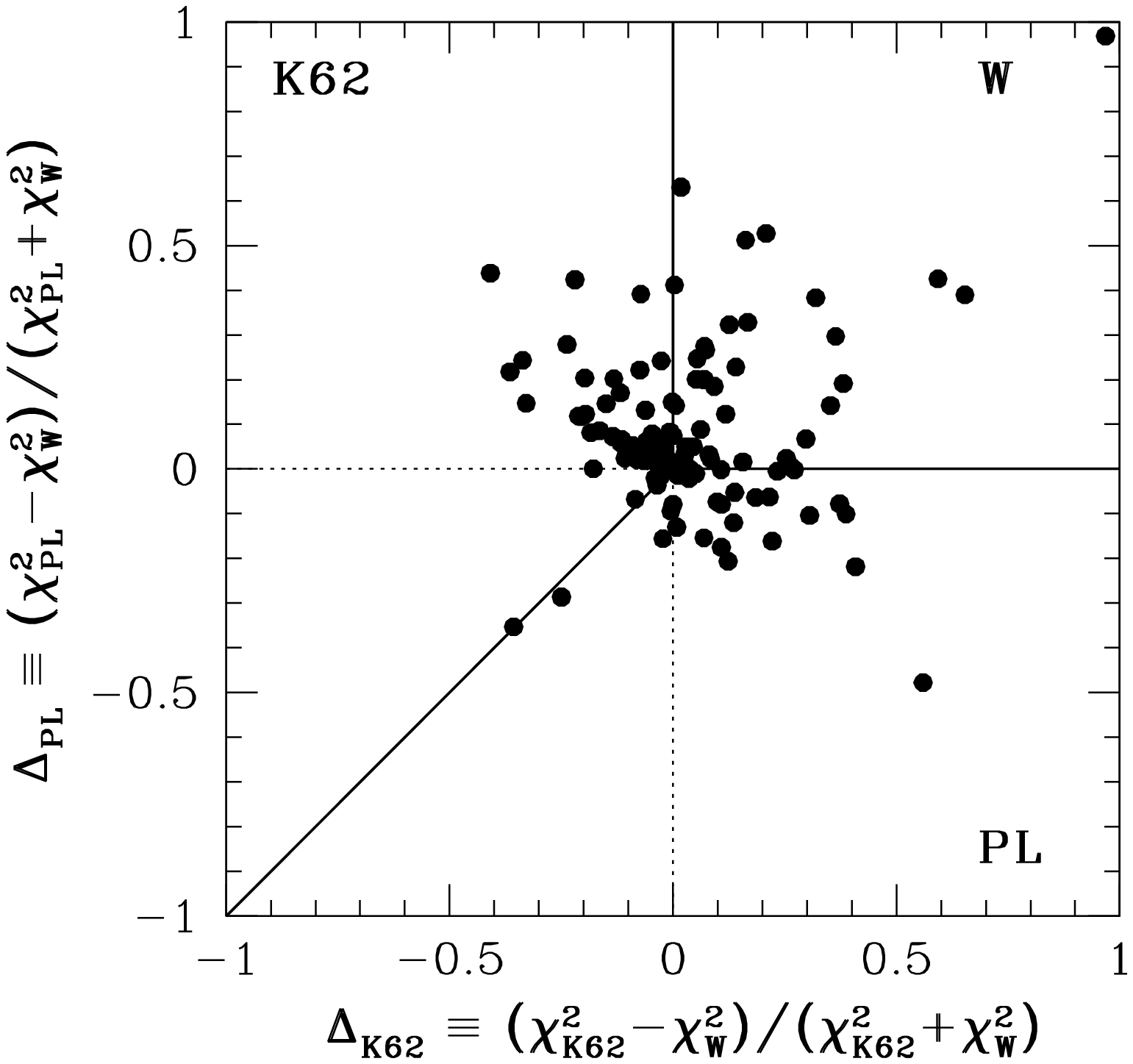}
\hfil}
\caption{
{\it Left:} Relative quality-of-fit statistic $\Delta$ (eq.~[\ref{eq:delchi}])
for \citeauthor{sersic} versus \citet{king66} models fitted to 124 reliable GC
candidates in NGC 5128, against $\Delta$ for \citeauthor{wil75} versus
\citet{king66} fits to the same clusters.
{\it Right:} Analogous comparison for power-law versus \citeauthor{wil75}
models, against that for \citet{king62} versus \citeauthor{wil75} models.
Labels and bold lines indicate which of the three models being
compared is the formal best fit for GCs falling in different regions of either
graph. An isotropic \citet{wil75} model nearly always fits at least as
well as any of the others we have tried, and often does significantly better.
\label{fig:chicomp}
}
\end{figure*}


We now return to the issue, raised in \S\ref{subsec:profs}, of deciding
which model family best describes the 
structure of GCs in NGC 5128 in general. Following \citet{mcl05}, we define
a statistic that compares the $\chi^2$ of the best fit of an ``alternate''
model for any object to the $\chi^2$ of the best-fit standard \citet{king66}
model:
\begin{equation}
\Delta \equiv (\chi_{\rm alternate}^2-\chi_{\rm K66}^2) /
         (\chi_{\rm alternate}^2+\chi_{\rm K66}^2) \ .
\label{eq:delchi}
\end{equation}
This statistic vanishes if the two models fit the
same cluster equally well. It is positive (with a maximum of $+1$) if the
alternate model is a worse fit than \citet{king66}, and negative (with a
minimum of $-1$) if the alternate model is a better fit. We have
found that for fairly modest values, 
$-0.2\la \Delta \la 0.2$, even though one model formally fits better than the
other the difference is generally not significant. For $\Delta$ more
towards the extremes of $\pm1$, the improvement afforded by the model with
lower $\chi^2$ is more substantial.

Figure \ref{fig:fitcomp} shows this $\Delta$ statistic for \citeauthor{wil75}-
and \citeauthor{sersic}-model fits (open squares and asterisks, respectively,
in all panels) versus \citet{king66} fits to 124 GCs in our current
sample. These numbers are plotted against the clusters' half-light radii
$R_h$, total model luminosities $L_V$, and the intrinsic average surface
brightness $\langle\mu_V\rangle_h\equiv26.402-2.5\,\log(L_V/2\pi R_h^2)$ as
estimated from the \citet{king66} fits.

The left-hand panel of Figure \ref{fig:fitcomp} shows immediately that many
clusters in NGC 5128 are better described by the distended shapes of
\citeauthor{wil75} or \citeauthor{sersic} models than by the concave
\citet{king66} profiles. However, the intrinsic size of a GC is clearly not a
good predictor of which model shape will fit it best.

The middle plot, of $\Delta$ vs.~$L_V$, shows that clusters brighter 
than $L_V\sim(1$--$2)\times10^5\,L_\odot$---that is, from just above
nominal turnover point of the GC luminosity function---are generally fit much
better by models with more extended haloes than allowed in the standard
\citet{king66} description. Fainter clusters are usually fit almost equally
well by \citeauthor{wil75} models as by \citet{king66}. \citeauthor{sersic} 
models tend to fare similarly, although there are a number of individual
cases in which \citeauthor{wil75} and \citet{king66} models both do better.

The graph of $\Delta$ versus $\langle\mu_V\rangle_h$, in the
right-hand panel of Figure \ref{fig:fitcomp}, is the most telling. This
shows clearly that clusters with high average surface brightnesses are
significantly better fit by models with larger haloes than \citet{king66}. At
some level this might seem almost a tautology, given that $\Delta$ correlates
with $L_V$ and that $\langle\mu_V\rangle_h$ obviously does as well.
But it is interesting at this point to compare with
the situation for Local Group clusters. \citet{mcl05} show that
153 globulars and young massive clusters in the Milky Way, the LMC and SMC,
and the Fornax dwarf spheroidal are systematically better fit by the extended
\citeauthor{wil75} haloes than by \citet{king66} models whenever the clusters'
surface-brightness profiles are accurately defined to large projected
radius (more than about $5\,R_h$) such that the intrinsic structural
differences between the model haloes become reliably observable in the first
place. There is no correlation in \citeauthor{mcl05} between the
\citeauthor{wil75}--\citeauthor{king66} $\Delta$ and cluster luminosity or
mass. In our case, clusters with brighter $\langle\mu_V\rangle_h$ are higher
above the background in NGC 5128 over a larger fraction of their volume, and
thus their far halo regions are necessarily better defined in purely
observational terms. It is when this happens that the large envelopes of
\citeauthor{wil75} models particularly (and those of \citeauthor{sersic}
models to a lesser extent) are preferred over the more
compact \citet{king66} structure. Fainter and lower surface-brightness
clusters may very well be similarly extended in general, but the data at large
radius in these cases are not good enough to show this definitively.

We therefore conclude from Figure \ref{fig:fitcomp} that
the haloes of GCs in NGC 5128 are {\it generically} more extended
than the classic \citet{king66} model allows. These globulars thus
appear to be very much like those in the Milky Way and some of its
satellite galaxies (see \citealt{barmby07} for a discussion of GCs in
M31, where the situation is less clear). Any correlation between
$\Delta$ and total cluster luminosity appears to derive from this more basic
fact; it does not imply that massive clusters have extended haloes just
because they are massive {\it per se}, nor that bright clusters have
systematically different intrinsic structures than faint ones.

The slight collective offset of the asterisks upwards from the squares in
Figure \ref{fig:fitcomp}, suggests that the globulars in NGC 5128 are
generally somewhat better fit by \citeauthor{wil75} models than by
\citeauthor{sersic} models. This is confirmed in the left-hand panel of Figure
\ref{fig:chicomp}, which shows the individual $\Delta$ statistics for
\citeauthor{wil75}-model fits to the same sample of 124
GCs as in Figure \ref{fig:fitcomp}, against the $\Delta$ for
\citeauthor{sersic}-model fits. Points in the upper-right hand quadrant
($\Delta_{\rm W}>0$ and $\Delta_{\rm S}>0$) of this plot represent clusters
for which a \citet{king66} model fits better than either of the other
two. Clusters that are formally fit best by a \citeauthor{sersic} model have
$\Delta_{\rm S}<0$ and $\Delta_{\rm S}<\Delta_{\rm W}$, and thus 
fall within the horizontal half-trapezoidal region (defined by bold lines)
in the lower part of the plot. When a \citeauthor{wil75} model fits best
($\Delta_{\rm W}<0$ and $\Delta_{\rm W}<\Delta_{\rm S}$), the clusters
populate the vertical half-trapezoidal area on the left-hand side.

About 60\% (75/124) of our clusters fall in this latter region of the
left-hand panel in Figure \ref{fig:chicomp} and are thus formally best fit by
\citeauthor{wil75} models. Fully 90\% (111/124) fall within the slightly more
generous limits $\Delta_{\rm W}<0.2$ and $\Delta_{\rm W}<\Delta_{\rm S}+0.2$,
which include cases where \citeauthor{wil75} models are just not significantly
worse than either \citeauthor{sersic} or \citet{king66} models. Presumably
these results reflect that GCs have relaxed, nearly isothermal core structures
that are captured by \citeauthor{wil75} and \citet{king66} models but not
\citeauthor{sersic} models in general, plus strong halo components that are
better matched by \citeauthor{wil75} or \citeauthor{sersic} models than by
those of \citet{king66}.

Points well inside the horizontal half-trapezoid in the lower part of the
left-hand panel of Figure \ref{fig:chicomp} represent clusters that are
noticeably better fit by \citeauthor{sersic} models than by either
\citet{king66} or \citeauthor{wil75} models. Although there are relatively few
such objects---only 12/124 have both $\Delta_{\rm S}<-0.1$ and
$(\Delta_{\rm S} - \Delta_{\rm W})<-0.1$---it might be thought that they could
be of particular interest. If, for example, they were all extremely massive,
this might then be supportive of suggestions that the brightest globulars were
stripped galaxy nuclei or possibly ultra-compact dwarfs. However, the average
total luminosity of the 12 clusters just mentioned is a completely normal
$\langle L_V\rangle\simeq 5\times10^4\,L_\odot$---nearly at the peak of the GC
luminosity function---and all have $L_V < 2\times10^5\,L_\odot$. 

Finally, the right-hand panel of Figure \ref{fig:chicomp} shows the
relative quality of fit for two other models that we fit to all of our ACS
cluster data: the analytical $I(R)$ models of \citet{king62}, and power laws
with constant-density cores (see the end of \S\ref{subsec:structure}). Now,
however, we have formed $\Delta$ statistics by comparing the best-fit $\chi^2$
of these models to those of the \citeauthor{wil75} fits rather than
\citet{king66}. Points that scatter to the upper-right quadrant of this plot
thus represent clusters for which the \citeauthor{wil75} model is the best
match of the three. Once again, 90\% (111/124) of GCs in NGC 5128 have both
$\Delta_{\rm K62}>-0.2$ and $\Delta_{\rm PL}>-0.2$, such that
\citeauthor{wil75} models fit them {\it at least} as well as either of the
other two models.
This is again in keeping with the situation for both old and young massive
clusters in the Milky Way, the Magellanic Clouds, and the Fornax dwarf
spheroidal: \citet{mcl05} show that the curvature inherent in
\citeauthor{wil75} model envelopes is never any worse, and often much better,
than an unlimited power law as a description of any observed extended halo
structure.

\subsection{Model dependence of cluster properties}
\label{subsec:modcomp}

Having argued that isotropic \citeauthor{wil75} models fit the majority
of GCs as well as or better than any other model that we have tried,
we next ask whether estimates of physical cluster parameters
can be significantly biased by fitting the ``wrong'' type of model.

\begin{figure*}
\centerline{\hfil
   \rotatebox{270}{\includegraphics[height=170mm]{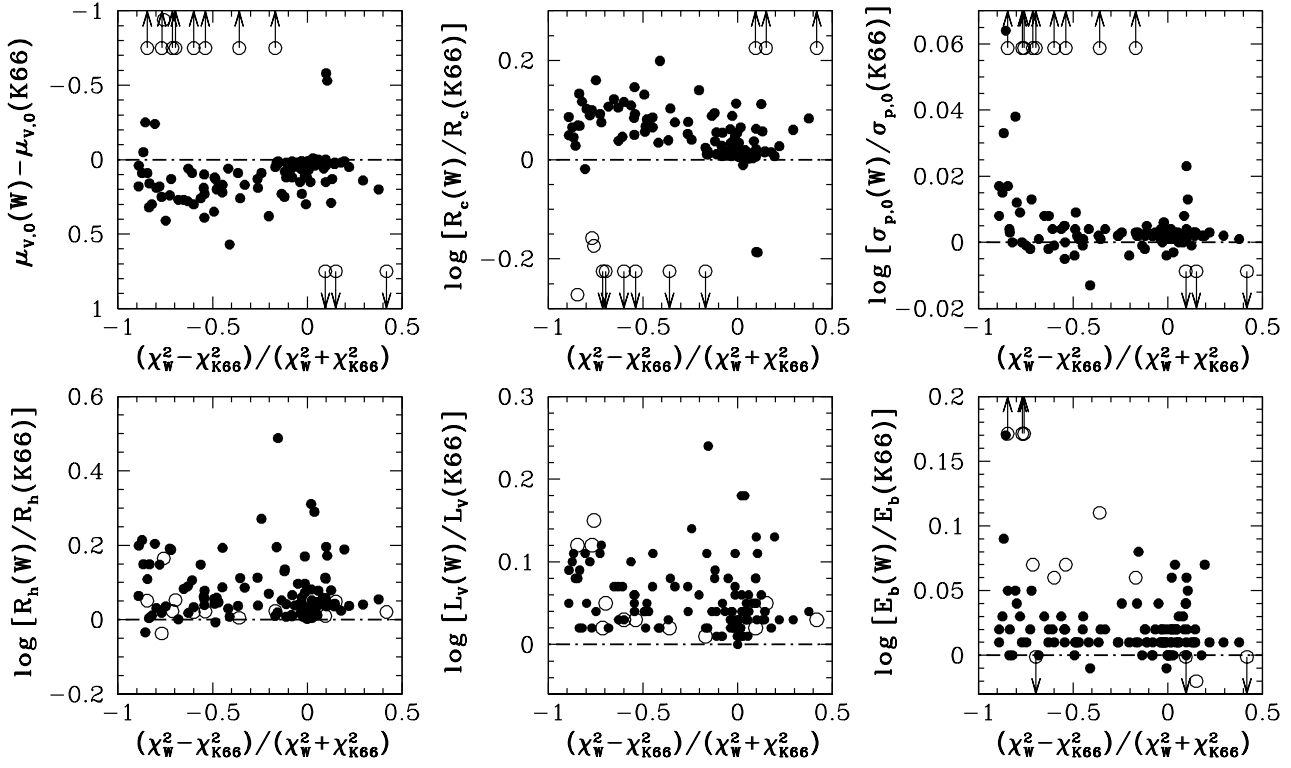}}
\hfil}
\caption{
{\it Left panels:} Comparison of isotropic \citeauthor{wil75} and
\citet{king66} model values for the intrinsic central surface brightnesses,
projected core radii, central line-of-sight velocity dispersions,
projected half-light (effective) radii, total luminosities, and global binding
energies for 124 reliable GC candidates in NGC 5128. Open circles represent 12
clusters with 
$|\mu_{V,0}({\rm Wilson})-\mu_{V,0}({\rm King})|>0.6$~mag~arcsec$^{-2}$,
which have particularly poorly constrained core structures. These
objects are listed in Table \ref{tab:poorfit}. Filled points represent the
remaining 112 clusters whose core parameters are better constrained.
\label{fig:parcompw}
}
\end{figure*}

Figure \ref{fig:parcompw} compares values from \citeauthor{wil75} and
\citet{king66} fits for the central surface
brightness $\mu_{V,0}$, projected core or half-power radius $\log\,R_c$,
central velocity dispersion $\log\,\sigma_{{\rm p},0}$, effective radius
$\log\,R_h$, total luminosity $\log\,L_V$, and global binding energy
$\log\,E_b$. The values of all parameters are taken from Tables
\ref{tab:n5128phot} and \ref{tab:n5128mass} above, and their differences
between the two models are plotted against the relative quality-of-fit
parameter $\Delta$ defined in equation (\ref{eq:delchi}). We again do this
for only 124 clusters; C168, F1GC34, F2GC14, C145, C152, C156, and C177 are
excluded as before.


\begin{figure*}
\centerline{\hfil
   \rotatebox{270}{\includegraphics[height=170mm]{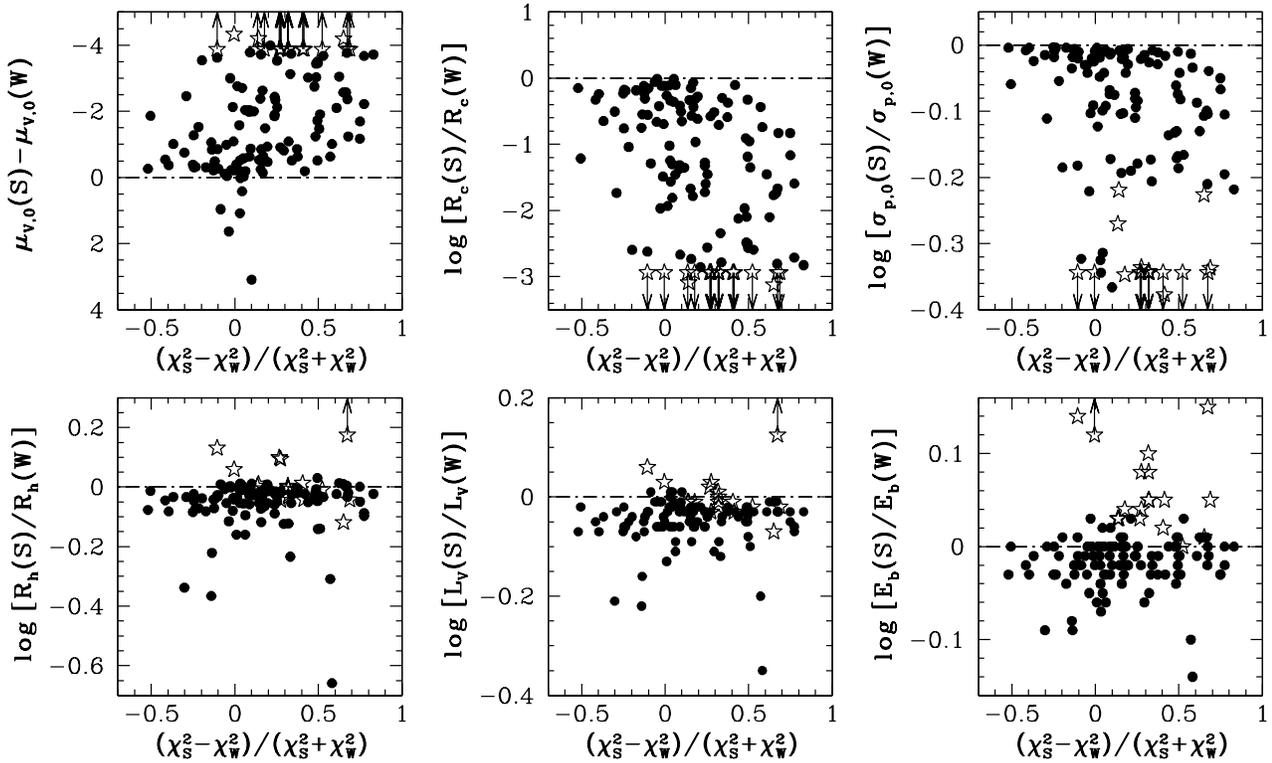}}
\hfil}
\caption{
Similar to Figure \ref{fig:parcompw}, but comparing cluster properties derived
from \citeauthor{sersic} versus \citeauthor{wil75} model fits. Open stars
represent 18 clusters with
$[\mu_{V,0}(${S\'ersic}$)-\mu_{V,0}({\rm Wilson})]<-4.0$~mag~arcsec$^{-2}$. 
\label{fig:parcomps}
}
\end{figure*}


The open circles in all panels of Figure \ref{fig:parcompw} refer to clusters
with differences of more than 0.6~mag~arcsec$^{-2}$ between the fitted
$\mu_{V,0}$ in the two models. There are 12 of these, listed for reference in
Table \ref{tab:poorfit}. Arrows are attached to the points for them
in the figure when the parameter differences fall outside the plotted range in
any panel. The innermost structures ($\mu_{V,0}$, $R_c$, and
$\sigma_{{\rm p},0}$) of these clusters are not 
well constrained in a model-independent sense, even if more
global properties like $R_h$, $L_V$, and $E_b$ are generally better
behaved. 

These 12 objects aside, the two models return encouragingly consistent
cluster parameters. The mean and rms scatter of the difference
$\delta \mu_{V,0} \equiv
    [\mu_{V,0}({\rm Wilson}) - \mu_{V,0}({\rm King\ 1966})]$ is
$0.10\pm0.15$~mag~arcsec$^{-2}$ (compared with an rms errorbar of 
$\pm\, 0.3$~mag~arcsec$^{-2}$), while the average
$\langle \delta(\log\,R_c) \rangle = 0.05\pm0.05$~dex (rms)
versus a typical errorbar of $\pm\, 0.12$. Thus, in order to
accomodate the more distended haloes of \citeauthor{wil75} models, the data
force these fits to (slightly) larger and fainter cores than in \citet{king66}
models. These effects combine to make the average
$\langle \delta(\log\,\sigma_{{\rm p},0}) \rangle \sim
       0.5\, \langle \delta(\log\,R_c) - 0.2\delta \mu_{V,0}\rangle$
a completely negligible $0.004$~dex. Similarly, the average
\citeauthor{wil75}-minus-\citeauthor{king66} offset in global binding energy
(which in either model is dominated by the core) is only
$\langle \delta(\log\,E_b) \rangle = 0.02\pm0.02$~dex (rms), compared
to an rms errorbar of $\pm\, 0.10$~dex.  

On the other hand, the projected half-light radius and total luminosity 
differences are slightly larger, and more systematic:
$\langle \delta(\log\,R_h) \rangle = 0.07\pm0.08$~dex and
$\langle \delta(\log\,L_V) \rangle = 0.05\pm0.04$~dex,
compared to average errorbars of $\pm\, 0.04$~dex in both cases. The
luminosity offset is the same as the $\simeq\! 0.15$-mag brighter integrated
magnitudes of \citeauthor{wil75} models, which was seen in Figure
\ref{fig:modmags} above. This and the systematically larger $R_h$ both result
from the intrinsically stronger haloes of \citeauthor{wil75} versus
\citet{king66} models.  

In most of these examples, the differences between the \citeauthor{wil75} and
\citet{king66} parameter estimates tend to be slightly larger for more
negative values of the relative $\chi^2$ statistic $\Delta$, i.e., for
clusters that are significantly better fit by the \citeauthor{wil75}
sphere. Clearly, the parameter values from this fit are then to be considered
the more reliable of the two. But even so, we conclude from
this discussion that analyses of structural correlations will {\it not}
be significantly affected by using \citet{king66} models when
\citeauthor{wil75} spheres are better fits---or {\it vice versa}.
The net effect of such an ``error'' on any given correlation will generally be
at the level of hundredths of a dex, which is within most
observational uncertainties. 


\begin{table}
\caption{NGC 5128 GCs with poorly constrained core parameters$^{~a}$
         \label{tab:poorfit}}
\begin{tabular}{@{}lclclc}
\hline
\multicolumn{1}{c}{Name}  &
\multicolumn{1}{c}{$\delta \mu_{V,0}$} &
\multicolumn{1}{c}{Name}  &
\multicolumn{1}{c}{$\delta \mu_{V,0}$} &
\multicolumn{1}{c}{Name}  &
\multicolumn{1}{c}{$\delta \mu_{V,0}$} \\
\multicolumn{1}{c}{(1)}  &
\multicolumn{1}{c}{(2)}  &
\multicolumn{1}{c}{(1)}  &
\multicolumn{1}{c}{(2)}  &
\multicolumn{1}{c}{(1)}  &
\multicolumn{1}{c}{(2)}  \\
\hline
AAT115339 & $+4.70$  &  C032   & $-1.37$  &  G284   & $-5.11$  \\
AAT118198 & $-2.60$  &  C116   & $+4.54$  &  K131   & $-3.41$  \\
C003      & $-0.94$  &  C129   & $+4.82$  &  PFF079 & $-3.53$  \\
C030      & $-1.03$  &  F2GC69 & $-7.43$  &  R223   & $-4.99$  \\
\hline
\end{tabular}

\medskip
$^{a}$~$\delta \mu_{V,0} \equiv
           \mu_{V,0}({\rm Wilson})-\mu_{v,0}({\rm King\ 1966})$

\end{table}


Figure \ref{fig:parcomps} compares estimates of the same cluster
parameters from \citeauthor{sersic} models versus
\citeauthor{wil75} models. The open stars in 
these graphs refer to 18 clusters with particularly large
$[\mu_{V,0}({\rm Sersic}) - \mu_{V,0}({\rm Wilson})] < -4$~mag~arcsec$^{-2}$.
Even apart from these, it is immediately obvious that \citeauthor{sersic} 
models typically imply much brighter and smaller cores, and rather
low central velocity dispersions, relative to \citeauthor{wil75}
(or, therefore, \citealt{king66}) models. This is mainly a combination
of model artifact and the PSF-blurring of the cores of our clusters.

With an assumed $I(R)\sim\exp(-R^{1/n})$, \citeauthor{sersic} models are
pushed towards high $n$ to fit the extended haloes of the observed GCs:
more than half (69/124) of the clusters represented in Figure
\ref{fig:parcomps} have $n>2$, which implies very strongly peaked 
central density profiles even in projection (e.g., see Figure
\ref{fig:morefits}; also, recall that the unprojected
density diverges and the velocity dispersion vanishes at $r=0$ for any
$n\ge 1$). These are probably allowed by the data here only because
the PSF flattens the models; realistically, most globular clusters do
not have such steep and non-isothermal cores.\footnotemark 
\footnotetext{We have, in fact, re-fit the 153 non--core-collapsed clusters
  in the Local Group sample of \citet{mcl05}---the cores of which are
  all completely resolved---with \citeauthor{sersic} models. Only 22\% of
  these (34/153) have $n>2$; and regardless of the value of $n$, only about
  5\% of all clusters are significantly better fit by a \citeauthor{sersic}
  model than by the isothermal cores of \citet{king66} or \citeauthor{wil75}
  models.} 

\citeauthor{sersic} models are therefore not a good choice to characterise
globular clusters whose cores are not fully resolved.
Nevertheless, the \citeauthor{sersic} values of
the more globally oriented structural parameters in Figure
\ref{fig:parcomps} ($R_h$, $L_V$, and $E_b$) are in better overall
agreement with those returned by \citeauthor{wil75} and  
\citet{king66} model fits. These quantities are evidently rather robust, in
general, against the details of the model adopted to measure them.
All in all, we conclude that \citeauthor{wil75} fits can safely be
used to investigate any GC parameter correlations, and it is indeed
the model that we employ for our discussion of the cluster fundamental
plane in \citetalias{mcl07}.  

\subsection{Clusters observed in \citetalias{har02}}
\label{subsec:STIS}

To build a cluster database for NGC 5128 that is as complete as
practical, we would like to fold the sample of 27 GCs observed in
\citetalias{har02} \citep{har02} into our current analysis.\footnotemark
\footnotetext{An entirely different sample of 21 clusters was also
  observed with HST, and fitted with \citet{king66} models, by 
  \citet{hol99}. However, their imaging program used the
  lower-resolution WF chips of the WFPC2 camera and focused on
  clusters at smaller galactocentric radius, against higher sky
  levels, than almost all of ours. We have not attempted to
  incorporate any results from these lower-quality data into our
  work.} 
As we mentioned at the beginning of \S\ref{sec:data}, \citet{king66}
models were fitted to STIS and WFPC2 surface-brightness profiles
for these clusters. However, a different modeling code was 
used \citep{hol99} to obtain a smaller set of cluster parameters
than we have derived for the ACS sample here. We do not re-fit the
original data, especially as 9 of the clusters there
have in fact been re-observed and 
re-modeled as part of this paper. Instead, we calculate the
full suite of physical properties in Tables \ref{tab:n5128phot},
\ref{tab:n5128mass}, and \ref{tab:n5128kappa} within the
context of \citet{king66} models only, starting from the few
parameters given in Table 2 of \citetalias{har02}.


\begin{table*}
\begin{minipage}{157mm}
\scriptsize
\caption{Predicted aperture velocity dispersions from 147 profiles
         of 131 GCs in NGC 5128 \label{tab:n5128vels}}
\begin{tabular}{@{}lcclccccccc}
\hline
\multicolumn{1}{c}{Name} &
\multicolumn{1}{c}{Detector} &
\multicolumn{1}{c}{$\Upsilon_V^{\rm pop}$} &
\multicolumn{1}{c}{Model} &
\multicolumn{1}{c}{$\log\,R_h$} &
\multicolumn{1}{c}{$\log\,R_h$} &
\multicolumn{1}{c}{$\log\,\sigma_{\rm ap}(R_h/8)$} &
\multicolumn{1}{c}{$\log\,\sigma_{\rm ap}(R_h/4)$} &
\multicolumn{1}{c}{$\log\,\sigma_{\rm ap}(R_h)$} &
\multicolumn{1}{c}{$\log\,\sigma_{\rm ap}(4\,R_h)$} &
\multicolumn{1}{c}{$\log\,\sigma_{\rm ap}(8\,R_h)$} \\
 &
 &
\multicolumn{1}{c}{[$M_\odot\,L_{\odot,V}^{-1}$]} &
 &
\multicolumn{1}{c}{[pc]} &
\multicolumn{1}{c}{[arcsec]} &
\multicolumn{1}{c}{[km~s$^{-1}$]} &
\multicolumn{1}{c}{[km~s$^{-1}$]} &
\multicolumn{1}{c}{[km~s$^{-1}$]} &
\multicolumn{1}{c}{[km~s$^{-1}$]} &
\multicolumn{1}{c}{[km~s$^{-1}$]} \\
\multicolumn{1}{c}{(1)} &
\multicolumn{1}{c}{(2)} &
\multicolumn{1}{c}{(3)} &
\multicolumn{1}{c}{(4)} &
\multicolumn{1}{c}{(5)} &
\multicolumn{1}{c}{(6)} &
\multicolumn{1}{c}{(7)} &
\multicolumn{1}{c}{(8)} &
\multicolumn{1}{c}{(9)} &
\multicolumn{1}{c}{(10)} &
\multicolumn{1}{c}{(11)} \\
\hline
   AAT111563  & WFC/F606  & $1.939^{+0.239}_{-0.240}$       & K66     &
                0.468  & $-0.798^{+0.005}_{-0.010}$  &
                $0.835^{+0.027}_{-0.030}$  & $0.833^{+0.027}_{-0.030}$  &
                $0.811^{+0.027}_{-0.030}$  & $0.762^{+0.027}_{-0.030}$  &
                $0.752^{+0.027}_{-0.030}$ \\
          ~~        & ~~        & ~~         & W     &
                0.486  & $-0.780^{+0.017}_{-0.014}$  &
                $0.838^{+0.027}_{-0.030}$  & $0.835^{+0.027}_{-0.030}$  &
                $0.809^{+0.027}_{-0.030}$  & $0.762^{+0.027}_{-0.030}$  &
                $0.749^{+0.027}_{-0.030}$ \\
          ~~        & ~~        & ~~         & S     &
                0.478  & $-0.787^{+0.010}_{-0.009}$  &
                $0.812^{+0.027}_{-0.031}$  & $0.828^{+0.027}_{-0.030}$  &
                $0.815^{+0.027}_{-0.030}$  & $0.764^{+0.027}_{-0.030}$  &
                $0.754^{+0.027}_{-0.030}$ \\
   AAT113992  & WFC/F606  & $2.914^{+0.431}_{-0.423}$       & K66     &
                0.560  & $-0.705^{+0.041}_{-0.023}$  &
                $0.893^{+0.032}_{-0.036}$  & $0.891^{+0.032}_{-0.036}$  &
                $0.869^{+0.032}_{-0.036}$  & $0.820^{+0.032}_{-0.036}$  &
                $0.811^{+0.032}_{-0.036}$ \\
          ~~        & ~~        & ~~         & W     &
                0.749  & $-0.516^{+0.144}_{-0.097}$  &
                $0.893^{+0.032}_{-0.036}$  & $0.888^{+0.032}_{-0.036}$  &
                $0.856^{+0.032}_{-0.037}$  & $0.807^{+0.033}_{-0.038}$  &
                $0.789^{+0.033}_{-0.038}$ \\
          ~~        & ~~        & ~~         & S     &
                0.528  & $-0.738^{+0.016}_{-0.015}$  &
                $0.863^{+0.032}_{-0.035}$  & $0.879^{+0.032}_{-0.035}$  &
                $0.871^{+0.032}_{-0.036}$  & $0.822^{+0.032}_{-0.036}$  &
                $0.816^{+0.032}_{-0.036}$ \\
   AAT115339  & WFC/F606  & $2.015^{+0.238}_{-0.235}$       & K66     &
                0.376  & $-0.890^{+0.003}_{-0.000}$  &
                $1.034^{+0.028}_{-0.029}$  & $1.022^{+0.027}_{-0.029}$  &
                $0.977^{+0.027}_{-0.029}$  & $0.914^{+0.026}_{-0.029}$  &
                $0.907^{+0.026}_{-0.029}$ \\
          ~~        & ~~        & ~~         & W     &
                0.397  & $-0.869^{+0.044}_{-0.028}$  &
                $0.970^{+0.026}_{-0.029}$  & $0.967^{+0.026}_{-0.029}$  &
                $0.940^{+0.026}_{-0.029}$  & $0.893^{+0.026}_{-0.029}$  &
                $0.879^{+0.026}_{-0.029}$ \\
          ~~        & ~~        & ~~         & S     &
                0.384  & $-0.881^{+0.009}_{-0.009}$  &
                $0.953^{+0.027}_{-0.029}$  & $0.967^{+0.027}_{-0.029}$  &
                $0.950^{+0.026}_{-0.029}$  & $0.898^{+0.026}_{-0.029}$  &
                $0.887^{+0.026}_{-0.029}$ \\
\hline
\end{tabular}

\medskip
  A machine-readable version of the full Table \ref{tab:n5128mass} is
  available online
  (http://www.astro.keele.ac.uk/$\sim$dem/clusters.html)
  or upon request from the first author.
  Only a short extract from it is shown here, for guidance
  regarding its form and content.

\end{minipage}
\end{table*}


It is necessary first to correct all of the intrinsic central surface
brightnesses previously published for these GCs. The earlier
model-fitting code of \citet{hol99} did not explicitly compute
$\mu_{V,0}$ for the best \citet{king66}
model, so it was estimated for every object in \citetalias{har02}
as the directly measured brightness of the central 1 pixel
of the cluster images. But the STIS pixel scale is
$0\farcs0508$~px$^{-1}$, so that $1~{\rm px} = 0.94~{\rm pc}$.
Unfortunately, this is comparable both to the half-width at half-maximum of 
the stellar PSF and to the typical GC core radius of $R_c\sim 1$~pc. The
intensity of the innermost pixel on a cluster image in this case is
therefore an average of the intrinsic surface-brightness profile over
the entire core and beyond, which is much fainter than the
brightness at exactly $R=0$. 

Here we use numerical calculations of \citet{king66}
models to derive a self-consistent $\mu_{V,0}$ from the published $W_0$,
angular $r_0$, and total apparent magnitudes
$V_{\rm tot}$ of  the clusters in \citetalias{har02}:
$\mu_{V,0}=V_{\rm tot}-2.5\,\log\,[{\cal I}_0/({\cal L}r_0^2)]$ for $r_0$ in
arcsec, where
${\cal I}_0$ and ${\cal L}$ are dimensionless functions of $W_0$
(or the concentration parameter $c$) described in, e.g., Appendix B of
\citet{mcl00}. Extinction-corrected $V_{\rm tot}$ are obtained from
the Washington $C T_1$ 
photometry given in Table 3 of \citetalias{har06} and our Table
\ref{tab:n5128colors} here, in exactly the same
way as described in connection with Figure \ref{fig:modmags} above. We
assign an errorbar of $\pm\, 0.3$ to every $W_0$ value, $\pm\, 10\%$
to every $r_0$, and $\pm\, 0.05$~mag to every $V_{\rm tot}$, and
propagate these to obtain uncertainties for the re-derived $\mu_{V,0}$.
Then, using the projected galactocentric radii given for all the clusters
in Table 1 of \citetalias{har02} and population-synthesis
mass-to-light ratios calculated for them in Table \ref{tab:n5128mtol}
above, we derive all of the parameters that we did for our ACS cluster
sample in \S\ref{subsec:derived}. 

Tables \ref{tab:har02fits}--\ref{tab:har02kappa} in Appendix
\ref{sec:STIStables} contain the results of this. Like Tables
\ref{tab:n5128fits}--\ref{tab:n5128kappa}, they are available in
machine-readable format, either online or upon request from the first
author.

As expected, the $\mu_{V,0}$ published in Table 2 of \citetalias{har02} are
all significantly fainter than our corrected values in Column (11) of Table 
\ref{tab:har02fits}: even after correcting the \citetalias{har02} numbers for
extinction ($A_V=0.341$~mag in all cases), we still have an average difference
of
$\langle \mu_{V,0}({\rm new}) - \mu_{V,0}({\rm old}) \rangle =
   -0.72 \pm 0.05$~mag~arcsec$^{-2}$,
due to the uncorrected effects of seeing on the old estimates.

Note that our effective radii $R_h$ in Column (6) of Table
\ref{tab:har02phot} are smaller than the quantities $r_h$ listed in Table
2 of \citetalias{har02}. This is because we record the projected half-light
radii, while \citetalias{har02} reported deprojected values. The ratio
$R_h/r_h$ ranges between $\simeq\! 0.73$--0.76 for the \citet{king66}
models that apply to real clusters.

The 9 clusters in common between \citetalias{har02} and the
current ACS sample (C007, C025, C029, C032, C037, C104, C105, G221, and
G293) are included in Tables \ref{tab:har02fits}--\ref{tab:har02kappa}.
We have compared the basic parameters $c$, $V_{\rm tot}$, $r_0$,
and $R_h$ given there to the results, in Tables \ref{tab:n5128fits} and
\ref{tab:n5128phot}, of our new \citet{king66} model fits to the ACS profiles
of these objects. The old and new concentrations and the integrated cluster
magnitudes agree quite well, but the radii $r_0$ and $R_h$ do not: both scales
are on average {\it smaller}, by a factor of about 0.7 (0.15 dex) in 
the mean, in our new analysis. As a result of this, the intrinsic central
surface brightnesses for these 9 objects in Table \ref{tab:n5128fits}
are typically {\it brighter} (by 0.8~$V$ mag~arcsec$^{-2}$ on average)
than even the revised estimates in Table \ref{tab:har02fits} based on the
\citetalias{har02} $r_0$ values. 

We have not been able to identify the cause of this discrepancy. Since
the very well resolved $R_h$ differ by as much as the core scales $r_0$ between
this paper and \citetalias{har02} for these 9 GCs, our PSF modeling and
convolution are not responsible. Also, direct inspection of the ACS data shows
clearly that the clusters are not as intrinsically faint at their centres
as the $\mu_{V,0}$ derived from the large $r_0$ in \citetalias{har02}
imply. Our \citet{king66} fits in \S\ref{sec:fits} above should be taken as
the definitive ones for the clusters observed both here and in
\citetalias{har02}. For the other 18 GCs in the earlier sample, we
caution that any structural or dynamical properties derived through
$r_0$ or $R_h$ could potentially be biased by overestimation of
these radii in \citetalias{har02}. This applies to most of the parameters
tabulated in Appendix \ref{sec:STIStables} and, e.g., to dynamical masses
estimated from velocity-dispersion measurements \citep{marho04,rej07}.

\section{Predicted Velocity Dispersions}
\label{sec:apvel}

Apart from studies of parameter correlations,
we expect that one eventual use for the material presented
in \S\ref{sec:fits} will be to facilitate the comparison of dynamical GC
mass-to-light ratios (derived from internal velocity-dispersion
measurements) to the population-synthesis model values that we calculated in
\S\ref{subsec:popsyn}. Such comparisons could potentially be 
used to constrain some of the stellar-population aspects of these
models; perhaps to identify clusters much younger than the 13 Gyr we
generally assume; and also to examine the extent to which long-term
dynamical cluster evolution (e.g., the evaporation of
low-mass stars driven by two-body relaxation) might affect $M/L$.

We have therefore calculated aperture velocity-dispersion profiles for all 131
objects in our ACS sample using the best-fit structural models from each of
the families we have explored in detail, and the 13-Gyr $\Upsilon_V^{\rm pop}$
listed in Table \ref{tab:n5128mtol}. Given any of our models with a fitted
$r_0$ and $W_0$ (or $c$) or \citeauthor{sersic} $n$, solving Poisson's and
Jeans' equations and projecting along the line of sight yields a dimensionless
$\widetilde{\sigma}_{\rm p}=\sigma_{\rm p}(R)/\sigma_{{\rm p},0}$ as a
function of projected clustercentric radius $R$. The predicted velocity
dispersion within a circular aperture of any radius $R_{\rm ap}$ then follows
from
\begin{equation}
\frac{\sigma_{\rm ap}^2}{\sigma_{{\rm p},0}^2} (R_{\rm ap}) \equiv
   \left[
      \int_{0}^{R_{\rm ap}} \widetilde{\sigma}_{\rm p}^2\, I(R)\, R\, dR
   \right]
   \left[
      \int_{0}^{R_{\rm ap}} I(R)\, R\, dR
   \right]^{-1} \ .
\label{eq:sigap}
\end{equation}

Using $\sigma_{{\rm p},0}$ from Table \ref{tab:n5128mass}, we have obtained
$\sigma_{\rm ap}$ for every 
structural model for every cluster, within each of five apertures:
$R_{\rm ap}=R_h/8,\, R_h/4,\, R_h,\, 4\,R_h$, and $8\,R_h$.
The results are listed in Table \ref{tab:n5128vels}. Including
$\sigma_{{\rm p},0}$ itself gives
$\sigma_{\rm ap}$ at $R_{\rm ap}=0$ for \citet{king66} and
\citeauthor{wil75} models, or at $R_{\rm ap}=R_c$ for
\citeauthor{sersic} models. The dispersion inside any other aperture
can then be obtained from interpolation on our numbers. 

By far the majority of a cluster is contained within $8\,R_h$
in any realistic model. Thus, the velocity dispersion in any larger aperture
differs from $\sigma_{\rm ap}(8\,R_h)$ by  $<\!0.01$~dex for all
\citet{king66} and \citeauthor{wil75} models, and by $<\!0.02$~dex for all
\citeauthor{sersic} models with $n\la  6$ (which includes about 95\% of the
GCs here). In any case, a typical GC half-light radius of 3--4~pc translates
to 0\farcs16--0\farcs22 in NGC 5128. A slit of width 1\arcsec, for
example, therefore usually corresponds to $R_{\rm ap}/R_h$ of just a few.

\citet{marho04} have measured the velocity dispersions for 14 relatively
bright GCs in NGC 5128, and \citet{rej07} have done so for 27 clusters
including some of the \citeauthor{marho04} objects. The combined sample
from these two studies includes 8 globulars modeled in this paper that do
not appear in Table \ref{tab:poorfit} above, and 9 objects observed only in
\citetalias{har02}, for which we have re-computed \citet{king66} parameters in
Appendix \ref{sec:STIStables}. Thus, we now compare the  observed
$\sigma_{\rm ap}$ of these 17 GCs to our \citet{king66} model predictions for
them. (We have calculated, though not tabulated, model velocity-dispersion
profiles for the \citetalias{har02} clusters in exactly the same way as for
the current ACS sample.)

For clusters observed by both \citet{marho04} and \citet{rej07}, we take
the root-mean-square of their independent velocity-dispersion measurements to
estimate a single observed $\sigma_{\rm ap}$. Then, since both
\citeauthor{marho04} and \citeauthor{rej07} used slits of width 1\arcsec\ for
their observations, we adopt an effective aperture radius of
$R_{\rm ap}=0\farcs71$---half the diagonal length of a 1\arcsec\ square---to
specify a predicted $\sigma_{\rm ap}$ for every cluster. With $D=3.8$~Mpc
for NGC 5128, this corresponds to $R_{\rm ap}\simeq 13.0$~pc, which is
significantly larger than the projected \citep{king66} half-light radii of all
17 GCs being considered here: $1.7 \la R_{\rm ap}/R_h \la 5.7$, with a median
of 2.8. As a result, even though the choice of $R_{\rm ap}$ is not
particularly rigorous, changing it by a factor of two affects our
comparison of dynamical and population-synthesis mass-to-light ratios at only
the $\simeq\!5\%$ level.  


\begin{figure}
\centerline{\hfil
   \includegraphics[width=75mm]{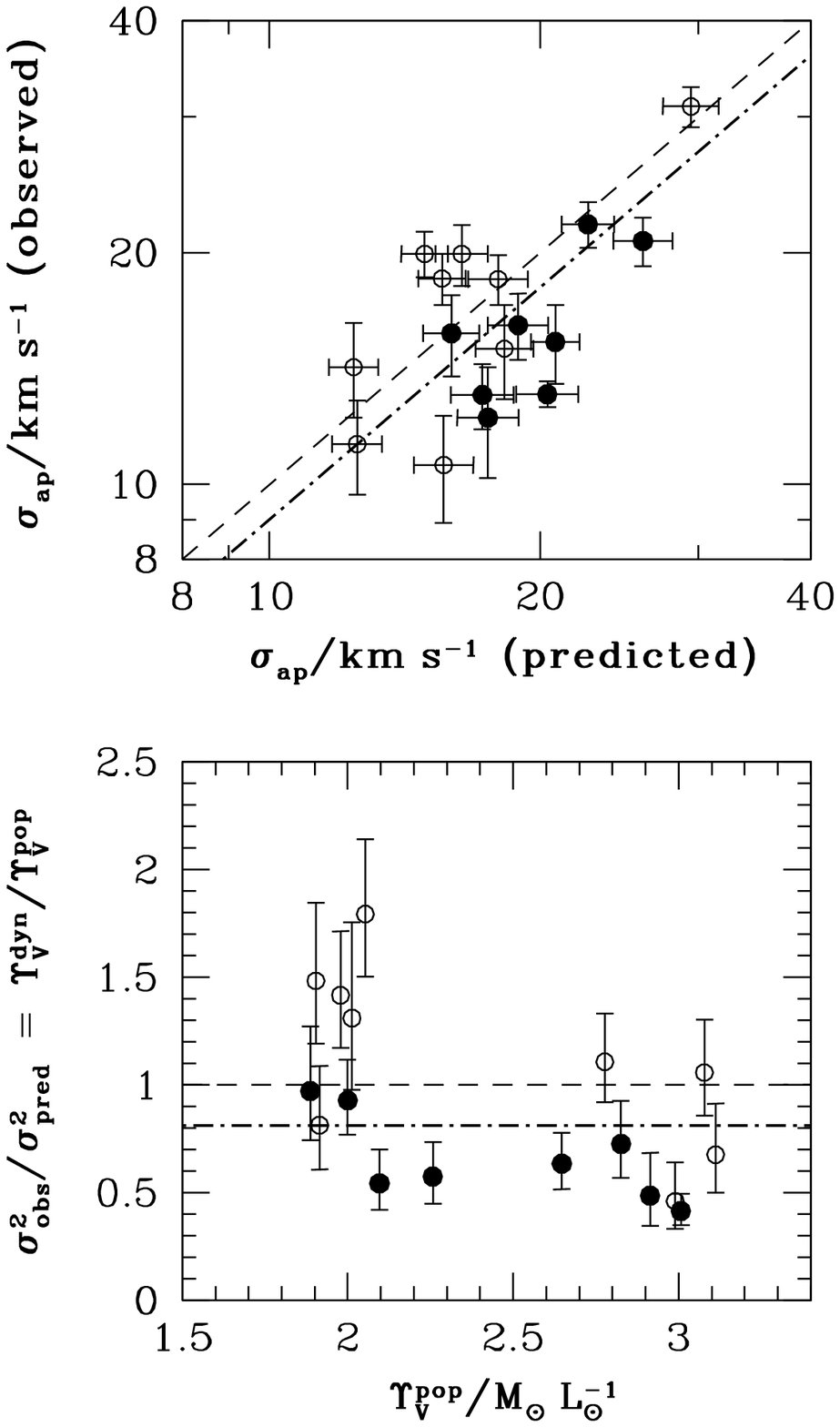}
\hfil}
\caption{
{\it Top:} Velocity dispersions observed for 17 GCs in NGC 5128 by
\citet{marho04} and \citet{rej07}, against dispersions within
apertures of radius $0\farcs71=13$~pc as predicted by our \citet{king66}
model fits and population-synthesis mass-to-light ratios.
{\it Bottom:} The implied ratio of dynamical and population-synthesis
mass-to-light ratios, as a function of model $\Upsilon_V^{\rm pop}$.
In both panels, filled circles denote 8 clusters in the ACS
sample of this paper:
C006, C007, C012, C018, C025, C029, C036, and C037.
Open circles correspond to 9 GCs observed only in
\citetalias{har02}, with revised \citet{king66} structural parameters given in
Appendix \ref{sec:STIStables} here:
C002, C011, C017, C021, C022, C023, C031, C041, and C044.
The bold, dash-dot lines indicate the
median ratios
$\sigma_{\rm ap}({\rm obs})/\sigma_{\rm ap}({\rm pred}) = 0.90$ and
$\Upsilon_V^{\rm dyn}/\Upsilon_V^{\rm pop} = 0.81$.
\label{fig:popdyn}
}
\end{figure}


The upper panel of Figure \ref{fig:popdyn} shows the observed versus model
$\sigma_{\rm ap}$ for the 17 objects in question. The lower panel then shows
the square of the ratio of observed to predicted dispersions,
$\sigma_{\rm obs}^2/\sigma_{\rm pred}^2$, which is
equal to the ratio of dynamical to population-synthesis $\Upsilon_V$.
The filled circles in both panels of the figure refer to
clusters with ACS data analysed in this paper; the open circles, to clusters
only in the \citetalias{har02} sample. The apparent mean offset
between these two sets of points could be the result of overestimating the
half-light radii of the GCs measured only in
\citetalias{har02} (as discussed at the end of \S\ref{subsec:STIS}),
since a spuriously large $R_h$ implies a low 
$\sigma_{\rm pred}^2 \propto \Upsilon_V^{\rm pop} L_V/R_h$.
In any event, the median ratio of $\sigma_{\rm obs}/\sigma_{\rm pred}$
for all 17 clusters is 0.90, and the rms scatter about this 
value is $\pm\, 0.20$. The median and rms scatter of
$\Upsilon_V^{\rm dyn}/\Upsilon_V^{\rm pop}$ is therefore $0.81\pm0.40$.

The main point to be taken from this is that our population-synthesis
estimates are relatively good guides, on
average, to the global dynamical mass-to-light ratios of globular clusters in
NGC 5128. This is also true in general of GCs in the Local Group that
have been modeled in the same way \citep{mcl05,barmby07}. In our
fundamental-plane analysis of \citetalias{mcl07} (and also
\citealt{barmby07}), we therefore use the cluster properties derived
from $\Upsilon_V^{\rm pop}$ in \S\ref{subsec:derived}, to explore accurately
a wide range of structural and dynamical correlations for much larger samples
of globulars than would be possible if we were restricted to
objects with direct velocity-dispersion measurements.

A more specific implication is that, as bright and as massive as the GCs
studied by \citet{marho04} and \citet{rej07} are (the 17 in Figure
\ref{fig:popdyn} have
$6.3 \times 10^5\,L_\odot \la L_V \la 2.6\times 10^6\,L_\odot$ and an
average $\langle \Upsilon_V \rangle \approx 2\ M_\odot\,L_\odot^{-1}$ in our
analysis), there is nothing in their mass-to-light ratios alone to suggest
that they are otherwise unusual. Both \citeauthor{marho04} and
\citeauthor{rej07} quote an average dynamical
$\langle \Upsilon_V^{\rm dyn} \rangle \approx 3\ M_\odot\,L_\odot^{-1}$
for their cluster samples, which is a factor of two higher than the average
for a representative sample of Milky Way GCs \citep[e.g.][]{mcl00,mcl05}. But
nearly half of the NGC 5128 clusters plotted in Figure \ref{fig:popdyn}
are {\it expected} to have $\Upsilon_V$ in this range. This is
simply because they are redder and, by inference, more
metal-rich than the majority of Galactic globulars; see Figure
\ref{fig:n5128mtol} above, and note also that $\Upsilon_V^{\rm pop}$ along
the $x$-axis of the lower panel in Figure \ref{fig:popdyn} is a rough
placeholder for cluster metallicity or Washington $(C-T_1)_0$ colour. The
other half of the GCs in Figure \ref{fig:popdyn} are relatively bluer and
more metal-poor, and they have an expected
$\Upsilon_V^{\rm pop}\approx 2\ M_\odot\,L_\odot^{-1}$ that is closer to the
Milky Way GCs. While 4 of these appear to have larger true
,$\Upsilon_V^{\rm dyn}$, 5 others do not. Moreover, the clusters with
the largest $\Upsilon_V^{\rm dyn}/\Upsilon_V^{\rm pop}$ appear only in the
sample of \citetalias{har02}, which again may have overestimated $R_h$
and thus 
$\Upsilon_V^{\rm dyn} \propto \sigma_{\rm obs}^2\,R_h/L_V$.

Aside from any possible issues with $R_h$ measurements, many of the dynamical
mass-to-light ratios that we have obtained for the 17 GCs in 
Figure \ref{fig:popdyn} are smaller than those computed by either
\citet{marho04} or \citet{rej07} for the same objects. This is usually because
both of these other groups estimate ``virial'' cluster masses using the
formula $M_{\rm vir} = 10\, \sigma_{\infty}^2 R_h/G$, where $\sigma_{\infty}$
is meant to be the velocity dispersion measured through an infinite aperture
and $R_h$ is the projected half-light radius. However, the coefficient in this
expression is model-dependent, ranging from 8--10 for \citet{king66}
models with concentrations $c < 2.5$. In addition, both \citeauthor{marho04}
and \citeauthor{rej07} assume that the aperture correction to relate their
measured $\sigma_{\rm obs}$ to the formal $\sigma_{\infty}$ is always
negligible. In fact, our \citet{king66} model calculations for the clusters in
Figure \ref{fig:popdyn} yield
$\sigma_{\infty}^2/\sigma_{\rm ap}^2(0\farcs71) \simeq 0.75$--0.96, so making
the correction (which we effectively do) can sometimes lower $M_{\rm vir}$ and
$\Upsilon_V^{\rm dyn}$ noticeably. 

Finally, we note that the ratio $\Upsilon_V^{\rm dyn}/\Upsilon_V^{\rm pop}$
does not correlate with the total luminosity of the clusters in Figure
\ref{fig:popdyn}. We conclude again that these objects are not special
or unusual in terms of their mass-to-light ratios only. \citet{marho04} and
\citet{rej07} go on to consider how the massive GCs in NGC 5128 relate to
Milky Way globulars, dwarf-galaxy nuclei, and ultra-compact dwarf galaxies, in
terms of structural correlations and the fundamental plane. We will discuss
this further in \citetalias{mcl07}.

\section{Summary}
\label{sec:summary}

We have used new HST imaging with the ACS/WFC camera to construct internal
surface-brightness profiles and derive structural parameters for a sample of 
131 globular cluster (GC) candidates in NGC 5128, the nearest giant elliptical
galaxy. We combined these with a sample of 27 clusters previously measured
with STIS or WFPC2 \citep[][$\equiv$~\citetalias{har02}]{har02}, for which
we have corrected and extended the structural parameters given in 
\citetalias{har02}. 

After identifying 4 ACS-observed objects that may not be not bona fide GCs
and another 3 that have especially ragged or contaminated surface-brightness
profiles, and accounting for 9 re-observations from the \citetalias{har02}
sample, we have now observed and modeled a total of 142 distinct, reliable
GC candidates in NGC 5128. The catalogue of parameters that we have produced
for them rivals the compilation of \citet{har96} for Milky Way globulars in
size and is based on more homogeneous data and analysis.
The final product is very similar in most respects to those of
\citet{mcl05} and \citet{barmby07} for a total of 196 old GCs and 50 massive
young star clusters in the Milky Way, the Magellanic Clouds, the Fornax dwarf
spheroidal, and M31.

We fitted the NGC 5128 clusters' surface-brightness profiles with three
different structural models: the standard \citet{king66} modified isothermal
sphere, an isotropic version of \citeauthor{wil75}'s (\citeyear{wil75})
alternate modification of an
isothermal sphere, and the $R^{1/n}$ surface-density profile of
\citet{sersic}. We estimated $V$-band mass-to-light ratios individually for
every cluster, by applying population-synthesis models given estimates of
cluster metallicities from their $(C-T_1)$ colours in the Washington filter
system and assuming a uniformly old age of 13 Gyr. We combined these with our
structural modeling to derive a wide range of physical properties
including central surface brightness and central potential, concentration
indices, core and half-light (effective) radii, total luminosity, and
{\it predicted} dynamical parameters including total 
mass, binding energy, central mass density, central escape velocity,
relaxation timescales and phase-space densities, and velocity dispersions
in series of circular apertures.

Extensive consistency checks and intercomparisons of the three different
models (\citealt{king66}, \citeauthor{wil75}, and \citeauthor{sersic}) have
shown that in most cases the \citeauthor{wil75} model fits at least as well as
the others and in many cases is superior to them, taking better account of the
detailed shape of the distended, low-intensity outer envelopes that many
clusters have. In this respect the NGC 5128 globulars are like those in the
Milky Way, the Magellanic Clouds, and the Fornax dwarf spheroidal
\citep{mcl05}, although the situation in M31 is not as clear \citep{barmby07}.

We compared aperture velocity dispersions predicted by our analysis for
17 GCs in NGC 5128, to actual measurements by \citet{marho04} and
\citet{rej07}. The expectations based on population-synthesis models are
in reasonable overall agreement with the observations.

In \citetalias{mcl07} \citep{mcl07}, we use the data derived here to discuss
a wide range of structural correlations for globular clusters in
NGC 5128, including comparisons with Milky Way GCs and with other
types of massive, cluster-like objects. \citet{barmby07} also use our
results, in a discussion of the combined fundamental plane of old GCs
in six galaxies. 

\section*{Acknowledgments}

We thank Claudia Maraston for computing $(V-F606W)$ cluster colours for us
from her population-synthesis models, and Andr\'es Jord\'an for helpful
discussions.
DEM and PB acknowledge a grant for HST program GO-10260 from the Space
Telescope Science Institute, which is operated by the Association of
Universities for Research in Astronomy, Inc., under NASA contract
NAS 5-26555.
WEH and GLHH thank the Natural Sciences and Engineering Research Council of
Canada for financial support, and
DAF thanks the Australian Research Council.


\appendix
\section{Parameters of Clusters Observed in Paper I}
\label{sec:STIStables}

Here we present the tables of fitted and derived properties
for \citet{king66} model descriptions of the 27 globular clusters observed
by STIS or WFPC2 and previously modeled in \citetalias{har02}
\citep{har02}. The construction of these 
tables is detailed in \S\ref{subsec:STIS}. Their contents and format
are essentially the same as those of Tables
\ref{tab:n5128fits}--\ref{tab:n5128kappa} for our main sample of
clusters observed with ACS (see the descriptions in
\S\ref{subsec:profs} and \S\ref{subsec:derived}), and they are also
available in machine-readable format, either online
(at http://www.astro.keele.ac.uk/$\sim$dem/clusters.html)
or upon request from the first author.

Some of the columns in Tables
\ref{tab:har02fits}--\ref{tab:har02kappa} are present only
to keep all of our tabulations of the fits to the combined
ACS+\citetalias{har02} cluster sample in a single, well-defined 
format. For example, Column (4) of Table \ref{tab:har02fits} contains the
``colour'' $V-V$, which is of course always identically zero but is the
analogue of the nontrivial $(V-F606)_0$ in Table \ref{tab:n5128fits}
(\S\ref{subsec:profs} above).


\begin{table*}
\begin{minipage}{160mm}
\scriptsize
\caption{Basic parameters of \citet{king66} model fits to 27 GCs from
 \citet[][$\equiv$~\citetalias{har02}]{har02} \label{tab:har02fits}}
\begin{tabular}{@{}lcccclrrrccrr}
\hline
\multicolumn{1}{c}{Name} &
\multicolumn{1}{c}{Detector} &
\multicolumn{1}{c}{$A_V$} &
\multicolumn{1}{c}{$V-V$} &
\multicolumn{1}{c}{$N_{\rm pts}$} &
\multicolumn{1}{c}{Model} &
\multicolumn{1}{c}{$\chi_{\rm min}^2$} &
\multicolumn{1}{c}{$I_{\rm bkg}$} &
\multicolumn{1}{c}{$W_0$} &
\multicolumn{1}{c}{$c$} &
\multicolumn{1}{c}{$\mu_{V,0}$} &
\multicolumn{1}{c}{$\log\,r_0$} &
\multicolumn{1}{c}{$\log\,r_0$} \\
    &
    &
\multicolumn{1}{c}{[mag]} &
\multicolumn{1}{c}{[mag]} &
    &
    &
    &
\multicolumn{1}{c}{[$L_\odot\,{\rm pc}^{-2}$]} &
    &
    &
\multicolumn{1}{c}{[mag arcsec$^{-2}$]} &
\multicolumn{1}{c}{[arcsec]} &
\multicolumn{1}{c}{[pc]}    \\
\multicolumn{1}{c}{(1)}  &
\multicolumn{1}{c}{(2)}  &
\multicolumn{1}{c}{(3)}  &
\multicolumn{1}{c}{(4)}  &
\multicolumn{1}{c}{(5)}  &
\multicolumn{1}{c}{(6)}  &
\multicolumn{1}{c}{(7)}  &
\multicolumn{1}{c}{(8)}  &
\multicolumn{1}{c}{(9)}  &
\multicolumn{1}{c}{(10)} &
\multicolumn{1}{c}{(11)} &
\multicolumn{1}{c}{(12)} &
\multicolumn{1}{c}{(13)} \\
\hline
        C002   & HAR02/V   & $0.341$  & $0.000\pm0.000$       & $5$       &
                 K66    & $1.00$  & $0.00\pm0.00$  & $8.50^{+0.30}_{-0.30}$  &
                 $1.98^{+0.08}_{-0.09}$  & $14.78^{+0.08}_{-0.10}$  &
                 $-1.367^{+0.041}_{-0.046}$  & $-0.101^{+0.041}_{-0.046}$ \\
        C007   & HAR02/V   & $0.341$  & $0.000\pm0.000$       & $5$       &
                 K66    & $1.00$  & $0.00\pm0.00$  & $8.00^{+0.30}_{-0.30}$  &
                 $1.83^{+0.09}_{-0.09}$  & $14.43^{+0.10}_{-0.11}$  &
                 $-1.119^{+0.041}_{-0.046}$  & $0.146^{+0.041}_{-0.046}$ \\
        C011   & HAR02/V   & $0.341$  & $0.000\pm0.000$       & $5$       &
                 K66    & $1.00$  & $0.00\pm0.00$  & $8.20^{+0.30}_{-0.30}$  &
                 $1.89^{+0.09}_{-0.09}$  & $15.04^{+0.09}_{-0.10}$  &
                 $-1.155^{+0.041}_{-0.046}$  & $0.110^{+0.041}_{-0.046}$ \\
        C017   & HAR02/V   & $0.341$  & $0.000\pm0.000$       & $5$       &
                 K66    & $1.00$  & $0.00\pm0.00$  & $6.60^{+0.30}_{-0.30}$  &
                 $1.41^{+0.09}_{-0.08}$  & $15.31^{+0.14}_{-0.16}$  &
                 $-0.914^{+0.041}_{-0.046}$  & $0.352^{+0.041}_{-0.046}$ \\
        C021   & HAR02/V   & $0.341$  & $0.000\pm0.000$       & $5$       &
                 K66    & $1.00$  & $0.00\pm0.00$  & $8.10^{+0.30}_{-0.30}$  &
                 $1.86^{+0.09}_{-0.09}$  & $14.91^{+0.09}_{-0.11}$  &
                 $-1.187^{+0.041}_{-0.046}$  & $0.078^{+0.041}_{-0.046}$ \\
        C022   & HAR02/V   & $0.341$  & $0.000\pm0.000$       & $5$       &
                 K66    & $1.00$  & $0.00\pm0.00$  & $7.30^{+0.30}_{-0.30}$  &
                 $1.62^{+0.09}_{-0.09}$  & $14.57^{+0.12}_{-0.13}$  &
                 $-1.229^{+0.041}_{-0.046}$  & $0.036^{+0.041}_{-0.046}$ \\
        C023   & HAR02/V   & $0.341$  & $0.000\pm0.000$       & $5$       &
                 K66    & $1.00$  & $0.00\pm0.00$  & $7.50^{+0.30}_{-0.30}$  &
                 $1.68^{+0.09}_{-0.09}$  & $13.26^{+0.12}_{-0.13}$  &
                 $-1.328^{+0.041}_{-0.046}$  & $-0.063^{+0.041}_{-0.046}$ \\
        C025   & HAR02/V   & $0.341$  & $0.000\pm0.000$       & $5$       &
                 K66    & $1.00$  & $0.00\pm0.00$  & $8.20^{+0.30}_{-0.30}$  &
                 $1.89^{+0.09}_{-0.09}$  & $15.19^{+0.09}_{-0.10}$  &
                 $-1.276^{+0.041}_{-0.046}$  & $-0.010^{+0.041}_{-0.046}$ \\
        C029   & HAR02/V   & $0.341$  & $0.000\pm0.000$       & $5$       &
                 K66    & $1.00$  & $0.00\pm0.00$  & $8.10^{+0.30}_{-0.30}$  &
                 $1.86^{+0.09}_{-0.09}$  & $15.10^{+0.09}_{-0.11}$  &
                 $-1.194^{+0.041}_{-0.046}$  & $0.072^{+0.041}_{-0.046}$ \\
        C031   & HAR02/V   & $0.341$  & $0.000\pm0.000$       & $5$       &
                 K66    & $1.00$  & $0.00\pm0.00$  & $7.70^{+0.30}_{-0.30}$  &
                 $1.74^{+0.09}_{-0.09}$  & $14.51^{+0.11}_{-0.12}$  &
                 $-1.310^{+0.041}_{-0.046}$  & $-0.044^{+0.041}_{-0.046}$ \\
        C032   & HAR02/V   & $0.341$  & $0.000\pm0.000$       & $5$       &
                 K66    & $1.00$  & $0.00\pm0.00$  & $8.80^{+0.30}_{-0.30}$  &
                 $2.07^{+0.08}_{-0.08}$  & $14.21^{+0.07}_{-0.09}$  &
                 $-1.523^{+0.041}_{-0.046}$  & $-0.258^{+0.041}_{-0.046}$ \\
        C037   & HAR02/V   & $0.341$  & $0.000\pm0.000$       & $5$       &
                 K66    & $1.00$  & $0.00\pm0.00$  & $8.10^{+0.30}_{-0.30}$  &
                 $1.86^{+0.09}_{-0.09}$  & $13.90^{+0.09}_{-0.11}$  &
                 $-1.509^{+0.041}_{-0.046}$  & $-0.243^{+0.041}_{-0.046}$ \\
        C040   & HAR02/V   & $0.341$  & $0.000\pm0.000$       & $5$       &
                 K66    & $1.00$  & $0.00\pm0.00$  & $7.50^{+0.30}_{-0.30}$  &
                 $1.68^{+0.09}_{-0.09}$  & $16.80^{+0.11}_{-0.13}$  &
                 $-0.967^{+0.041}_{-0.046}$  & $0.299^{+0.041}_{-0.046}$ \\
        C041   & HAR02/V   & $0.341$  & $0.000\pm0.000$       & $5$       &
                 K66    & $1.00$  & $0.00\pm0.00$  & $8.10^{+0.30}_{-0.30}$  &
                 $1.86^{+0.09}_{-0.09}$  & $14.64^{+0.09}_{-0.11}$  &
                 $-1.377^{+0.041}_{-0.046}$  & $-0.111^{+0.041}_{-0.046}$ \\
        C044   & HAR02/V   & $0.341$  & $0.000\pm0.000$       & $5$       &
                 K66    & $1.00$  & $0.00\pm0.00$  & $7.00^{+0.30}_{-0.30}$  &
                 $1.53^{+0.09}_{-0.09}$  & $15.15^{+0.13}_{-0.14}$  &
                 $-1.174^{+0.041}_{-0.046}$  & $0.091^{+0.041}_{-0.046}$ \\
        C100   & HAR02/V   & $0.341$  & $0.000\pm0.000$       & $5$       &
                 K66    & $1.00$  & $0.00\pm0.00$  & $7.75^{+0.30}_{-0.30}$  &
                 $1.76^{+0.09}_{-0.09}$  & $16.92^{+0.11}_{-0.12}$  &
                 $-1.066^{+0.041}_{-0.046}$  & $0.200^{+0.041}_{-0.046}$ \\
        C101   & HAR02/V   & $0.341$  & $0.000\pm0.000$       & $5$       &
                 K66    & $1.00$  & $0.00\pm0.00$  & $6.50^{+0.30}_{-0.30}$  &
                 $1.39^{+0.08}_{-0.08}$  & $18.03^{+0.14}_{-0.16}$  &
                 $-0.959^{+0.041}_{-0.046}$  & $0.307^{+0.041}_{-0.046}$ \\
        C102   & HAR02/V   & $0.341$  & $0.000\pm0.000$       & $5$       &
                 K66    & $1.00$  & $0.00\pm0.00$  & $9.20^{+0.30}_{-0.30}$  &
                 $2.17^{+0.07}_{-0.08}$  & $19.07^{+0.07}_{-0.09}$  &
                 $-1.222^{+0.041}_{-0.046}$  & $0.043^{+0.041}_{-0.046}$ \\
        C103   & HAR02/V   & $0.341$  & $0.000\pm0.000$       & $5$       &
                 K66    & $1.00$  & $0.00\pm0.00$  & $7.60^{+0.30}_{-0.30}$  &
                 $1.71^{+0.09}_{-0.09}$  & $15.35^{+0.11}_{-0.12}$  &
                 $-1.276^{+0.041}_{-0.046}$  & $-0.010^{+0.041}_{-0.046}$ \\
        C104   & HAR02/V   & $0.341$  & $0.000\pm0.000$       & $5$       &
                 K66    & $1.00$  & $0.00\pm0.00$  & $7.20^{+0.30}_{-0.30}$  &
                 $1.59^{+0.09}_{-0.09}$  & $16.09^{+0.12}_{-0.14}$  &
                 $-1.252^{+0.041}_{-0.046}$  & $0.014^{+0.041}_{-0.046}$ \\
        C105   & HAR02/V   & $0.341$  & $0.000\pm0.000$       & $5$       &
                 K66    & $1.00$  & $0.00\pm0.00$  & $7.20^{+0.30}_{-0.30}$  &
                 $1.59^{+0.09}_{-0.09}$  & $21.24^{+0.12}_{-0.14}$  &
                 $-0.719^{+0.041}_{-0.046}$  & $0.546^{+0.041}_{-0.046}$ \\
        C106   & HAR02/V   & $0.341$  & $0.000\pm0.000$       & $5$       &
                 K66    & $1.00$  & $0.00\pm0.00$  & $6.80^{+0.30}_{-0.30}$  &
                 $1.47^{+0.09}_{-0.08}$  & $16.27^{+0.14}_{-0.15}$  &
                 $-1.456^{+0.041}_{-0.046}$  & $-0.191^{+0.041}_{-0.046}$ \\
        G019   & HAR02/V   & $0.341$  & $0.000\pm0.000$       & $5$       &
                 K66    & $1.00$  & $0.00\pm0.00$  & $7.40^{+0.30}_{-0.30}$  &
                 $1.65^{+0.09}_{-0.09}$  & $16.70^{+0.12}_{-0.13}$  &
                 $-0.996^{+0.041}_{-0.046}$  & $0.270^{+0.041}_{-0.046}$ \\
        G221   & HAR02/V   & $0.341$  & $0.000\pm0.000$       & $5$       &
                 K66    & $1.00$  & $0.00\pm0.00$  & $7.10^{+0.30}_{-0.30}$  &
                 $1.56^{+0.09}_{-0.09}$  & $15.70^{+0.13}_{-0.14}$  &
                 $-1.222^{+0.041}_{-0.046}$  & $0.043^{+0.041}_{-0.046}$ \\
        G277   & HAR02/V   & $0.341$  & $0.000\pm0.000$       & $5$       &
                 K66    & $1.00$  & $0.00\pm0.00$  & $6.90^{+0.30}_{-0.30}$  &
                 $1.50^{+0.09}_{-0.09}$  & $15.59^{+0.13}_{-0.15}$  &
                 $-1.174^{+0.041}_{-0.046}$  & $0.091^{+0.041}_{-0.046}$ \\
        G293   & HAR02/V   & $0.341$  & $0.000\pm0.000$       & $5$       &
                 K66    & $1.00$  & $0.00\pm0.00$  & $7.80^{+0.30}_{-0.30}$  &
                 $1.77^{+0.09}_{-0.09}$  & $14.70^{+0.10}_{-0.12}$  &
                 $-1.444^{+0.041}_{-0.046}$  & $-0.178^{+0.041}_{-0.046}$ \\
        G302   & HAR02/V   & $0.341$  & $0.000\pm0.000$       & $5$       &
                 K66    & $1.00$  & $0.00\pm0.00$  & $7.15^{+0.30}_{-0.30}$  &
                 $1.57^{+0.09}_{-0.09}$  & $15.33^{+0.13}_{-0.14}$  &
                 $-1.268^{+0.041}_{-0.046}$  & $-0.002^{+0.041}_{-0.046}$ \\
\hline
\end{tabular}

\medskip
  A machine-readable version of Table \ref{tab:har02fits} is
  available online
  (http://www.astro.keele.ac.uk/$\sim$dem/clusters.html)
  or upon request from the first author.

\end{minipage}
\end{table*}



\begin{table*}
\begin{minipage}{157mm}
\scriptsize
\caption{Derived structural and photometric parameters from \citet{king66}
         model fits to 27 GCs from
         \citet[][$\equiv$~\citetalias{har02}]{har02} \label{tab:har02phot}}
\begin{tabular}{@{}lclcrccccccc}
\hline
\multicolumn{1}{c}{Name} &
\multicolumn{1}{c}{Detector} &
\multicolumn{1}{c}{Model} &
\multicolumn{1}{c}{$\log\,r_{\rm tid}$} &
\multicolumn{1}{c}{$\log\,R_c$} &
\multicolumn{1}{c}{$\log\,R_h$} &
\multicolumn{1}{c}{$\log\,(R_h/R_c)$  } &
\multicolumn{1}{c}{$\log\,I_0$ } &
\multicolumn{1}{c}{$\log\,j_0$} &
\multicolumn{1}{c}{$\log\,L_{V}$ } &
\multicolumn{1}{c}{$V_{\rm tot}$} &
\multicolumn{1}{c}{$\log\,I_h$} \\
 &
 &
 &
\multicolumn{1}{c}{[pc]} &
\multicolumn{1}{c}{[pc]} &
\multicolumn{1}{c}{[pc]} &
 &
\multicolumn{1}{c}{[$L_{\odot, V}\,{\rm pc}^{-2}$]} &
\multicolumn{1}{c}{[$L_{\odot, V}\,{\rm pc}^{-3}$]} &
\multicolumn{1}{c}{[$L_{\odot, V}$]} &
\multicolumn{1}{c}{[mag]} &
\multicolumn{1}{c}{[$L_{\odot, V}\,{\rm pc}^{-2}$]}    \\
\multicolumn{1}{c}{(1) } &
\multicolumn{1}{c}{(2) } &
\multicolumn{1}{c}{(3) } &
\multicolumn{1}{c}{(4) } &
\multicolumn{1}{c}{(5) } &
\multicolumn{1}{c}{(6) } &
\multicolumn{1}{c}{(7) } &
\multicolumn{1}{c}{(8) } &
\multicolumn{1}{c}{(9) } &
\multicolumn{1}{c}{(10)} &
\multicolumn{1}{c}{(11)} &
\multicolumn{1}{c}{(12)} \\
\hline
        C002   & HAR02/V       & K66  & $1.88^{+0.04}_{-0.05}$  &
                 $-0.111^{+0.040}_{-0.044}$  & $0.779^{+0.056}_{-0.057}$  &
                 $0.890^{+0.100}_{-0.097}$  & $4.65^{+0.04}_{-0.03}$  &
                 $4.46^{+0.08}_{-0.07}$  & $5.87^{+0.02}_{-0.02}$  &
                 $18.05^{+0.05}_{-0.05}$  & $3.51^{+0.09}_{-0.09}$ \\
        C007   & HAR02/V       & K66  & $1.98^{+0.04}_{-0.05}$  &
                 $0.133^{+0.039}_{-0.044}$  & $0.865^{+0.049}_{-0.046}$  &
                 $0.732^{+0.093}_{-0.085}$  & $4.79^{+0.04}_{-0.04}$  &
                 $4.35^{+0.09}_{-0.08}$  & $6.39^{+0.02}_{-0.02}$  &
                 $16.75^{+0.05}_{-0.05}$  & $3.86^{+0.07}_{-0.08}$ \\
        C011   & HAR02/V       & K66  & $2.00^{+0.04}_{-0.05}$  &
                 $0.099^{+0.039}_{-0.044}$  & $0.892^{+0.053}_{-0.051}$  &
                 $0.793^{+0.097}_{-0.091}$  & $4.54^{+0.04}_{-0.04}$  &
                 $4.14^{+0.09}_{-0.08}$  & $6.12^{+0.02}_{-0.02}$  &
                 $17.43^{+0.05}_{-0.05}$  & $3.54^{+0.08}_{-0.09}$ \\
        C017   & HAR02/V       & K66  & $1.76^{+0.04}_{-0.04}$  &
                 $0.326^{+0.037}_{-0.042}$  & $0.742^{+0.009}_{-0.007}$  &
                 $0.416^{+0.051}_{-0.045}$  & $4.43^{+0.06}_{-0.06}$  &
                 $3.80^{+0.11}_{-0.09}$  & $6.18^{+0.02}_{-0.02}$  &
                 $17.27^{+0.05}_{-0.05}$  & $3.90^{+0.00}_{-0.01}$ \\
        C021   & HAR02/V       & K66  & $1.94^{+0.04}_{-0.05}$  &
                 $0.066^{+0.039}_{-0.044}$  & $0.828^{+0.051}_{-0.049}$  &
                 $0.762^{+0.095}_{-0.088}$  & $4.60^{+0.04}_{-0.04}$  &
                 $4.23^{+0.09}_{-0.08}$  & $6.08^{+0.02}_{-0.02}$  &
                 $17.52^{+0.05}_{-0.05}$  & $3.63^{+0.08}_{-0.08}$ \\
        C022   & HAR02/V       & K66  & $1.65^{+0.05}_{-0.05}$  &
                 $0.018^{+0.038}_{-0.043}$  & $0.567^{+0.029}_{-0.024}$  &
                 $0.549^{+0.072}_{-0.062}$  & $4.73^{+0.05}_{-0.05}$  &
                 $4.41^{+0.10}_{-0.09}$  & $5.97^{+0.02}_{-0.02}$  &
                 $17.80^{+0.05}_{-0.05}$  & $4.04^{+0.03}_{-0.04}$ \\
        C023   & HAR02/V       & K66  & $1.62^{+0.05}_{-0.05}$  &
                 $-0.079^{+0.039}_{-0.043}$  & $0.517^{+0.035}_{-0.030}$  &
                 $0.596^{+0.078}_{-0.069}$  & $5.25^{+0.05}_{-0.05}$  &
                 $5.03^{+0.09}_{-0.08}$  & $6.34^{+0.02}_{-0.02}$  &
                 $16.89^{+0.05}_{-0.05}$  & $4.50^{+0.04}_{-0.05}$ \\
        C025   & HAR02/V       & K66  & $1.88^{+0.04}_{-0.05}$  &
                 $-0.022^{+0.039}_{-0.044}$  & $0.771^{+0.053}_{-0.051}$  &
                 $0.793^{+0.097}_{-0.091}$  & $4.49^{+0.04}_{-0.04}$  &
                 $4.20^{+0.09}_{-0.08}$  & $5.82^{+0.02}_{-0.02}$  &
                 $18.18^{+0.05}_{-0.05}$  & $3.48^{+0.08}_{-0.09}$ \\
        C029   & HAR02/V       & K66  & $1.94^{+0.04}_{-0.05}$  &
                 $0.059^{+0.039}_{-0.044}$  & $0.821^{+0.051}_{-0.049}$  &
                 $0.762^{+0.095}_{-0.088}$  & $4.52^{+0.04}_{-0.04}$  &
                 $4.16^{+0.09}_{-0.08}$  & $5.99^{+0.02}_{-0.02}$  &
                 $17.74^{+0.05}_{-0.05}$  & $3.55^{+0.08}_{-0.08}$ \\
        C031   & HAR02/V       & K66  & $1.70^{+0.05}_{-0.05}$  &
                 $-0.060^{+0.039}_{-0.044}$  & $0.588^{+0.041}_{-0.036}$  &
                 $0.647^{+0.085}_{-0.075}$  & $4.76^{+0.05}_{-0.04}$  &
                 $4.51^{+0.09}_{-0.08}$  & $5.91^{+0.02}_{-0.02}$  &
                 $17.94^{+0.05}_{-0.05}$  & $3.94^{+0.05}_{-0.06}$ \\
        C032   & HAR02/V       & K66  & $1.81^{+0.03}_{-0.04}$  &
                 $-0.266^{+0.040}_{-0.044}$  & $0.724^{+0.055}_{-0.060}$  &
                 $0.990^{+0.099}_{-0.100}$  & $4.87^{+0.04}_{-0.03}$  &
                 $4.84^{+0.08}_{-0.07}$  & $5.86^{+0.02}_{-0.02}$  &
                 $18.08^{+0.05}_{-0.05}$  & $3.61^{+0.10}_{-0.09}$ \\
        C037   & HAR02/V       & K66  & $1.62^{+0.04}_{-0.05}$  &
                 $-0.256^{+0.039}_{-0.044}$  & $0.507^{+0.051}_{-0.049}$  &
                 $0.762^{+0.095}_{-0.088}$  & $5.00^{+0.04}_{-0.04}$  &
                 $4.95^{+0.09}_{-0.08}$  & $5.85^{+0.02}_{-0.02}$  &
                 $18.11^{+0.05}_{-0.05}$  & $4.04^{+0.08}_{-0.08}$ \\
        C040   & HAR02/V       & K66  & $1.98^{+0.05}_{-0.05}$  &
                 $0.282^{+0.039}_{-0.043}$  & $0.878^{+0.035}_{-0.030}$  &
                 $0.596^{+0.078}_{-0.069}$  & $3.84^{+0.05}_{-0.05}$  &
                 $3.25^{+0.09}_{-0.08}$  & $5.64^{+0.02}_{-0.02}$  &
                 $18.62^{+0.05}_{-0.05}$  & $3.09^{+0.04}_{-0.05}$ \\
        C041   & HAR02/V       & K66  & $1.75^{+0.04}_{-0.05}$  &
                 $-0.124^{+0.039}_{-0.044}$  & $0.639^{+0.051}_{-0.049}$  &
                 $0.762^{+0.095}_{-0.088}$  & $4.71^{+0.04}_{-0.04}$  &
                 $4.52^{+0.09}_{-0.08}$  & $5.81^{+0.02}_{-0.02}$  &
                 $18.19^{+0.05}_{-0.05}$  & $3.74^{+0.08}_{-0.08}$ \\
        C044   & HAR02/V       & K66  & $1.62^{+0.04}_{-0.05}$  &
                 $0.070^{+0.038}_{-0.043}$  & $0.557^{+0.020}_{-0.016}$  &
                 $0.487^{+0.062}_{-0.054}$  & $4.50^{+0.06}_{-0.05}$  &
                 $4.13^{+0.10}_{-0.09}$  & $5.80^{+0.02}_{-0.02}$  &
                 $18.23^{+0.05}_{-0.05}$  & $3.89^{+0.01}_{-0.02}$ \\
        C100   & HAR02/V       & K66  & $1.96^{+0.05}_{-0.05}$  &
                 $0.185^{+0.039}_{-0.044}$  & $0.846^{+0.043}_{-0.038}$  &
                 $0.661^{+0.086}_{-0.077}$  & $3.79^{+0.05}_{-0.04}$  &
                 $3.30^{+0.09}_{-0.08}$  & $5.45^{+0.02}_{-0.02}$  &
                 $19.11^{+0.05}_{-0.05}$  & $2.96^{+0.06}_{-0.07}$ \\
        C101   & HAR02/V       & K66  & $1.69^{+0.04}_{-0.04}$  &
                 $0.280^{+0.037}_{-0.042}$  & $0.680^{+0.007}_{-0.005}$  &
                 $0.401^{+0.049}_{-0.043}$  & $3.35^{+0.06}_{-0.06}$  &
                 $2.76^{+0.11}_{-0.09}$  & $4.99^{+0.02}_{-0.02}$  &
                 $20.24^{+0.05}_{-0.05}$  & $2.83^{+0.01}_{-0.01}$ \\
        C102   & HAR02/V       & K66  & $2.21^{+0.03}_{-0.04}$  &
                 $0.037^{+0.040}_{-0.045}$  & $1.159^{+0.049}_{-0.058}$  &
                 $1.122^{+0.094}_{-0.098}$  & $2.93^{+0.04}_{-0.03}$  &
                 $2.59^{+0.08}_{-0.07}$  & $4.62^{+0.02}_{-0.02}$  &
                 $21.18^{+0.05}_{-0.05}$  & $1.50^{+0.10}_{-0.08}$ \\
        C103   & HAR02/V       & K66  & $1.70^{+0.05}_{-0.05}$  &
                 $-0.026^{+0.039}_{-0.044}$  & $0.595^{+0.038}_{-0.033}$  &
                 $0.621^{+0.082}_{-0.072}$  & $4.42^{+0.05}_{-0.04}$  &
                 $4.14^{+0.09}_{-0.08}$  & $5.63^{+0.02}_{-0.02}$  &
                 $18.67^{+0.05}_{-0.05}$  & $3.64^{+0.05}_{-0.06}$ \\
        C104   & HAR02/V       & K66  & $1.60^{+0.05}_{-0.05}$  &
                 $-0.006^{+0.038}_{-0.043}$  & $0.521^{+0.026}_{-0.021}$  &
                 $0.527^{+0.069}_{-0.059}$  & $4.12^{+0.05}_{-0.05}$  &
                 $3.82^{+0.10}_{-0.09}$  & $5.30^{+0.02}_{-0.02}$  &
                 $19.48^{+0.05}_{-0.05}$  & $3.46^{+0.02}_{-0.03}$ \\
        C105   & HAR02/V       & K66  & $2.13^{+0.05}_{-0.05}$  &
                 $0.527^{+0.038}_{-0.043}$  & $1.054^{+0.026}_{-0.021}$  &
                 $0.527^{+0.069}_{-0.059}$  & $2.06^{+0.05}_{-0.05}$  &
                 $1.23^{+0.10}_{-0.09}$  & $4.30^{+0.02}_{-0.02}$  &
                 $21.97^{+0.05}_{-0.05}$  & $1.40^{+0.02}_{-0.03}$ \\
        C106   & HAR02/V       & K66  & $1.28^{+0.04}_{-0.04}$  &
                 $-0.214^{+0.038}_{-0.043}$  & $0.236^{+0.014}_{-0.011}$  &
                 $0.450^{+0.057}_{-0.049}$  & $4.05^{+0.06}_{-0.05}$  &
                 $3.96^{+0.10}_{-0.09}$  & $4.75^{+0.02}_{-0.02}$  &
                 $20.85^{+0.05}_{-0.05}$  & $3.48^{+0.00}_{-0.01}$ \\
        G019   & HAR02/V       & K66  & $1.92^{+0.05}_{-0.05}$  &
                 $0.252^{+0.039}_{-0.043}$  & $0.824^{+0.032}_{-0.027}$  &
                 $0.572^{+0.075}_{-0.066}$  & $3.88^{+0.05}_{-0.05}$  &
                 $3.32^{+0.10}_{-0.09}$  & $5.60^{+0.02}_{-0.02}$  &
                 $18.72^{+0.05}_{-0.05}$  & $3.16^{+0.03}_{-0.04}$ \\
        G221   & HAR02/V       & K66  & $1.60^{+0.05}_{-0.05}$  &
                 $0.023^{+0.038}_{-0.043}$  & $0.530^{+0.023}_{-0.018}$  &
                 $0.506^{+0.065}_{-0.057}$  & $4.28^{+0.06}_{-0.05}$  &
                 $3.95^{+0.10}_{-0.09}$  & $5.50^{+0.02}_{-0.02}$  &
                 $18.98^{+0.05}_{-0.05}$  & $3.64^{+0.02}_{-0.03}$ \\
        G277   & HAR02/V       & K66  & $1.59^{+0.04}_{-0.04}$  &
                 $0.069^{+0.038}_{-0.043}$  & $0.537^{+0.017}_{-0.013}$  &
                 $0.468^{+0.059}_{-0.051}$  & $4.33^{+0.06}_{-0.05}$  &
                 $3.95^{+0.10}_{-0.09}$  & $5.61^{+0.02}_{-0.02}$  &
                 $18.72^{+0.05}_{-0.05}$  & $3.73^{+0.01}_{-0.01}$ \\
        G293   & HAR02/V       & K66  & $1.59^{+0.05}_{-0.05}$  &
                 $-0.193^{+0.039}_{-0.044}$  & $0.482^{+0.044}_{-0.039}$  &
                 $0.674^{+0.088}_{-0.078}$  & $4.68^{+0.05}_{-0.04}$  &
                 $4.57^{+0.09}_{-0.08}$  & $5.59^{+0.02}_{-0.02}$  &
                 $18.75^{+0.05}_{-0.05}$  & $3.83^{+0.06}_{-0.07}$ \\
        G302   & HAR02/V       & K66  & $1.57^{+0.05}_{-0.05}$  &
                 $-0.022^{+0.038}_{-0.043}$  & $0.495^{+0.024}_{-0.020}$  &
                 $0.517^{+0.067}_{-0.058}$  & $4.43^{+0.06}_{-0.05}$  &
                 $4.15^{+0.10}_{-0.09}$  & $5.57^{+0.02}_{-0.02}$  &
                 $18.82^{+0.05}_{-0.05}$  & $3.78^{+0.02}_{-0.03}$ \\
\hline
\end{tabular}

\medskip
  A machine-readable version of Table \ref{tab:har02phot} is
  available online
  (http://www.astro.keele.ac.uk/$\sim$dem/clusters.html)
  or upon request from the first author.

\end{minipage}
\end{table*}



\begin{table*}
\begin{minipage}{170mm}
\scriptsize
\caption{Derived dynamical parameters from \citet{king66} model fits to 27
         GCs from
         \citet[][$\equiv$~\citetalias{har02}]{har02} \label{tab:har02mass}}
\begin{tabular}{@{}lcclrrrrrrrrr}
\hline
\multicolumn{1}{c}{Name} &
\multicolumn{1}{c}{Detector} &
\multicolumn{1}{c}{$\Upsilon_V^{\rm pop}$} &
\multicolumn{1}{c}{Model} &
\multicolumn{1}{c}{$\log\,M_{\rm tot}$} &
\multicolumn{1}{c}{$\log\,E_b$} &
\multicolumn{1}{c}{$\log\,\Sigma_0$} &
\multicolumn{1}{c}{$\log\,\rho_0$ } &
\multicolumn{1}{c}{$\log\,\Sigma_h$} &
\multicolumn{1}{c}{$\log\,\sigma_{{\rm p},0}$} &
\multicolumn{1}{c}{$\log\,v_{{\rm esc},0}$} &
\multicolumn{1}{c}{$\log\,t_{\rm rh}$} &
\multicolumn{1}{c}{$\log\,f_0$} \\
 &
 &
\multicolumn{1}{c}{[$M_\odot\,L_{\odot,V}^{-1}$]} &
 &
\multicolumn{1}{c}{[$M_\odot$]} &
\multicolumn{1}{c}{[erg]} &
\multicolumn{1}{c}{[$M_\odot\,{\rm pc}^{-2}$]} &
\multicolumn{1}{c}{[$M_\odot\,{\rm pc}^{-3}$]} &
\multicolumn{1}{c}{[$M_\odot\,{\rm pc}^{-2}$]} &
\multicolumn{1}{c}{[km~s$^{-1}$]} &
\multicolumn{1}{c}{[km~s$^{-1}$]} &
\multicolumn{1}{c}{[yr]} &
~~   \\
\multicolumn{1}{c}{(1)}  &
\multicolumn{1}{c}{(2)}  &
\multicolumn{1}{c}{(3)}  &
\multicolumn{1}{c}{(4)}  &
\multicolumn{1}{c}{(5)}  &
\multicolumn{1}{c}{(6)}  &
\multicolumn{1}{c}{(7)}  &
\multicolumn{1}{c}{(8)}  &
\multicolumn{1}{c}{(9)}  &
\multicolumn{1}{c}{(10)} &
\multicolumn{1}{c}{(11)} &
\multicolumn{1}{c}{(12)} &
\multicolumn{1}{c}{(13)} \\
\hline
        C002   & HAR02/V  & $2.013^{+0.237}_{-0.234}$       & K66  &
                 $6.18^{+0.05}_{-0.06}$  & $52.03^{+0.07}_{-0.08}$  &
                 $4.95^{+0.06}_{-0.06}$  & $4.76^{+0.10}_{-0.09}$  &
                 $3.82^{+0.11}_{-0.11}$  & $1.159^{+0.024}_{-0.027}$  &
                 $1.783^{+0.024}_{-0.027}$  & $9.74^{+0.10}_{-0.10}$  &
                 $0.069^{+0.097}_{-0.090}$ \\
        C007   & HAR02/V  & $1.998^{+0.237}_{-0.234}$       & K66  &
                 $6.69^{+0.05}_{-0.06}$  & $52.95^{+0.07}_{-0.08}$  &
                 $5.09^{+0.07}_{-0.07}$  & $4.65^{+0.10}_{-0.10}$  &
                 $4.16^{+0.09}_{-0.10}$  & $1.350^{+0.024}_{-0.027}$  &
                 $1.963^{+0.025}_{-0.028}$  & $10.09^{+0.09}_{-0.08}$  &
                 $-0.615^{+0.095}_{-0.086}$ \\
        C011   & HAR02/V  & $3.078^{+0.458}_{-0.450}$       & K66  &
                 $6.61^{+0.06}_{-0.07}$  & $52.77^{+0.09}_{-0.10}$  &
                 $5.03^{+0.07}_{-0.08}$  & $4.63^{+0.10}_{-0.10}$  &
                 $4.03^{+0.10}_{-0.11}$  & $1.304^{+0.030}_{-0.034}$  &
                 $1.922^{+0.030}_{-0.035}$  & $10.11^{+0.09}_{-0.09}$  &
                 $-0.499^{+0.097}_{-0.091}$ \\
        C017   & HAR02/V  & $1.904^{+0.242}_{-0.241}$       & K66  &
                 $6.46^{+0.06}_{-0.06}$  & $52.60^{+0.08}_{-0.09}$  &
                 $4.71^{+0.08}_{-0.08}$  & $4.08^{+0.12}_{-0.11}$  &
                 $4.18^{+0.05}_{-0.06}$  & $1.261^{+0.028}_{-0.031}$  &
                 $1.842^{+0.031}_{-0.033}$  & $9.81^{+0.03}_{-0.04}$  &
                 $-0.928^{+0.083}_{-0.076}$ \\
        C021   & HAR02/V  & $2.053^{+0.240}_{-0.236}$       & K66  &
                 $6.40^{+0.05}_{-0.06}$  & $52.41^{+0.07}_{-0.08}$  &
                 $4.91^{+0.06}_{-0.07}$  & $4.54^{+0.10}_{-0.09}$  &
                 $3.94^{+0.09}_{-0.10}$  & $1.226^{+0.024}_{-0.027}$  &
                 $1.841^{+0.025}_{-0.027}$  & $9.91^{+0.09}_{-0.09}$  &
                 $-0.356^{+0.095}_{-0.087}$ \\
        C022   & HAR02/V  & $1.978^{+0.237}_{-0.234}$       & K66  &
                 $6.27^{+0.05}_{-0.06}$  & $52.38^{+0.07}_{-0.08}$  &
                 $5.03^{+0.07}_{-0.07}$  & $4.71^{+0.11}_{-0.10}$  &
                 $4.34^{+0.06}_{-0.07}$  & $1.262^{+0.025}_{-0.028}$  &
                 $1.860^{+0.027}_{-0.030}$  & $9.46^{+0.06}_{-0.05}$  &
                 $-0.305^{+0.089}_{-0.081}$ \\
        C023   & HAR02/V  & $2.777^{+0.401}_{-0.393}$       & K66  &
                 $6.78^{+0.06}_{-0.07}$  & $53.46^{+0.08}_{-0.10}$  &
                 $5.70^{+0.08}_{-0.08}$  & $5.47^{+0.11}_{-0.11}$  &
                 $4.95^{+0.07}_{-0.08}$  & $1.549^{+0.029}_{-0.033}$  &
                 $2.151^{+0.031}_{-0.035}$  & $9.61^{+0.07}_{-0.06}$  &
                 $-0.395^{+0.092}_{-0.085}$ \\
        C025   & HAR02/V  & $2.911^{+0.430}_{-0.422}$       & K66  &
                 $6.28^{+0.06}_{-0.07}$  & $52.24^{+0.08}_{-0.10}$  &
                 $4.95^{+0.07}_{-0.08}$  & $4.67^{+0.10}_{-0.10}$  &
                 $3.94^{+0.10}_{-0.11}$  & $1.202^{+0.030}_{-0.034}$  &
                 $1.819^{+0.030}_{-0.034}$  & $9.79^{+0.09}_{-0.09}$  &
                 $-0.155^{+0.097}_{-0.091}$ \\
        C029   & HAR02/V  & $2.833^{+0.413}_{-0.405}$       & K66  &
                 $6.45^{+0.06}_{-0.07}$  & $52.51^{+0.08}_{-0.09}$  &
                 $4.97^{+0.07}_{-0.08}$  & $4.61^{+0.10}_{-0.10}$  &
                 $4.01^{+0.10}_{-0.11}$  & $1.254^{+0.030}_{-0.034}$  &
                 $1.869^{+0.030}_{-0.034}$  & $9.93^{+0.09}_{-0.09}$  &
                 $-0.370^{+0.097}_{-0.090}$ \\
        C031   & HAR02/V  & $3.112^{+0.462}_{-0.454}$       & K66  &
                 $6.41^{+0.06}_{-0.07}$  & $52.65^{+0.09}_{-0.10}$  &
                 $5.25^{+0.08}_{-0.08}$  & $5.00^{+0.11}_{-0.11}$  &
                 $4.43^{+0.08}_{-0.09}$  & $1.334^{+0.030}_{-0.034}$  &
                 $1.940^{+0.031}_{-0.035}$  & $9.56^{+0.08}_{-0.07}$  &
                 $-0.217^{+0.094}_{-0.087}$ \\
        C032   & HAR02/V  & $3.064^{+0.455}_{-0.447}$       & K66  &
                 $6.34^{+0.06}_{-0.07}$  & $52.44^{+0.09}_{-0.10}$  &
                 $5.36^{+0.07}_{-0.07}$  & $5.32^{+0.10}_{-0.10}$  &
                 $4.10^{+0.12}_{-0.11}$  & $1.285^{+0.031}_{-0.035}$  &
                 $1.915^{+0.030}_{-0.034}$  & $9.74^{+0.10}_{-0.10}$  &
                 $0.255^{+0.100}_{-0.094}$ \\
        C037   & HAR02/V  & $2.253^{+0.270}_{-0.264}$       & K66  &
                 $6.20^{+0.05}_{-0.06}$  & $52.33^{+0.07}_{-0.08}$  &
                 $5.35^{+0.06}_{-0.07}$  & $5.31^{+0.10}_{-0.09}$  &
                 $4.39^{+0.09}_{-0.10}$  & $1.288^{+0.025}_{-0.027}$  &
                 $1.903^{+0.025}_{-0.028}$  & $9.34^{+0.09}_{-0.09}$  &
                 $0.225^{+0.095}_{-0.087}$ \\
        C040   & HAR02/V  & $2.138^{+0.251}_{-0.245}$       & K66  &
                 $5.97^{+0.05}_{-0.06}$  & $51.49^{+0.07}_{-0.08}$  &
                 $4.17^{+0.07}_{-0.07}$  & $3.58^{+0.11}_{-0.10}$  &
                 $3.42^{+0.06}_{-0.07}$  & $0.965^{+0.024}_{-0.027}$  &
                 $1.567^{+0.026}_{-0.028}$  & $9.80^{+0.07}_{-0.06}$  &
                 $-0.533^{+0.090}_{-0.082}$ \\
        C041   & HAR02/V  & $2.990^{+0.444}_{-0.436}$       & K66  &
                 $6.29^{+0.06}_{-0.07}$  & $52.38^{+0.09}_{-0.10}$  &
                 $5.18^{+0.07}_{-0.08}$  & $5.00^{+0.11}_{-0.10}$  &
                 $4.21^{+0.10}_{-0.11}$  & $1.267^{+0.030}_{-0.034}$  &
                 $1.882^{+0.031}_{-0.035}$  & $9.59^{+0.09}_{-0.09}$  &
                 $-0.018^{+0.097}_{-0.090}$ \\
        C044   & HAR02/V  & $1.915^{+0.241}_{-0.239}$       & K66  &
                 $6.08^{+0.06}_{-0.06}$  & $52.01^{+0.08}_{-0.09}$  &
                 $4.78^{+0.08}_{-0.08}$  & $4.41^{+0.11}_{-0.11}$  &
                 $4.17^{+0.05}_{-0.06}$  & $1.167^{+0.027}_{-0.030}$  &
                 $1.758^{+0.029}_{-0.032}$  & $9.36^{+0.05}_{-0.04}$  &
                 $-0.318^{+0.086}_{-0.079}$ \\
        C100   & HAR02/V  & $1.898^{+0.244}_{-0.243}$       & K66  &
                 $5.73^{+0.06}_{-0.06}$  & $51.03^{+0.07}_{-0.08}$  &
                 $4.07^{+0.07}_{-0.07}$  & $3.58^{+0.10}_{-0.10}$  &
                 $3.24^{+0.08}_{-0.09}$  & $0.867^{+0.026}_{-0.030}$  &
                 $1.475^{+0.027}_{-0.031}$  & $9.65^{+0.08}_{-0.07}$  &
                 $-0.239^{+0.093}_{-0.085}$ \\
        C101   & HAR02/V  & $2.409^{+0.305}_{-0.298}$       & K66  &
                 $5.38^{+0.06}_{-0.06}$  & $50.48^{+0.08}_{-0.09}$  &
                 $3.73^{+0.08}_{-0.08}$  & $3.15^{+0.12}_{-0.11}$  &
                 $3.22^{+0.05}_{-0.06}$  & $0.746^{+0.028}_{-0.030}$  &
                 $1.327^{+0.031}_{-0.033}$  & $9.26^{+0.03}_{-0.03}$  &
                 $-0.323^{+0.083}_{-0.075}$ \\
        C102   & HAR02/V  & $2.217^{+0.263}_{-0.258}$       & K66  &
                 $4.96^{+0.05}_{-0.06}$  & $49.26^{+0.07}_{-0.08}$  &
                 $3.28^{+0.06}_{-0.06}$  & $2.94^{+0.09}_{-0.09}$  &
                 $1.85^{+0.11}_{-0.09}$  & $0.395^{+0.025}_{-0.027}$  &
                 $1.037^{+0.024}_{-0.027}$  & $9.81^{+0.09}_{-0.10}$  &
                 $0.542^{+0.099}_{-0.093}$ \\
        C103   & HAR02/V  & $3.027^{+0.450}_{-0.442}$       & K66  &
                 $6.11^{+0.06}_{-0.07}$  & $52.04^{+0.09}_{-0.10}$  &
                 $4.90^{+0.08}_{-0.08}$  & $4.62^{+0.11}_{-0.11}$  &
                 $4.12^{+0.08}_{-0.09}$  & $1.177^{+0.030}_{-0.034}$  &
                 $1.781^{+0.031}_{-0.036}$  & $9.45^{+0.07}_{-0.07}$  &
                 $-0.128^{+0.093}_{-0.086}$ \\
        C104   & HAR02/V  & $1.896^{+0.244}_{-0.244}$       & K66  &
                 $5.58^{+0.06}_{-0.06}$  & $51.05^{+0.08}_{-0.09}$  &
                 $4.40^{+0.08}_{-0.08}$  & $4.10^{+0.11}_{-0.11}$  &
                 $3.74^{+0.06}_{-0.07}$  & $0.938^{+0.027}_{-0.030}$  &
                 $1.533^{+0.029}_{-0.032}$  & $9.10^{+0.05}_{-0.05}$  &
                 $0.066^{+0.088}_{-0.081}$ \\
        C105   & HAR02/V  & $2.827^{+0.412}_{-0.404}$       & K66  &
                 $4.76^{+0.06}_{-0.07}$  & $48.87^{+0.09}_{-0.10}$  &
                 $2.52^{+0.08}_{-0.08}$  & $1.68^{+0.11}_{-0.11}$  &
                 $1.85^{+0.06}_{-0.07}$  & $0.261^{+0.030}_{-0.034}$  &
                 $0.867^{+0.031}_{-0.035}$  & $9.58^{+0.05}_{-0.05}$  &
                 $-0.323^{+0.090}_{-0.082}$ \\
        C106   & HAR02/V  & $2.709^{+0.383}_{-0.375}$       & K66  &
                 $5.18^{+0.06}_{-0.07}$  & $50.54^{+0.09}_{-0.10}$  &
                 $4.48^{+0.08}_{-0.08}$  & $4.39^{+0.12}_{-0.11}$  &
                 $3.91^{+0.06}_{-0.07}$  & $0.875^{+0.030}_{-0.033}$  &
                 $1.462^{+0.032}_{-0.035}$  & $8.52^{+0.04}_{-0.04}$  &
                 $0.540^{+0.086}_{-0.079}$ \\
        G019   & HAR02/V  & $1.904^{+0.242}_{-0.241}$       & K66  &
                 $5.88^{+0.06}_{-0.06}$  & $51.36^{+0.08}_{-0.09}$  &
                 $4.16^{+0.07}_{-0.08}$  & $3.60^{+0.11}_{-0.10}$  &
                 $3.44^{+0.06}_{-0.07}$  & $0.945^{+0.026}_{-0.030}$  &
                 $1.545^{+0.028}_{-0.031}$  & $9.68^{+0.06}_{-0.06}$  &
                 $-0.455^{+0.090}_{-0.082}$ \\
        G221   & HAR02/V  & $2.303^{+0.281}_{-0.274}$       & K66  &
                 $5.86^{+0.05}_{-0.06}$  & $51.60^{+0.07}_{-0.08}$  &
                 $4.64^{+0.08}_{-0.08}$  & $4.31^{+0.11}_{-0.11}$  &
                 $4.00^{+0.05}_{-0.06}$  & $1.073^{+0.026}_{-0.028}$  &
                 $1.666^{+0.028}_{-0.030}$  & $9.24^{+0.05}_{-0.05}$  &
                 $-0.128^{+0.087}_{-0.079}$ \\
        G277   & HAR02/V  & $1.994^{+0.237}_{-0.234}$       & K66  &
                 $5.90^{+0.05}_{-0.06}$  & $51.68^{+0.07}_{-0.08}$  &
                 $4.63^{+0.08}_{-0.08}$  & $4.25^{+0.11}_{-0.11}$  &
                 $4.03^{+0.05}_{-0.06}$  & $1.088^{+0.026}_{-0.028}$  &
                 $1.676^{+0.028}_{-0.030}$  & $9.26^{+0.04}_{-0.04}$  &
                 $-0.238^{+0.085}_{-0.078}$ \\
        G293   & HAR02/V  & $1.882^{+0.248}_{-0.248}$       & K66  &
                 $5.86^{+0.06}_{-0.06}$  & $51.68^{+0.08}_{-0.09}$  &
                 $4.95^{+0.07}_{-0.07}$  & $4.84^{+0.10}_{-0.10}$  &
                 $4.10^{+0.08}_{-0.09}$  & $1.120^{+0.027}_{-0.031}$  &
                 $1.728^{+0.028}_{-0.032}$  & $9.16^{+0.08}_{-0.07}$  &
                 $0.264^{+0.094}_{-0.086}$ \\
        G302   & HAR02/V  & $1.921^{+0.239}_{-0.238}$       & K66  &
                 $5.85^{+0.05}_{-0.06}$  & $51.62^{+0.08}_{-0.08}$  &
                 $4.71^{+0.08}_{-0.08}$  & $4.43^{+0.11}_{-0.11}$  &
                 $4.06^{+0.05}_{-0.06}$  & $1.086^{+0.026}_{-0.029}$  &
                 $1.679^{+0.028}_{-0.031}$  & $9.17^{+0.05}_{-0.05}$  &
                 $-0.050^{+0.088}_{-0.080}$ \\
\hline
\end{tabular}

\medskip
  A machine-readable version of Table \ref{tab:har02mass} is
  available online
  (http://www.astro.keele.ac.uk/$\sim$dem/clusters.html)
  or upon request from the first author.

\end{minipage}
\end{table*}



\begin{table*}
\begin{minipage}{85mm}
\scriptsize
\caption{Galactocentric radii and $\kappa$-space parameters from
         \citet{king66} model fits to 27 GCs from
         \citet[][$\equiv$~\citetalias{har02}]{har02} \label{tab:har02kappa}}
\begin{tabular}{@{}lcrlrrr}
\hline
\multicolumn{1}{c}{Name} &
\multicolumn{1}{c}{Detector} &
\multicolumn{1}{c}{$R_{\rm gc}$} &
\multicolumn{1}{c}{Model} &
\multicolumn{1}{c}{$\kappa_{m,1}$} &
\multicolumn{1}{c}{$\kappa_{m,2}$} &
\multicolumn{1}{c}{$\kappa_{m,3}$} \\
   &
   &
\multicolumn{1}{c}{[kpc]} &
   &
   &
~~ \\
\multicolumn{1}{c}{(1)} &
\multicolumn{1}{c}{(2)} &
\multicolumn{1}{c}{(3)} &
\multicolumn{1}{c}{(4)} &
\multicolumn{1}{c}{(5)} &
\multicolumn{1}{c}{(6)} &
\multicolumn{1}{c}{(7)} \\
\hline
        C002   & HAR02/V     & 14.20       & K66  &
                 $0.069^{+0.049}_{-0.052}$  & $4.971^{+0.120}_{-0.120}$  &
                 $0.416^{+0.017}_{-0.017}$ \\
        C007   & HAR02/V      & 9.20       & K66  &
                 $0.399^{+0.048}_{-0.050}$  & $5.372^{+0.097}_{-0.108}$  &
                 $0.388^{+0.016}_{-0.014}$ \\
        C011   & HAR02/V      & 6.60       & K66  &
                 $0.354^{+0.055}_{-0.059}$  & $5.213^{+0.116}_{-0.126}$  &
                 $0.399^{+0.017}_{-0.016}$ \\
        C017   & HAR02/V      & 6.20       & K66  &
                 $0.186^{+0.042}_{-0.045}$  & $5.365^{+0.064}_{-0.073}$  &
                 $0.345^{+0.005}_{-0.003}$ \\
        C021   & HAR02/V      & 7.20       & K66  &
                 $0.198^{+0.048}_{-0.050}$  & $5.106^{+0.102}_{-0.111}$  &
                 $0.394^{+0.017}_{-0.015}$ \\
        C022   & HAR02/V      & 5.70       & K66  &
                 $0.065^{+0.044}_{-0.045}$  & $5.564^{+0.067}_{-0.077}$  &
                 $0.359^{+0.011}_{-0.008}$ \\
        C023   & HAR02/V      & 5.80       & K66  &
                 $0.435^{+0.051}_{-0.053}$  & $6.318^{+0.083}_{-0.097}$  &
                 $0.366^{+0.013}_{-0.009}$ \\
        C025   & HAR02/V      & 8.50       & K66  &
                 $0.124^{+0.055}_{-0.058}$  & $5.110^{+0.116}_{-0.125}$  &
                 $0.399^{+0.017}_{-0.016}$ \\
        C029   & HAR02/V     & 21.00       & K66  &
                 $0.233^{+0.054}_{-0.057}$  & $5.184^{+0.111}_{-0.121}$  &
                 $0.393^{+0.017}_{-0.015}$ \\
        C031   & HAR02/V      & 6.10       & K66  &
                 $0.181^{+0.053}_{-0.056}$  & $5.694^{+0.093}_{-0.107}$  &
                 $0.374^{+0.014}_{-0.011}$ \\
        C032   & HAR02/V     & 12.50       & K66  &
                 $0.209^{+0.054}_{-0.060}$  & $5.323^{+0.133}_{-0.130}$  &
                 $0.433^{+0.015}_{-0.017}$ \\
        C037   & HAR02/V     & 11.90       & K66  &
                 $0.058^{+0.049}_{-0.050}$  & $5.652^{+0.103}_{-0.111}$  &
                 $0.393^{+0.017}_{-0.015}$ \\
        C040   & HAR02/V     & 23.20       & K66  &
                 $-0.136^{+0.045}_{-0.045}$  & $4.445^{+0.073}_{-0.083}$  &
                 $0.366^{+0.012}_{-0.010}$ \\
        C041   & HAR02/V     & 23.20       & K66  &
                 $0.123^{+0.055}_{-0.058}$  & $5.440^{+0.112}_{-0.123}$  &
                 $0.394^{+0.017}_{-0.015}$ \\
        C044   & HAR02/V     & 20.20       & K66  &
                 $-0.077^{+0.043}_{-0.046}$  & $5.354^{+0.064}_{-0.073}$  &
                 $0.352^{+0.008}_{-0.005}$ \\
        C100   & HAR02/V      & 8.20       & K66  &
                 $-0.297^{+0.049}_{-0.051}$  & $4.230^{+0.088}_{-0.101}$  &
                 $0.376^{+0.015}_{-0.012}$ \\
        C101   & HAR02/V      & 6.20       & K66  &
                 $-0.585^{+0.041}_{-0.044}$  & $4.182^{+0.064}_{-0.071}$  &
                 $0.344^{+0.004}_{-0.002}$ \\
        C102   & HAR02/V      & 5.40       & K66  &
                 $-0.743^{+0.044}_{-0.050}$  & $2.584^{+0.123}_{-0.110}$  &
                 $0.452^{+0.011}_{-0.014}$ \\
        C103   & HAR02/V      & 5.90       & K66  &
                 $-0.036^{+0.052}_{-0.056}$  & $5.306^{+0.089}_{-0.103}$  &
                 $0.370^{+0.013}_{-0.011}$ \\
        C104   & HAR02/V      & 8.90       & K66  &
                 $-0.426^{+0.045}_{-0.048}$  & $4.828^{+0.068}_{-0.080}$  &
                 $0.357^{+0.010}_{-0.007}$ \\
        C105   & HAR02/V      & 9.10       & K66  &
                 $-1.007^{+0.049}_{-0.052}$  & $2.517^{+0.076}_{-0.088}$  &
                 $0.357^{+0.009}_{-0.007}$ \\
        C106   & HAR02/V      & 9.20       & K66  &
                 $-0.717^{+0.046}_{-0.050}$  & $5.038^{+0.070}_{-0.079}$  &
                 $0.348^{+0.006}_{-0.004}$ \\
        G019   & HAR02/V      & 9.00       & K66  &
                 $-0.202^{+0.046}_{-0.048}$  & $4.467^{+0.073}_{-0.085}$  &
                 $0.363^{+0.011}_{-0.009}$ \\
        G221   & HAR02/V      & 9.40       & K66  &
                 $-0.230^{+0.043}_{-0.045}$  & $5.153^{+0.063}_{-0.072}$  &
                 $0.354^{+0.009}_{-0.006}$ \\
        G277   & HAR02/V      & 8.60       & K66  &
                 $-0.203^{+0.041}_{-0.043}$  & $5.187^{+0.060}_{-0.067}$  &
                 $0.350^{+0.007}_{-0.005}$ \\
        G293   & HAR02/V      & 9.50       & K66  &
                 $-0.197^{+0.050}_{-0.052}$  & $5.292^{+0.091}_{-0.105}$  &
                 $0.379^{+0.015}_{-0.012}$ \\
        G302   & HAR02/V      & 7.90       & K66  &
                 $-0.236^{+0.044}_{-0.046}$  & $5.225^{+0.065}_{-0.076}$  &
                 $0.355^{+0.009}_{-0.006}$ \\
\hline
\end{tabular}

\medskip
  A machine-readable version of Table \ref{tab:har02kappa} is
  available online
  (http://www.astro.keele.ac.uk/$\sim$dem/clusters.html)
  or upon request from the first author.

\end{minipage}
\end{table*}


\label{lastpage}

\end{document}